\pdfoutput=1

\documentclass[11pt,twoside,a4paper,cmspaper,final,collab]{cms-tdr}
\def\svnVersion{476910P}\def\cmsCernNoTag{CERN-EP-2018-229}\def\cmsCernDate{\today}\def\cmsMessage{Published in the Journal of Instrumentation as \href{http://dx.doi.org/10.1088/1748-0221/13/10/P10005}{\doi{10.1088/1748-0221/13/10/P10005.}}}

\begin{document}\cmsNoteHeader{TAU-16-003}

\hyphenation{had-ron-i-za-tion}
\hyphenation{cal-or-i-me-ter}
\hyphenation{de-vices}
\RCS$HeadURL: svn+ssh://svn.cern.ch/reps/tdr2/papers/TAU-16-003/trunk/TAU-16-003.tex $
\RCS$Id: TAU-16-003.tex 476112 2018-09-25 16:07:09Z anayak $

\newlength\cmsFigWidth
\setlength\cmsFigWidth{0.4\textwidth}
\providecommand{\cmsLeft}{left\xspace}
\providecommand{\cmsRight}{right\xspace}

\newcommand{\Plepton}{\ensuremath{\ell}}
\newcommand{\Ppizero}{\Pgpz}
\newcommand{\Pnut}{\ensuremath{\nu_{\Pgt}}}
\renewcommand{\Pgg}{\ensuremath{\gamma}\xspace}
\newcommand{\Pggx}{\ensuremath{\gamma^{*}}\xspace}
\newcommand{\PWprime}{\ensuremath{\PW'}\xspace}
\newcommand{\PZprime}{\cPZpr}
\newcommand{\oneProngZeroPizero}{\ensuremath{\mathrm{h}^{\pm}}\xspace}
\newcommand{\oneProngOnePizero}{\ensuremath{\mathrm{h}^{\pm}\Ppizero}\xspace}
\newcommand{\oneProngTwoPizero}{\ensuremath{\mathrm{h}^{\pm}\Ppizero\Ppizero}\xspace}
\newcommand{\threeProngZeroPizero}{\ensuremath{\mathrm{h}^{\pm}\mathrm{h}^{\mp}\mathrm{h}^{\pm}}\xspace}
\newcommand{\Pgtm}{\ensuremath{\tau}^{-}}
\newcommand{\hminus}{\ensuremath{\mathrm{h}^{-}}\xspace}
\newcommand{\hplus}{\ensuremath{\mathrm{h}^{+}}\xspace}
\newcommand{\mvis}{\ensuremath{m_\text{vis}}}

\cmsNoteHeader{TAU-16-003}
\title{Performance of reconstruction and identification of \texorpdfstring{$\Pgt$}{tau} leptons decaying to hadrons and \texorpdfstring{$\Pgngt$}{tau neutrino} in $\Pp\Pp$ collisions at $\sqrt{s}=13\TeV$}

\date{\today}

\abstract{
The algorithm developed by the CMS Collaboration to reconstruct and identify $\Pgt$
leptons produced in proton-proton collisions at $\sqrt{s}=7$ and
8\TeV, via their decays to hadrons and a neutrino, has been
significantly improved.
The changes include a revised reconstruction of $\Pgpz$ candidates, and
improvements in multivariate discriminants to separate $\Pgt$ leptons
from jets and electrons.
The algorithm is extended to reconstruct
$\Pgt$ leptons in highly Lorentz-boosted pair production, and in
the high-level trigger.
The performance of the algorithm is studied using  proton-proton
collisions
recorded during 2016 at $\sqrt{s}=13\TeV$, corresponding to an integrated
luminosity of 35.9\fbinv. The performance is evaluated in terms of the
efficiency for a genuine $\Pgt$ lepton to pass the identification
criteria and of the probabilities for jets, electrons, and muons to be
misidentified as $\Pgt$ leptons.
The results are found to be very close to those expected from Monte
Carlo simulation.
}

\hypersetup{%
pdfauthor={CMS Collaboration},%
pdftitle={Performance of reconstruction and identification of tau leptons decaying to hadrons and tau neutrino in pp collisions at sqrt(s)=13 TeV},%
pdfsubject={CMS tau lepton reconstruction},%
pdfkeywords={CMS, physics, software, tau lepton, reconstruction, performance}}

\maketitle

\tableofcontents
\clearpage
\section{Introduction}
\label{sec:introduction}

Searches for new phenomena that consider signatures with $\Pgt$ leptons have gained great interest in proton-proton ($\Pp\Pp$) collisions at the CERN LHC.
The most prominent one among these is the decay of Higgs bosons (H) to pairs of $\Pgt$ leptons, which constitutes an especially sensitive channel for probing Higgs boson couplings to fermions.
The observation of the standard model (SM) Higgs boson decaying to a pair of $\Pgt$ leptons has recently been reported~\cite{Khachatryan:2016vau,Sirunyan:2017khh}.
Moreover, searches with $\Pgt$ leptons in the final state have high sensitivity to the production of both neutral and charged Higgs bosons expected in the minimal supersymmetric standard model (MSSM)~\cite{Fayet:1974pd,Fayet:1977yc}, in which enhancements in the couplings to $\Pgt$ leptons can be substantial at large tan$\beta$, where tan$\beta$ is the ratio of vacuum expectation values of the two Higgs doublets in the MSSM. Examples of such searches can be found in Refs.~\cite{HIG-13-021,HIG-14-023,Aad:2012cfr}.
In addition, searches for particles beyond the SM, such as new or heavy Higgs bosons~\cite{HIG-14-034,HIG-14-040,HIG-14-005,Aaboud:2017sjh}, leptoquarks~\cite{EXO-14-008}, supersymmetric particles~\cite{Khachatryan:2016trj,Chatrchyan:2013dsa,Aaboud:2017nhr,Aaboud:2018kya}, or gauge bosons~\cite{EXO-11-031,EXO-12-011,Aaboud:2018vgh} benefit significantly from any improvements made in $\Pgt$ lepton reconstruction and identification.

The $\Pgt$ lepton, with a mass of $m_{\Pgt} = 1776.86\pm0.12\MeV$~\cite{PDG}, is the only lepton sufficiently massive to decay into hadrons and a neutrino.
About one third of the time, $\Pgt$ leptons decay into an electron or a muon, and two neutrinos.
The neutrinos escape undetected, but the ${\Pe}$ and ${\Pgm}$ are reconstructed and identified through the usual techniques available for such leptons~\cite{Khachatryan:2015hwa,Sirunyan:2018fpa}.
These decay final states are denoted as $\Pgt_{\Pe}$ and $\Pgt_{\Pgm}$, respectively.
Almost all the remaining decay final states of $\Pgt$ leptons contain hadrons, typically with a combination of charged and neutral mesons, and a $\Pnut$.

The decays of $\Pgt$ leptons into hadrons and neutrinos, denoted by $\tauh$, are reconstructed and identified using the hadrons-plus-strips (HPS) algorithm~\cite{Chatrchyan:2012zz,TAU-14-001}, which was developed and used in CMS when the LHC operated at $\sqrt{s}=7$ and 8\TeV.
The HPS algorithm reconstructs the $\tauh$ modes by combining information from charged hadrons, which are reconstructed using their associated tracks in the inner tracker, and $\Ppizero$ candidates, obtained by clustering photon and electron candidates from photon conversions in rectangular regions of pseudorapidity and azimuth, $\eta{\times}\phi$ regions, called ``strips''.
The major challenge in the identification of $\tauh$ is to distinguish these objects from quark and gluon jets, which are copiously produced in $\Pp\Pp$ collisions.
The primary method for reducing backgrounds from jets misidentified as $\tauh$ candidates exploits the fact that there are fewer particles present in $\tauh$ decays,
and that their energies are deposited in narrow regions of ($\eta$, $\phi$) compared to those from energetic quark or gluon jets.
In certain analyses, the misidentification (MisID) of electrons or muons as $\tauh$ candidates
can also constitute a sizeable background.

The $\tauh$ identification algorithm improved for analyzing data at $\sqrt{s}=13\TeV$ contains the following new features:
\begin{enumerate}
\item A modification of the strip reconstruction algorithm, to the so-called dynamic strip reconstruction, that changes the size of a strip in a dynamic way that collects the $\Ppizero$ decay products more effectively;
\item improvements in the multivariate-analysis (MVA) based discriminant~\cite{TAU-14-001} that reduces the background from jets, by combining information on isolation, lifetime of the $\Pgt$ lepton, and energy distribution in the shower; and
\item improvements in the MVA-based discriminant that suppresses electrons misidentified as $\tauh$ candidates.
\end{enumerate}

This paper is organized as follows. After a brief introduction of the CMS detector in Section~\ref{sec:detector}, we discuss the data and the event simulations used to evaluate the performance of the HPS algorithm in Section~\ref{sec:datasamples_and_MonteCarloSimulation}. The reconstruction and identification of physical objects (other than $\tauh$) is briefly described in Section~\ref{sec:Selection}.
Section~\ref{sec:tauId} describes the HPS algorithm used for 13\TeV data and its simulation.
The extended version of the algorithm used to reconstruct $\tauh$ pairs produced in topologies with high Lorentz-boosts is presented in Section~\ref{sec:boostedTauReco}, while the specialized version developed for trigger purposes is discussed in Section~\ref{sec:Tau_Trigger}.
The selection of events used to evaluate the performance of the $\tauh$ reconstruction algorithm, as well as systematic uncertainties common to all measurements are discussed in Section~\ref{sec:validation}.
The performance evaluation of the improved algorithm using selected data samples is given thereafter: Section~\ref{sec:Tau_identification_efficiency} describes the $\tauh$ identification efficiency, while Sections~\ref{sec:jetToTauFakeRate} and~\ref{sec:lToTauFakeRate} summarize the respective jet~$\mapsto\tauh$ and $\Pe/\Pgm\mapsto\tauh$ misidentification probabilities. The $\tauh$ energy scale is discussed in Section~\ref{sec:TauEnergyScale}. Finally, Section~\ref{sec:Tau_Trigger_Performance} presents the performance of $\tauh$ identification in the high-level trigger, and a brief summary in Section~\ref{sec:summary} concludes this paper.

\section{The CMS detector}
\label{sec:detector}

The central feature of the CMS apparatus is a superconducting solenoid of 6\unit{m} internal diameter,
providing a magnetic field of 3.8\unit{T}.
A silicon pixel and strip tracker,
a lead tungstate crystal electromagnetic calorimeter (ECAL), and a brass and scintillator hadron calorimeter (HCAL),
each composed of a barrel and two endcap sections, reside within the field of the solenoid.
Extensive forward calorimetry complements the coverage provided by the barrel and endcaps.
Muons are measured in gas-ionization detectors embedded in the steel flux-return yoke outside the solenoid.

The CMS tracker is a cylindrical detector, constructed from 1\,440 silicon-pixel and 15\,148 silicon-strip detector modules that cover the range of $\abs{\eta} < 2.5$.
Tracks of charged hadrons are reconstructed with typical efficiencies of 80--90\%, depending on transverse momentum (\pt) and $\eta$~\cite{tracker,Chatrchyan:2014fea}.
The silicon tracker presents a significant amount of material in front of the ECAL, mostly due to the mechanical structure, the associated services, and the cooling system.
A minimum of 0.4 radiation lengths ($X_{0}$) of material is present at $\abs{\eta} \approx 0$, which rises to ${\approx}2.0\,X_{0}$ at $\abs{\eta} \approx 1.4$, and decreases to ${\approx}1.3\,X_{0}$ at $\abs{\eta} \approx 2.5$.
Photons originating from $\Ppizero$ decays therefore have a high probability to convert into $\Pem\Pep$ pairs within the volume of the tracker.

The ECAL is a homogeneous and hermetic calorimeter made of PbWO$_4$ scintillating crystals.
It is composed of a central barrel, covering the region $\abs{\eta} < 1.48$,
and two endcaps, covering $1.48 < \abs{\eta} < 3.0$.
The small radiation length ($X_{0} = 0.89$\unit{cm}) and small Moli\`ere radius (2.3\unit{cm}) of the PbWO$_4$ crystals
provide a compact calorimeter with excellent two-shower separation.
The ECAL is ${>}25\,X_{0}$ thick.

The HCAL is a sampling calorimeter made of brass and plastic scintillator, with a coverage up to $\abs{\eta} = 3.0$.
The scintillation light is converted by wavelength-shifting fibres and channelled to photodetectors via clear fibres.
The thickness of the HCAL is in the range 7--11 interaction lengths, depending on $\eta$.

The muon detection system is made up of four planes of gas-ionization detectors, where each plane consists of several layers of aluminium drift tubes (DTs) in the barrel region and cathode strip chambers (CSCs) in the endcap region, complemented by resistive-plate chambers (RPCs) that are used only in the trigger.

A two-tiered trigger system~\cite{Khachatryan:2016bia} is employed to select interesting events from the LHC bunch crossing rate of up to 40\unit{MHz}. The first level~(L1), composed of custom-made hardware processors, uses information from the calorimeters and muon detectors to select events at a rate of $\approx$100\unit{kHz}, within a fixed time interval of less than 4\mus. The second level, known as the high-level trigger~(HLT), consists of a farm of processors running a version of the full
event reconstruction software, optimized for fast processing, and reduces the
event rate to $\approx$1\unit{kHz} before data storage.

A more detailed description of the CMS detector, together with a definition of the coordinate system and kinematic variables, can be found in Ref.~\cite{Chatrchyan:2008zzk}.

\section{Data and simulated events}
\label{sec:datasamples_and_MonteCarloSimulation}

The performance of $\tauh$ reconstruction and identification algorithms are evaluated in $\Pp\Pp$ collisions recorded by CMS during 2016 at $\sqrt{s}$ = 13\TeV, corresponding to an integrated luminosity of 35.9\fbinv.
The Monte Carlo (MC) simulated signal samples contain $\PH \to \Pgt\Pgt$, $\PZprime \to \Plepton\Plepton$, $\PWprime \to \Plepton\Pgn$, and $\cPZ/\Pggx \to \Plepton\Plepton$ events, where $\Plepton$ refers to $\Pe$, $\Pgm$, or $\Pgt$ leptons.
Simulated signal contributions from $\PH \to \Pgt\Pgt$, $\PZprime \to \Plepton\Plepton$ (with masses up to 4\TeV), $\PWprime \to \Plepton\Pgn$ (with masses up to 5.8\TeV), and MSSM $\PH \to\Pgt\Pgt$ (with masses up to 3.2\TeV) are used to optimize the identification of $\tauh$ candidates over a wide range of their $\pt$ values.
The $\PH \to \Pgt\Pgt$ events are generated at next-to-leading order (NLO) in perturbative quantum chromodynamics (QCD) using \POWHEG~v2~\cite{Nason:2004rx,Frixione:2007vw,Alioli:2010xd,Alioli:2008tz,Nason:2009ai}, while $\PZprime$ and $\PWprime$ boson events are generated using leading-order (LO) \PYTHIA~8.212~\cite{Pythia8}.
In simulation, the reconstructed $\tauh$ candidate is taken as matched to the generated $\tauh$ when both objects lie within a cone of $\Delta R = \sqrt{\smash[b]{(\Delta\eta)^{2} + (\Delta\phi)^{2}}} <0.3 $, where $\Delta\phi$ and $\Delta\eta$ are the distances respectively in $\phi$ and $\eta$ between the reconstructed and generated candidates.

The $\PW$+jets and $\cPZ/\Pggx \to \Plepton\Plepton$ events are generated at LO in perturbative QCD using \MGvATNLO v2.2.2~\cite{Alwall:2014hca} with the MLM jet merging scheme~\cite{Alwall:2007fs},
while the single top quark and $\cPqt\cPaqt$ events are generated at NLO in perturbative QCD using \POWHEG~\cite{Alioli:2009je,Re:2010bp,Campbell:2014kua}.
The diboson $\PW\PW$, $\PW\cPZ$, and $\cPZ\cPZ$ events are generated at NLO using \MGvATNLO with the FXFX jet merging scheme~\cite{Frederix:2012ps} or \POWHEG~\cite{Nason:2013ydw}, while events comprised uniquely of jets produced through the strong interaction, referred to as QCD multijet events, are generated at LO with \PYTHIA.
The \PYTHIA generator, with the CUETP8M1 underlying-event tune~\cite{Khachatryan:2015pea}, is used to model the parton shower and hadronization processes, as well as $\Pgt$ lepton decays in all events.
The $\cPZ/\Pggx \to \Plepton\Plepton$ and $\PW$+jets samples are normalized
according to cross sections computed at next-to-next-to-leading order (NNLO) in perturbative QCD accuracy~\cite{FEWZ,Melnikov:2006di,Gavin:2010az,Gavin:2012sy,Li:2012wna}, while the  $\ttbar$ sample is normalised to the cross section computed at NNLO supplemented by soft-gluon resummation with next-to-next-to-leading logarithmic accuracy~\cite{Czakon:2011xx,Czakon:2013goa}.
The cross sections for single top quark and diboson production are computed at NLO in perturbative QCD accuracy~\cite{Campbell:2011bn}.
The production of off-shell W bosons ($m_{\PW}>200\GeV$), with subsequent $\PW\to\Pgt\Pgn$ or $\PW\to\Pgm\Pgn$ decays, is simulated at LO with the \PYTHIA generator. The differential cross section is reweighted as a function of the invariant mass of the W boson decay products, incorporating NNLO QCD and NLO electroweak corrections~\cite{Bondarenko:2013nu,Gavin:2012sy,Andersen:2014efa}.
The \textsc{NNPDF3.0} parton distribution functions~\cite{Ball:2014uwa} are used in all the calculations.

Additional $\Pp\Pp$ collisions that overlap temporally the interactions of interest, referred to as pileup (PU),
 are generated using \PYTHIA, and overlaid on all MC events
according to the luminosity profile of the analyzed data.
The generated events are passed through a detailed simulation of the CMS detector based on \GEANTfour~\cite{geant},
and are reconstructed using the same CMS reconstruction software as used for data.

\section{Event reconstruction}
\label{sec:Selection}

The particles emerging from $\Pp\Pp$ collisions, such as charged and neutral hadrons, photons, electrons, and muons, are reconstructed and identified by combining the information from the CMS subdetectors using a particle-flow (PF) algorithm~\cite{CMS-PRF-14-001}.
These particles are further grouped to reconstruct higher-level objects, such as jets, missing transverse momentum, $\tauh$ candidates, and to quantify lepton isolation.

The trajectories of charged particles are reconstructed from their hits in the silicon tracker~\cite{Chatrchyan:2014fea}, and are referred to as tracks.

Electrons are reconstructed from their trajectories in the tracker and from clusters of energy deposition in the ECAL~\cite{Khachatryan:2015hwa}.
Electron identification relies on the energy distribution in the electromagnetic
shower and on other observables based on tracker and calorimeter information.
The selection criteria depend on the $\pt$ and $\abs{\eta}$ of the electron, and on a categorization according to observables
sensitive to the amount of bremsstrahlung emitted along the trajectory in the tracker.

Muons are reconstructed by combining tracks reconstructed in both the inner tracker and the outer muon spectrometer~\cite{Sirunyan:2018fpa}.
The identification of muons is based on the quality criteria of reconstructed muon tracks, and through requirements of
minimal energy deposition along the muon track in the calorimeters.

The isolation of individual electrons or muons ($I_\text{rel}^{\Pe/\Pgm}$) is measured relative to their transverse momenta $\pt^{\Pe/\Pgm}$
by summing over the scalar $\pt$ values of charged and neutral hadrons, as well as photons, in a cone of
$\Delta R<0.3$ for electrons or 0.4 for muons around the direction of the lepton at the interaction vertex:
\begin{linenomath}
\begin{equation}
\label{eq:relIso}
I_\text{rel}^{\Pe/\Pgm} = \left( \sum  \pt^\text{charged} + \max\left[ 0, \sum \pt^\text{neutral}
                                 +  \sum \pt^{\Pgg} - \pt^\text{PU}  \right] \right) /  \pt^{\Pe/\Pgm}.
\end{equation}
\end{linenomath}
The primary $\Pp\Pp$ interaction vertex is defined as the reconstructed vertex with largest value
of summed $\pt^2$ of jets, clustered using all tracks assigned to the vertex, and of the associated missing transverse momentum, taken as the negative vector sum of the \ptvec of those jets.
To suppress the contribution from PU, the charged hadrons are required to originate from the primary vertex.
The neutral contribution to the isolation
from PU (referred to as $\pt^\text{PU}$) is estimated through a jet area method~\cite{Cacciari:2007fd} for electrons. For muons, the $\pt^\text{PU}$ contribution is estimated using the sum of the scalar \pt of charged hadrons not originating from the primary vertex, scaled down by a factor of 0.5 (to accommodate the assumed ratio for the production of neutral and charged hadrons).

Jets are clustered from PF particles using the infrared and collinear-safe
anti-\kt algorithm~\cite{Cacciari:2008gp,Cacciari:2011ma} with a distance
parameter of 0.4. The jet momentum is defined by the vectorial sum of all
particle momenta in the jet. The simulation is found to provide results for jet \pt within 5 to
10\% of their true values over the whole \pt spectrum and detector acceptance.
To suppress contributions from PU, charged hadrons not originating from the
primary vertex are discarded, and an offset correction is applied to correct
the remaining PF contributions. Jet energy corrections are obtained from simulation to bring the measured response of jets to that of particle level jets on
average, and are confirmed with in situ measurements through momentum balance
in dijet, $\Pgg$+jet, \cPZ+jet, and multijet events~\cite{Khachatryan:2016kdb}.
The combined secondary vertex~v2 (CSVv2) $\cPqb$~tagging algorithm~\cite{Sirunyan:2017ezt} with a medium working point (WP) is used to identify jets originating from $\cPqb$~quarks.
The working point corresponds to an identification efficiency of about 70\% for $\cPqb$~quark jets with $\pt>30\GeV$, and a probability for light-quark
 or gluon jets to be misidentified as $\cPqb$~quarks of $\approx$1\%.

The missing transverse momentum vector, \ptvecmiss, is defined as
the projection of the negative vector sum of the momenta of all reconstructed particles in an event on the plane perpendicular to the beams.
The \ptvecmiss is corrected by propagating to it all the corrections made to the momenta of jets. Its magnitude is referred to as \ptmiss.

\section{Reconstruction and identification of \texorpdfstring{$\tauh$}{tau[h]}}
\label{sec:tauId}

The basic features of the HPS algorithm are identical to those used during the previous data taking at $\sqrt{s}=7$ and 8\TeV~\cite{TAU-14-001}, except for the improvements in $\Ppizero$ reconstruction described below in Section~\ref{sec:tauIdAlgorithm_dynamicStripReco}.
Sections~\ref{sec:tauIdDiscriminators_jets},~\ref{sec:antiElectronDiscrMVABased}, and~\ref{sec:antiMuonDiscrCutBased} discuss the discriminants used to distinguish reconstructed $\tauh$ candidates from jets, electrons, and muons, respectively.

\subsection{The hadrons-plus-strips algorithm}
\label{sec:tauIdAlgorithm_hps}

Starting from the constituents of reconstructed jets, the HPS algorithm reconstructs the different decays of the $\Pgt$ lepton into hadrons.
The final states include charged hadrons, as well as neutral pions, as shown in Table~\ref{tab:tau_decays}. The $\Ppizero$ mesons promptly decay into pairs of photons, which have a high probability of converting into $\Pep\Pem$ pairs as they traverse the tracker material.
The large magnetic field of the CMS solenoid leads to a spatial separation of the $\Pep\Pem$ pairs in the ($\phi$, $\eta$) plane.
To reconstruct the full energy of the neutral pions, the electron and photon candidates falling within a certain region of $\Delta\eta{\times}\Delta\phi$ are clustered together, with the resulting object referred to as a ``strip''. The strip momentum is defined by the vectorial sum of all its constituent momenta.
The procedure is described in Section~\ref{sec:tauIdAlgorithm_dynamicStripReco}, together with the improvements introduced to the previous algorithm.

\begin{table}[!hbtp]
        \centering
                \topcaption
                [Tau lepton decays and their branching fractions.]
                {Weak decays of $\Pgt$ leptons and their branching fractions ($\mathcal{B}$) in \%~\cite{PDG} are given, rounded to one decimal place. Also, where appropriate, we indicate the known intermediate resonances of all the listed hadrons. Charged hadrons are denoted by the symbol \oneProngZeroPizero. Although for simplicity we show just $\Pgtm$ decays in the table, the values are also valid for the charge-conjugate processes.}
                \begin{tabular}{llccc}
                        & Decay mode & Resonance & \multicolumn{2}{c}{$\mathcal{B}$ (\%)} \\
                        \hline
                        \multicolumn{2}{l}{Leptonic decays} & & 35.2 & \\
                        & $\Pgtm\to\Pem\Pagne\Pgngt$ & & & 17.8 \\
                        & $\Pgtm\to\Pgmm\Pagngm\Pgngt$ & & & 17.4 \\
                        \hline
                        \multicolumn{2}{l}{Hadronic decays} & & 64.8 & \\
                        & $\Pgtm\to\hminus\Pgngt$ & & & 11.5 \\
                        & $\Pgtm\to\hminus\Pgpz\Pgngt$ & \Pgr & & 25.9 \\
                        & $\Pgtm\to\hminus\Pgpz\Pgpz\Pgngt$ & \Pai & & 9.5 \\
                        & $\Pgtm\to\hminus\hplus\hminus\Pgngt$ & \Pai & & 9.8 \\
                        & $\Pgtm\to\hminus\hplus\hminus\Pgpz\Pgngt$ & & & 4.8 \\
                        & Other & & & 3.3 \\
                        \hline
                \end{tabular}
        \label{tab:tau_decays}
\end{table}

Charged particles used in the reconstruction of $\tauh$ candidates are required to have $\pt>0.5\GeV$, and must be compatible with originating from the primary vertex of the event, where the criterion on the transverse impact parameter is not highly restrictive ($d_{xy}<0.1\unit{cm}$), to minimize the rejection of genuine $\Pgt$ leptons with long lifetimes.
The requirement of $\pt>0.5\GeV$ on the charged particles ensures that the corresponding tracks have sufficient quality, and pass a minimal requirement on the number of layers with hits in the tracking detector.

Based on the set of charged particles and strips contained in a jet, the HPS algorithm generates all possible combinations of hadrons for the following decay modes: $\oneProngZeroPizero$, $\oneProngOnePizero$, $\oneProngTwoPizero$, and $\threeProngZeroPizero$.
The reconstructed mass of the ``visible'' hadronic constituents of the $\tauh$ candidate (\ie, the decay products, excluding neutrinos) is required to be compatible either with the $\Pgr$, or the $\Pai$ resonances in the $\oneProngOnePizero$ and in the $\oneProngTwoPizero$ or $\threeProngZeroPizero$ decay modes, respectively, as discussed in Section~\ref{sec:tauIdAlgorithm_dmreco}.
The $\oneProngOnePizero$ and $\oneProngTwoPizero$ modes are consolidated into the $\oneProngOnePizero$ mode, and are analyzed together.
The combinations of charged particles and strips considered by the HPS algorithm represent all the hadronic $\Pgt$ lepton decay modes in Table~\ref{tab:tau_decays}, except $\Pgtm\to\hminus\hplus\hminus\Pgpz\Pgngt$ with $\mathcal{B}=4.8\%$. This decay is not considered in the current version of the algorithm, because of its greater contamination by jets.
The $\tauh$ candidates of charge other than $\pm$1 are rejected, as are those with charged particles or strips outside the signal cone, defined by $R_\text{sig} = (3.0\GeVns)/\pt$, where the \pt is that of the hadronic system, with cone size limited to the range 0.05--0.10.
Finally, only the $\tauh$ candidate with largest \pt is kept for further analysis, resulting in a single $\tauh$ candidate per jet.

\subsubsection{Dynamic strip reconstruction}
\label{sec:tauIdAlgorithm_dynamicStripReco}

Photon and electron constituents of jets, which seed the $\tauh$ reconstruction, are clustered into $\Delta\eta{\times}\Delta\phi$ strips, and used to collect all energy depositions in the ECAL that arise from neutral pions produced in $\tauh$ decays.
The size of the $\Delta\eta{\times}\Delta\phi$ window is set to a fixed value of $0.05{\times}0.20$ in the ($\eta$, $\phi$) plane in the previous version of the HPS algorithm~\cite{TAU-14-001}.
However, this fixed strip size is not always adequate to contain all electrons and photons originating from the $\tauh$ decays, meaning that some of the particles from $\tauh$ lepton decay can contribute to the isolation region and thereby reduce the isolation efficiency for genuine $\tauh$ candidates.

Our studies of $\tauh$ reconstruction have led to the following observations:
\begin{enumerate}
\item A charged pion from $\tauh$ decays undergoing nuclear interactions in the tracker material can produce secondary particles with lower \pt. This can result in cascades of low-\pt electrons and photons that can appear outside of the strip window, and affect the isolation of a $\tauh$ candidate, despite these particles originating from remnants of the $\tauh$ decay.
\item Photons from $\Ppizero$ decays have a large probability to convert into $\Pep\Pem$ pairs and, after multiple scattering and bremsstrahlung, some of the remaining electrons and photons can end up outside a fixed size window, also affecting the isolation.
\end{enumerate}
Naively, these decay products can be integrated into the strip by suitably increasing its size.
Conversely, if the $\tauh$ has large \pt, the decay products tend to be boosted in the direction of the $\tauh$ candidate momentum.
In this case, a smaller than previously considered strip size can reduce background contributions to that strip, while taking full account of all decay products.

\begin{figure}[!htbp]
\centering
\includegraphics[width=0.48\textwidth]{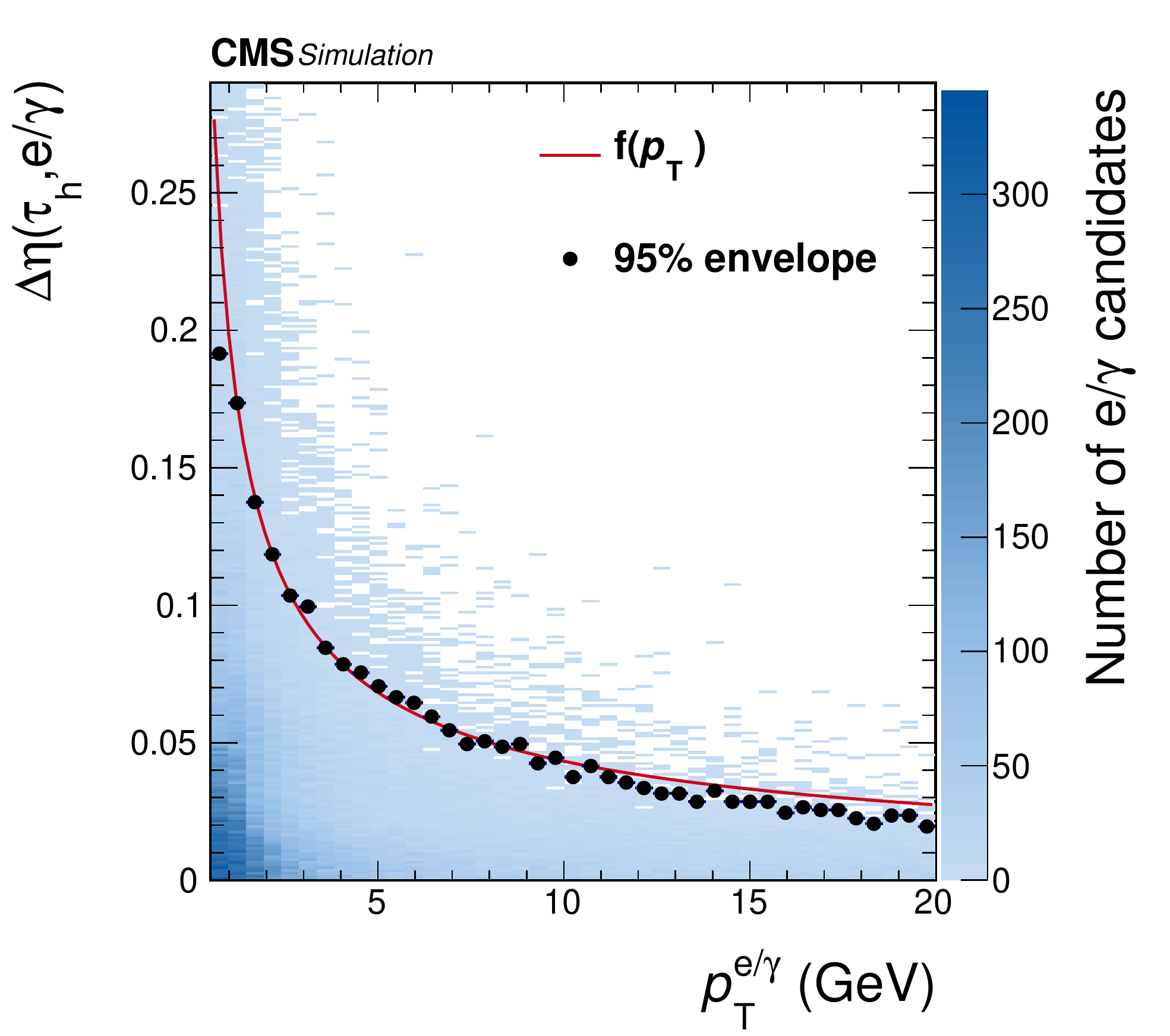}
\includegraphics[width=0.48\textwidth]{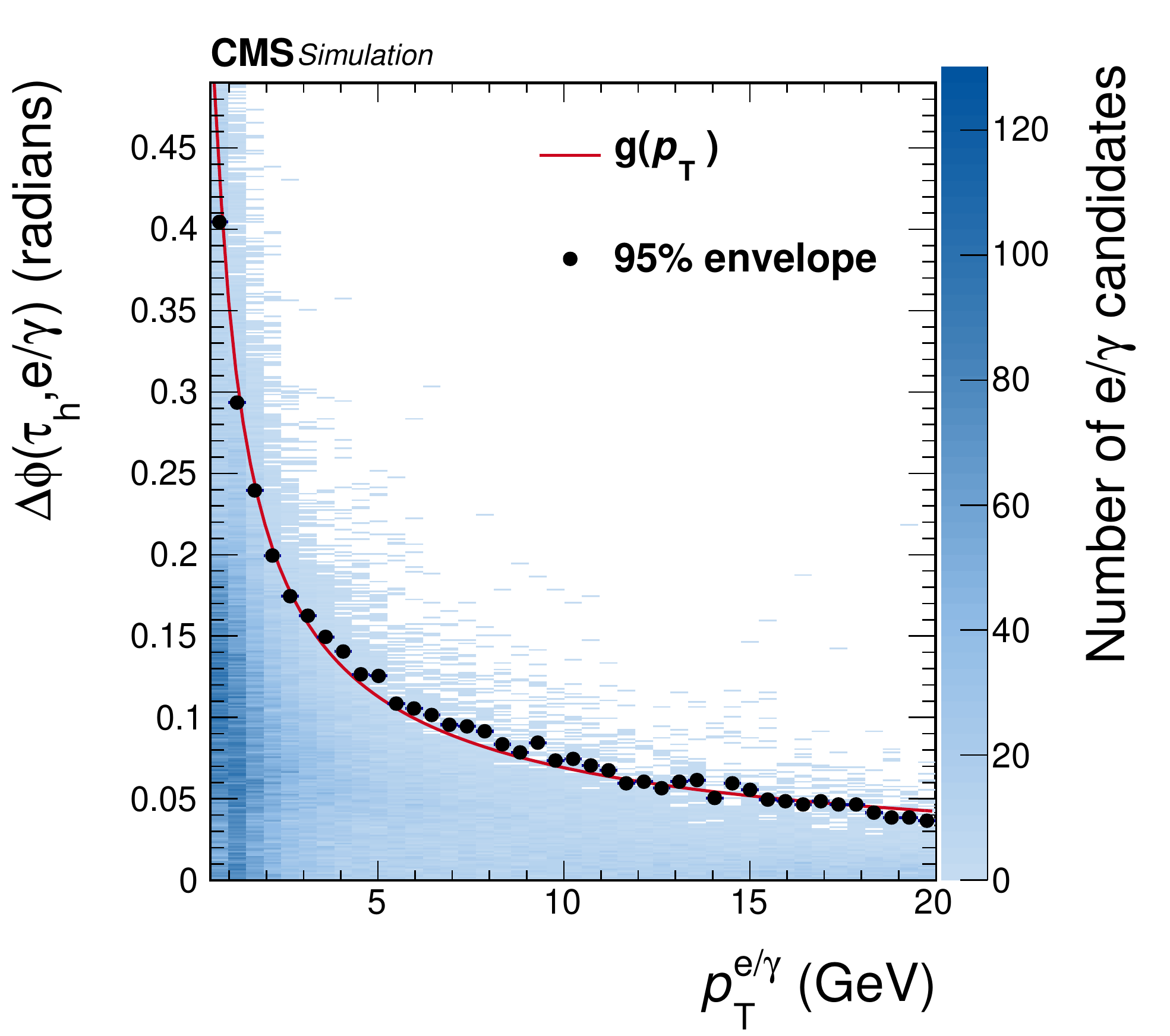}
\caption{
  Distance in $\eta$ (left) and in $\phi$ (right) between the $\tauh$ and $\Pe/\Pgg$ candidates for $\tauh$ decay products, as a function of the \pt of the $\Pe/\Pgg$ candidate, in simulated $\tauh$ decays.
  The points show the 95\% envelope for a given bin, and the solid red lines represent the fitted functions $f$ and $g$ given in Eq.~(\ref{eq:dynamicStripReco_window2}).
}
\label{fig:dynamicStripReco_window}
\end{figure}

Based on these considerations, the strip clustering of the HPS algorithm has been changed as follows:
\begin{enumerate}
\item The electron or photon ($\Pe/\Pgg$) with the highest \pt not yet included in any strip is used to seed a new strip,
with initial position set to the $\eta$ and $\phi$ values of the new $\Pe/\Pgg$ seed.

\item The \pt of the second-highest $\Pe/\Pgg$ deposition within
\begin{linenomath}
\begin{equation}
\begin{aligned}
\Delta\eta & = f(\pt^{\Pe/\Pgg}) + f(\pt^{\text{strip}}) \quad \text{and} \\
\Delta\phi & = g(\pt^{\Pe/\Pgg}) + g(\pt^{\text{strip}})
\end{aligned}
\label{eq:dynamicStripReco_window1}
\end{equation}
\end{linenomath}
of the strip position is merged into the strip.
The dimensionless functions $f$ and $g$ are determined from single $\Pgt$ lepton events, generated in MC with uniform \pt in the range from 20 to 400\GeV and $\abs{\eta}<2.3$,
such that 95\% of all electrons and photons that arise from $\tauh$ decays are contained within one strip.
The functional form is based on the $\Delta\eta$ and $\Delta\phi$ between the $\tauh$ and the $\Pe/\Pgg$ candidate, studied as a function of the \pt of the $\Pe/\Pgg$ candidate.
As shown in Fig.~\ref{fig:dynamicStripReco_window}, the 95\% envelope of points in each bin is fitted using the analytic form $a/(\pt)^b$, resulting in:
\begin{linenomath}
\begin{equation}
\begin{aligned}
f(\pt) & = 0.20\, \pt^{-0.66} \quad \text{and} \\
g(\pt) & = 0.35\, \pt^{-0.71},
\end{aligned}
\label{eq:dynamicStripReco_window2}
\end{equation}
\end{linenomath}
where the \pt is in \GeV.
The upper limits on the strip size are set to 0.3 in $\Delta\phi$ and 0.15 in $\Delta \eta$, and the lower limits are set to 0.05 for both $\Delta\phi$ and $\Delta \eta$.
The size of the window depends on the \pt values of both the strip and the merged $\Pe/\Pgg$ candidate.
The size is defined by the maximum separation between the two objects,
assuming they have opposite charges and are produced back-to-back in their rest frame.
Although, strictly speaking, this reasoning applies only to the $\phi$ direction, it is also used for the $\eta$ direction.

\item The strip position is recomputed using the $\pt$-weighted average of all $\Pe/\Pgg$ constituents in the strip:
\begin{linenomath}
\begin{equation}
\begin{aligned}
\eta_{\text{strip}} & = \frac{1}{\pt^{\text{strip}}}  \sum \pt^{\Pe/\Pgg} \, \eta_{\Pe/\Pgg}, \\
\phi_{\text{strip}} & = \frac{1}{\pt^{\text{strip}}}  \sum \pt^{\Pe/\Pgg} \, \phi_{\Pe/\Pgg}.
\end{aligned}
\label{eq:dynamicStripReco_position}
\end{equation}
\end{linenomath}
\item The construction of the strip ends when there is no other $\Pe/\Pgg$ candidate within the $\Delta\eta{\times}\Delta\phi$ window.
In this case, the clustering proceeds by selecting a new strip, seeded by the $\Pe/\Pgg$ candidate of highest $\pt$ that is not as yet associated with any strip.

\end{enumerate}
As defined above, the size of the strip does not depend on the cone-size of the $\tauh$ signal. The $\pt$-weighted center ($\eta$, $\phi$) of the strip is required to be within the signal cone, while part of the strip can lie outside of it.

\subsubsection{Mass constraints on decay modes}
\label{sec:tauIdAlgorithm_dmreco}
Strips are combined with charged particles to form
$\tauh$ decay hypotheses. Then, to check the compatibility of each hypothesis
with the signatures expected from different $\tauh$ decay modes, the reconstructed mass of the visible hadronic constituents of the $\tauh$ candidate (that we refer to as $m_{\tauh}$) is required to lie within a mass window corresponding either to the $\Pgr$ or $\Pai$ meson. The widths and positions of the mass windows are optimized for each decay mode to maximize the ratio of the $\tauh$ reconstruction efficiency to the $\text{jet}\mapsto\tauh$ misidentification probability, with results that can be summarized as follows:
\begin{enumerate}
\item{$0.3\GeV-\Delta m_{\tauh} < m_{\tauh} < 1.3\GeV\,\sqrt{\smash[b]{\pt^{\tauh}/(100\GeVns)}}+\Delta m_{\tauh}$ for $\oneProngOnePizero$, with the mass window enlarged for $\tauh$ candidates of high $\pt$ to account for resolution, and the upper limit on the mass window constrained to lie between 1.3 and 4.2\GeV,}
\item{$0.4\GeV-\Delta m_{\tauh} < m_{\tauh} < 1.2\GeV\,\sqrt{\smash[b]{\pt^{\tauh}/(100\GeVns)}}+\Delta m_{\tauh}$ for $\oneProngTwoPizero$, with the upper limit on the mass window restricted to lie between 1.2 and 4.0\GeV, and}
\item{$0.8 < m_{\tauh} < 1.5\GeV$ for the $\threeProngZeroPizero$ channels,}
\end{enumerate}
where $\Delta m_{\tauh}$ is the change in the mass of the $\tauh$ candidate brought about by the addition of the $\Pe/\Pgg$ candidates to its strip. It is calculated as follows:
\begin{linenomath}
\begin{equation}
\Delta m_{\tauh} = \sqrt{\left( \frac{\partial m_{\tauh}}{\partial \eta_{\text{strip}}} \,  f(\pt^{\text{strip}}) \right)^{2}
 + \left( \frac{\partial m_{\tauh}}{\partial \phi_{\text{strip}}} \,  g(\pt^{\text{strip}}) \right)^{2}},
\label{eq:dynamicStripReco_masscorr}
\end{equation}
\end{linenomath}
with
\begin{linenomath}
\begin{equation*}
\begin{aligned}
\frac{\partial m_{\tauh}}{\partial \eta_{\text{strip}}} & = \frac{p_{z}^{\text{strip}} \, E_{\tauh} - E_{\text{strip}} \, p_{z}^{\tauh}}{m_{\tauh}} \quad \text{and} \\
\frac{\partial m_{\tauh}}{\partial \phi_{\text{strip}}} & = \frac{-\left( p_{y}^{\tauh} - p_{y}^{\text{strip}} \right) \, p_{x}^{\text{strip}}
  + \left( p_{x}^{\tauh} - p_{x}^{\text{strip}} \right) \, p_{y}^{\text{strip}}}{m_{\tauh}},
\end{aligned}
\end{equation*}
\end{linenomath}
where $p_{\tauh}=(E_{\tauh},p_{x}^{\tauh},p_{y}^{\tauh},p_{z}^{\tauh})$ and $p_{\text{strip}}=(E_{\text{strip}},p_{x}^{\text{strip}},p_{y}^{\text{strip}},p_{z}^{\text{strip}})$ are the four-momenta of the $\tauh$ and of the strip, respectively.

\subsection{Discrimination of \texorpdfstring{$\tauh$}{tau[h]} candidates against jets}
\label{sec:tauIdDiscriminators_jets}
Requiring $\tauh$ candidates to pass certain specific isolation requirements provides a strong handle for reducing the jet $\mapsto \tauh$ misidentification probability.
The two $\tauh$ isolation discriminants developed previously~\cite{TAU-14-001}, namely the isolation sum and the MVA-based discriminants, have now been reoptimized.
A cone with $\Delta R = 0.5$ was originally used in the definition of isolation for all event types.
However, in processes with a high number of final-state objects, such as for Higgs boson production in association with top quarks ($\ttbar\PH$), the isolation is affected by the presence of nearby objects.
Studies using such $\ttbar\PH$ events with $\PH\to\Pgt\Pgt$ decays led to the conclusion that a smaller isolation cone improves the $\tauh$ efficiency in such events.
A smaller isolation cone of radius $\Delta R = 0.3$ is therefore now used in these types of events.

\subsubsection{Isolation sum discriminants}
\label{sec:tauIdDiscrCutBased}
The isolation of $\tauh$ candidates is computed by summing the scalar \pt of charged particles ($\sum \pt^{\text{charged}}$) and photons ($\sum \pt^{\Pgg}$) reconstructed using the PF algorithm within the isolation cone centered on the direction of the $\tauh$ candidate. Charged-hadron and photon constituents of $\tauh$ candidates are excluded from the \pt sum, defining thereby the isolation as:
\begin{linenomath}
\begin{equation}
I_{\tauh} = \sum \pt^{\text{charged}} (d_z < 0.2\unit{cm}) + \max\left( 0, \sum \pt^{\Pgg} - \Delta\beta \sum \pt^{\text{charged}} (d_z > 0.2\unit{cm}) \right).
\label{eq:tauIsolationDeltaBeta}
\end{equation}
\end{linenomath}
The contribution from PU is suppressed by requiring the charged particles to originate from the production vertex of the $\tauh$ candidate within a distance of $d_z<0.2\unit{cm}$.
The PU contribution to the \pt sum of photons in the isolation cone is estimated by summing the scalar \pt of charged particles not originating from the vertex of the $\tauh$ candidate ($\sum \pt^{\text{charged}}$ with $d_z > 0.2\unit{cm}$), but appearing within a cone of $\Delta R = 0.8$ around the $\tauh$ direction multiplied by a so-called $\Delta \beta$ factor, which accounts for the ratio of energies carried by charged hadrons and photons in inelastic $\Pp\Pp$ collisions, as well as for the different cone sizes used to estimate the PU contributions.

Previously, an empirical factor of 0.46 was used as the $\Delta \beta$~\cite{TAU-14-001}.
However, this is found to overestimate the PU contribution to the isolation in data taken in 2015 and 2016.
And a new $\Delta\beta$ factor of 0.2 is therefore chosen.
This value corresponds approximately to the ratio of neutral to charged pion production rates (0.5), corrected for the difference in the size of the isolation cone ($\Delta R = 0.5$) and the cone used to compute the $\Delta \beta$ correction ($\Delta R = 0.8$): $0.5 \times (0.5^{2}/0.8^{2}) \approx 0.195$.

The loose, medium, and tight working points of the isolation sum discriminants are defined by requiring $I_{\tauh}$ to be less than 2.5, 1.5, or 0.8\GeV, respectively. These thresholds are chosen such that the resulting efficiencies for the three working points cover the range required for the analyses.

In dynamic strip reconstruction, a photon candidate outside the signal cone can still contribute to the signal.
This effectively increases the $\text{jet}\mapsto\tauh$ misidentification probability because of the decrease in the value of $I_{\tauh}$ for misidentified $\tauh$ candidates.
An additional handle is therefore exploited to reduce the jet $\mapsto\tauh$ misidentification probability using the scalar \pt sum of $\Pe/\Pgg$ candidates included in strips, but located outside of the signal cone, which is defined as
\begin{linenomath}
\begin{equation}
\pt^{\text{strip, outer}} = \sum \pt^{\Pe/\Pgg} (\Delta R > R_{\text{sig}}).
\label{eq:pTouterCut}
\end{equation}
\end{linenomath}
A reduction of about 20\% in the jet~$\mapsto\tauh$ misidentification probability is achieved by requiring $\pt^{\text{strip, outer}}$ to be less than 10\% of $\pt^{\tauh}$, for similar values of efficiency.

\begin{figure}[!htbp]
\centering
\includegraphics[width=0.48\textwidth]{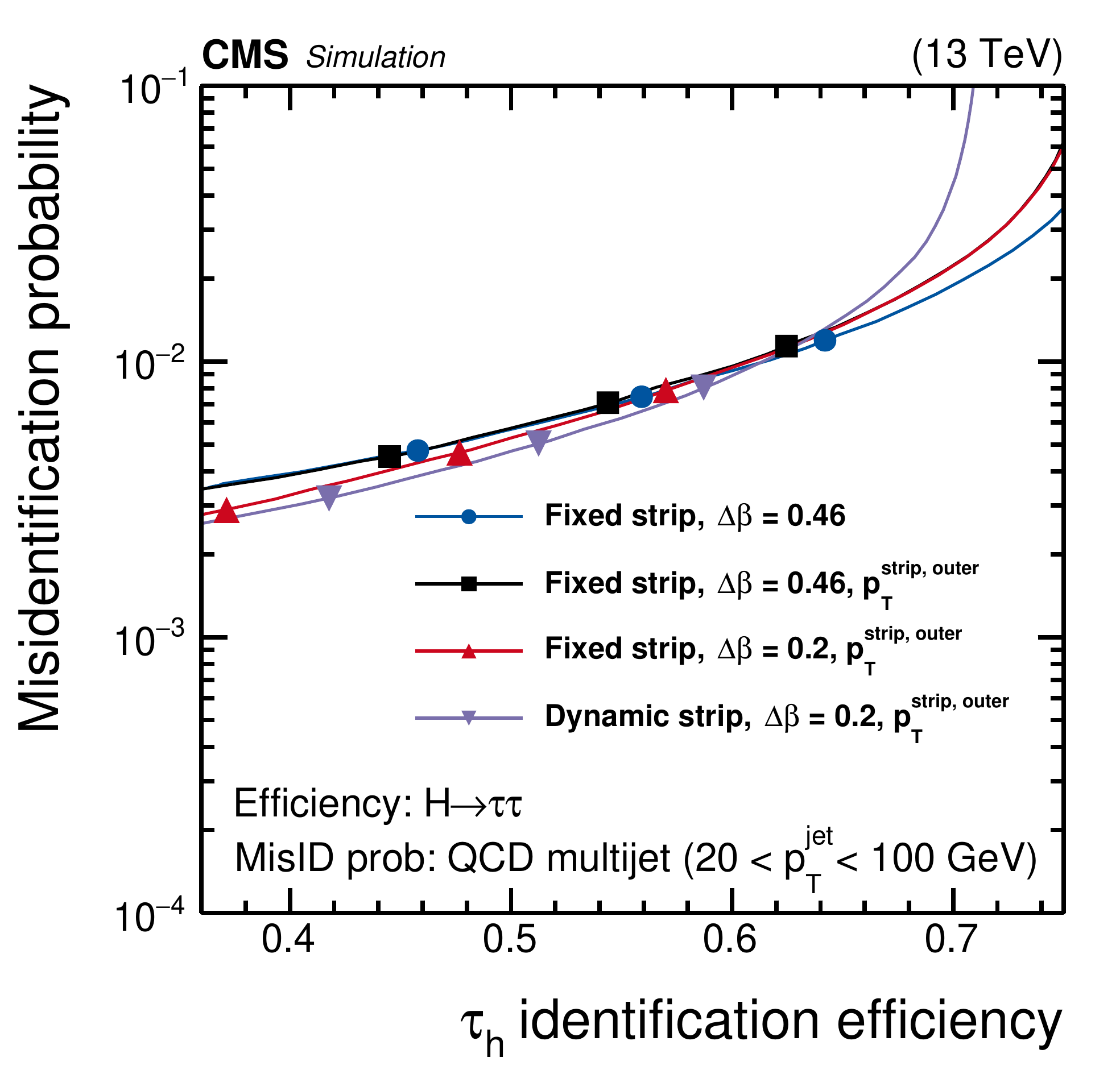}
\includegraphics[width=0.48\textwidth]{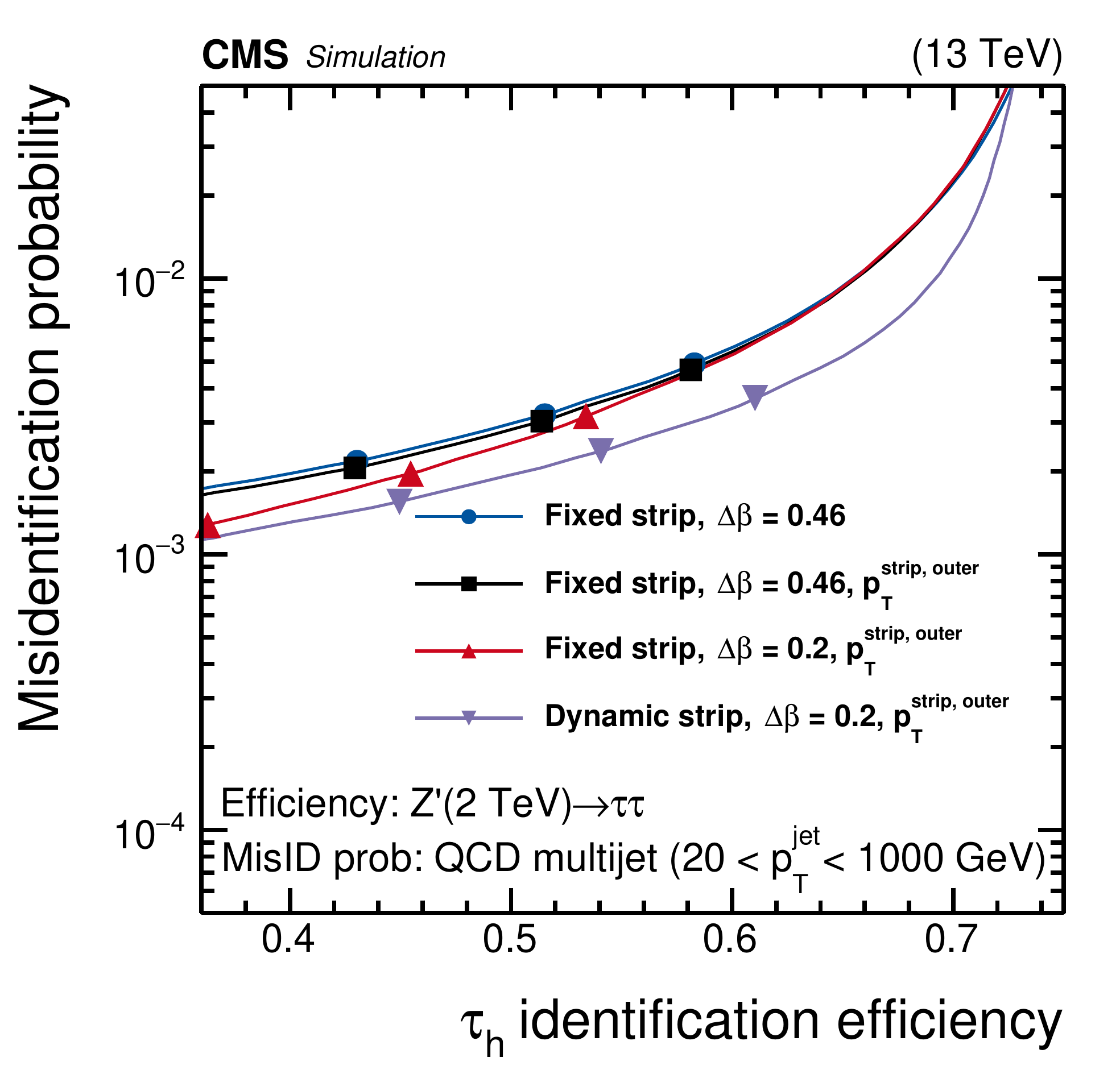}

\caption{
  Misidentification probabilities as a function of the $\tauh$ identification efficiencies, evaluated for $\PH\to\Pgt\Pgt$ (left) and $\cPZ^{'}(2\TeV) \to \Pgt\Pgt$ (right), and for QCD multijet MC events.
Four configurations of the reconstruction and isolation method are compared.
The three points on each curve correspond, from left to right, to the tight, medium, and loose WPs. The solid curves are obtained by imposing cutoffs on  $I_{\tauh}$ that decrease linearly from small to large efficiency.
}
\label{fig:dynamicStripReco_perf}
\end{figure}

A comparison of the expected performance of the isolation sum discriminant for the previous and current versions of the HPS algorithm is shown in Fig.~\ref{fig:dynamicStripReco_perf}.
The efficiency is calculated for generated $\tauh$ candidates with $\pt>20\GeV$, $\abs{\eta}<2.3$, having a decay mode of $\oneProngZeroPizero$, $\oneProngOnePizero$, $\oneProngTwoPizero$, or $\threeProngZeroPizero$, and matching to a reconstructed $\tauh$ candidate with $\pt>18\GeV$. The misidentification probability is calculated for jets with $\pt>20\GeV$, $\abs{\eta}<2.3$, and matching to a reconstructed $\tauh$ candidate with $\pt>18\GeV$.
The different sources of improvement in performance of the algorithm with fixed strip size are shown separately for $\Delta\beta = 0.46$, $\Delta\beta = 0.46$ with $\pt^{\text{strip, outer}} < 0.1\,\pt^{\tauh}$, and for $\Delta\beta=0.2$ with $\pt^{\text{strip, outer}} < 0.1\,\pt^{\tauh}$.
The signal process is modelled using MC events for $\PH\to\Pgt\Pgt$ (for low-\pt $\tauh$) and $\cPZ^{'} \to \Pgt\Pgt$ decays, with $m_{\cPZ^{'}}$ = 2\TeV (for high-\pt $\tauh$).
The QCD multijet MC events are used as background, with jet \pt values up to 100
and 1000\GeV, respectively,
such that the \pt coverage is similar to that in signal events.
The improvement brought about by the dynamic strip reconstruction for high-\pt $\Pgt$ leptons can be observed by comparing the two plots in Fig.~\ref{fig:dynamicStripReco_perf}.
At low-\pt (Fig.~\ref{fig:dynamicStripReco_perf}, left), the performance for $\tauh$ candidates for medium and tight WPs improves slightly.
However, in the high-efficiency region, the misidentification probability starts to increase faster than the efficiency in the current algorithm.
This is caused by choosing the working points of the algorithm through changes in the requirements on $I_{\tauh}$.
To reach a higher efficiency, the requirement on $I_{\tauh}$ is relaxed, which in turn leads to an increase in the misidentification probability.
However, the $\pt^{\text{strip, outer}}$ requirement prevents the efficiency from rising at a similar rate, leading thereby to the observed behaviour of the response in the high-efficiency region.

\subsubsection{MVA-based discriminants}
\label{sec:tauIdDiscrMVABased}
The MVA-based $\tauh$ identification discriminants combine the isolation and other differential variables sensitive to the $\Pgt$ lifetime, to provide the best possible discrimination between $\tauh$ decays and quark or gluon jets.
A classifier based on boosted decision trees (BDT) is used to achieve a reduction in the jet~$\mapsto \tauh$ misidentification probability.
The MVA identification method and the variables used as input to the BDT are discussed in Ref.~\cite{TAU-14-001}.

In addition to those discussed in Ref.~\cite{TAU-14-001}, the following variables are included in the classifier to improve its performance:
\begin{enumerate}
\item{Differential variables such as $\pt^{\text{strip, outer}}$ in Eq.~(\ref{eq:pTouterCut}), and \pt-weighted $\Delta R$, $\Delta\eta$, and $\Delta\phi$ (relative to the $\tauh$ axis) of photons and electrons in strips within or outside of the signal cone;}
\item{$\Pgt$ lifetime information, based on the signed three-dimensional impact parameter of the leading track of the $\tauh$ candidate and its significance (the impact parameter length divided by its uncertainty); and}
\item{multiplicity of photon and electron candidates with $\pt > 0.5\GeV$ in the signal and isolation cones.}
\end{enumerate}
The charged and neutral-particle isolation sums and the $\Delta\beta$ correction, as defined in Eq.~(\ref{eq:tauIsolationDeltaBeta}), are used as separate variables in the BDT classifier, and correspond to the most powerful discriminating variables.
Other significant variables are the two- and three-dimensional impact parameters of the leading track and their significances, as well as the flight length and its significance for the $\tauh$ candidates decaying into three charged hadrons and a neutrino. The multiplicity of photon and electron candidates in the jet seeding the $\tauh$ candidate is found to contribute to the decision of the BDT classifier at levels similar to those of the lifetime variables.

The BDT is trained using simulated $\tauh$ candidates selected with $\pt>20\GeV$ and $\abs{\eta}<2.3$ in $\cPZ/\Pggx\to\Pgt\Pgt$, $\PH\to\Pgt\Pgt$, $\PZprime\to\Pgt\Pgt$, and $\PW^{\prime}\to\Pgt\Pgn$ events (with the mass ranges of $\PH$, $\PZprime$, and $\PWprime$ detailed in Section~\ref{sec:datasamples_and_MonteCarloSimulation}).
The QCD multijet, $\PW$+jets, and $\ttbar$ events are used to model quark and gluon jets.
These events are reweighted to provide identical two-dimensional distributions in \pt and $\eta$ for $\tauh$ candidates in signal and in background sources,
which makes the MVA training insensitive to differences in \pt and $\eta$ distributions of $\Pgt$ leptons and jets in the training samples.

The working points of the MVA-isolation discriminant, corresponding to different $\tauh$ identification efficiencies, are defined through requirements on the BDT discriminant.
For a given working point, the threshold on the BDT discriminant is adjusted as a function of \pt of the $\tauh$ candidate to ensure uniform efficiency over $\pt^{\tauh}$.
The working points for the reconstructed $\tauh$ candidates are chosen to have isolation efficiencies between 40 and 90\%, in steps of 10\%, for the reconstructed $\tauh$ candidates.

\begin{figure}[!htbp]
\centering
\includegraphics[width=0.48\textwidth]{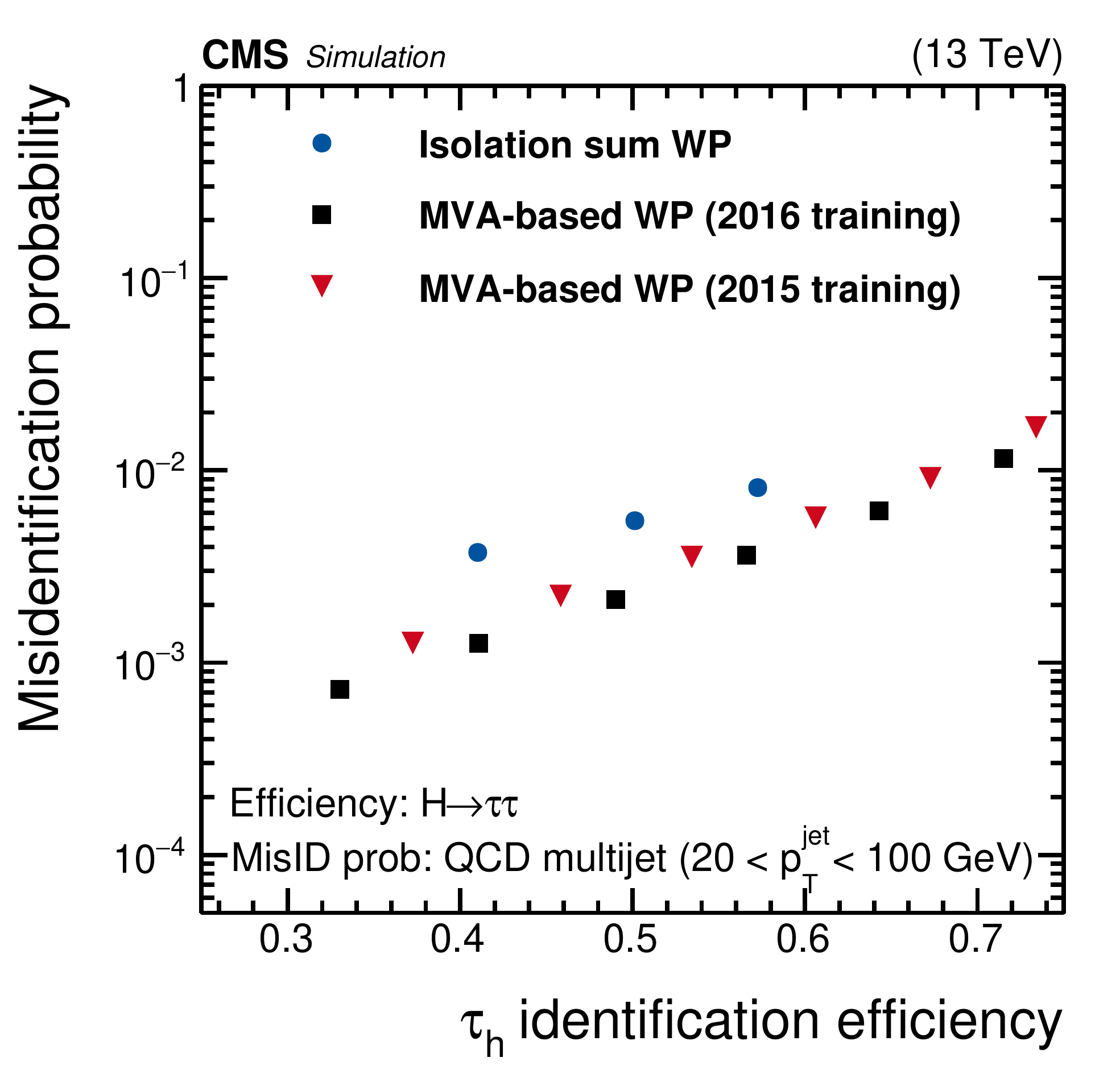}
\includegraphics[width=0.48\textwidth]{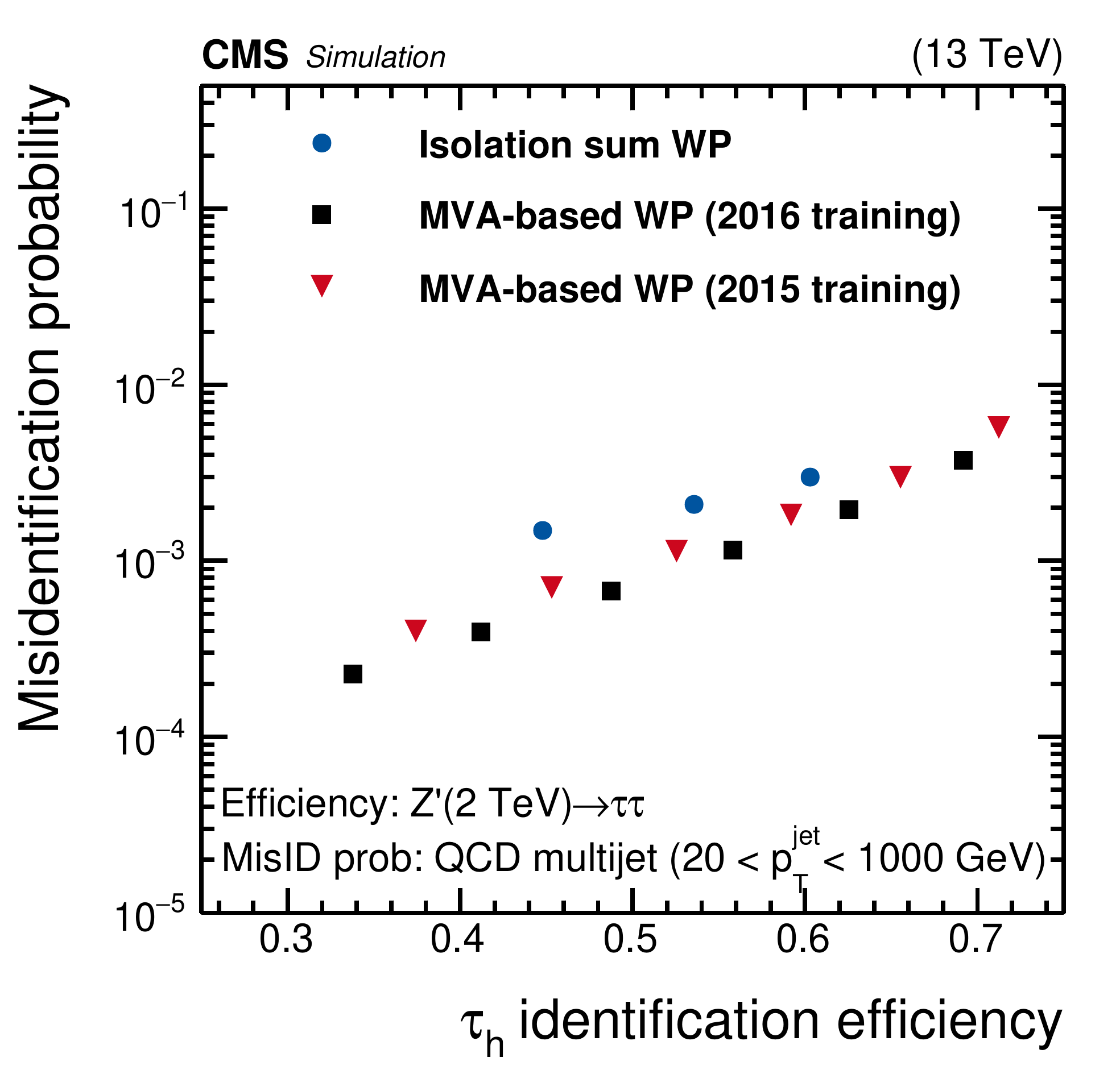}

\caption{Misidentification probabilities for $\tauh$ as a function of their identification efficiency, evaluated using $\PH\to\Pgt\Pgt$ (left), $\cPZ^{'}(2\TeV) \to \Pgt\Pgt$ (right), and QCD multijet MC events. The MVA-based discriminants trained on their corresponding MC events are compared to each other, as well as to the isolation-sum discriminants. The points correspond to different working points of the discriminants. The three points for the isolation-sum discriminants from left to right correspond to the tight, medium, and loose WPs. Similarly, the six points of the MVA-based discriminants define the WP as very-very tight, very tight, tight, medium, loose, and very loose, respectively.
}
\label{fig:tauId_perf_mvarun2DBoldDM}
\end{figure}

The expected jet~$\mapsto\tauh$ misidentification probability is shown in Fig.~\ref{fig:tauId_perf_mvarun2DBoldDM}, as a function of expected $\tauh$ identification efficiency.
It demonstrates a reduction in the misidentification probability by a factor of 2 for MVA-based discriminants, at efficiencies similar to those obtained using isolation-sum discriminants.
We compare two sets of MVA-based discriminants that were trained using MC samples that correspond to different conditions during data taking.
The working points of the MVA-based discriminants are shifted relative to each other, but follow the same performance curve. This confirms the stability of the MVA-based discriminants.
The expected $\tauh$ selection efficiencies and jet~$\mapsto\tauh$ misidentification probabilities for low to medium \pt, for the most commonly used working point (tight) of the training in 2016 are 49\% and 0.21\%, respectively.
For high \pt, the expected misidentification probability drops to 0.07\%, while the $\tauh$ selection efficiency remains constant, as desired.

\begin{figure}[!htbp]
\centering
\includegraphics[width=0.48\textwidth]{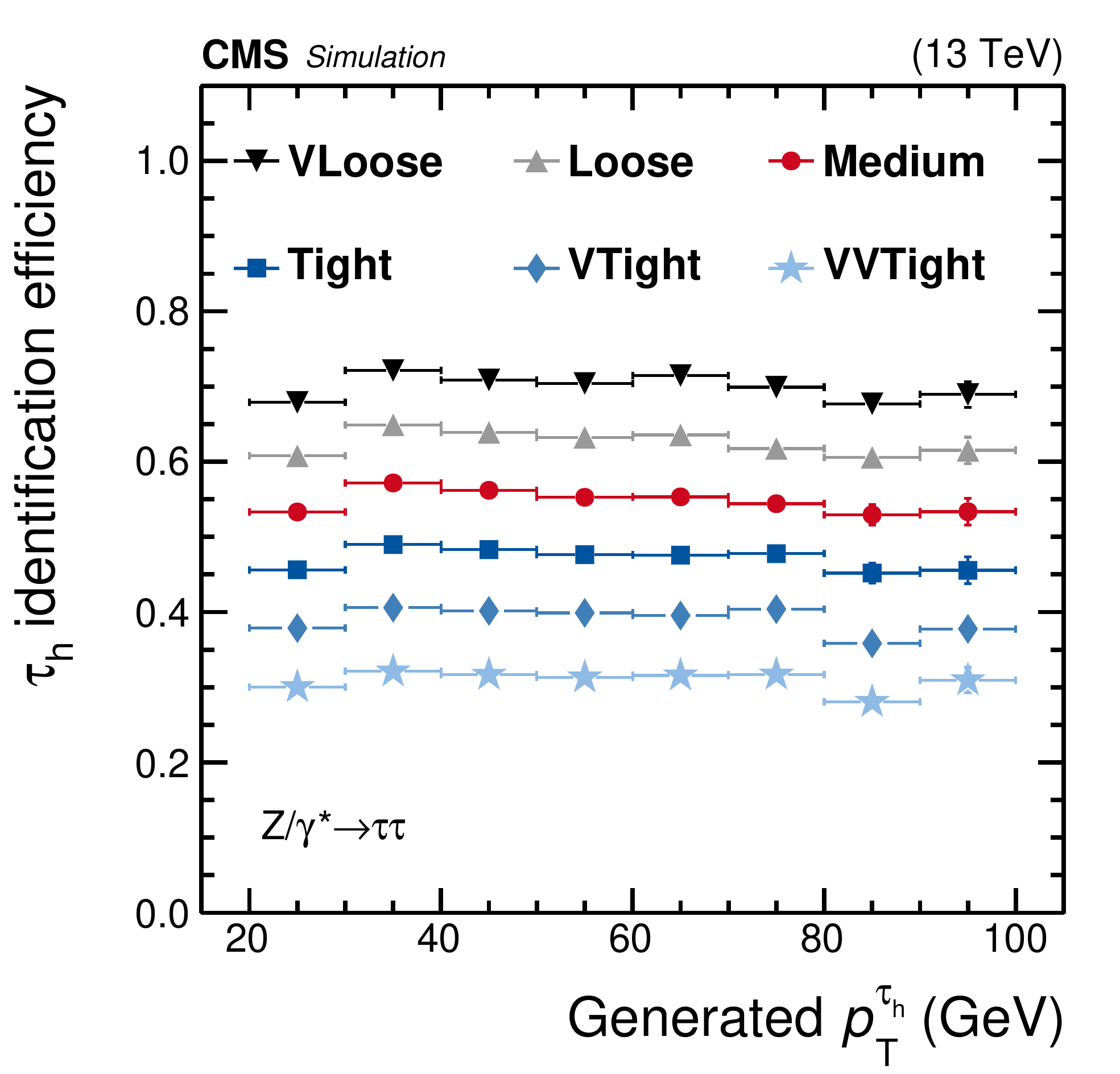}
\includegraphics[width=0.48\textwidth]{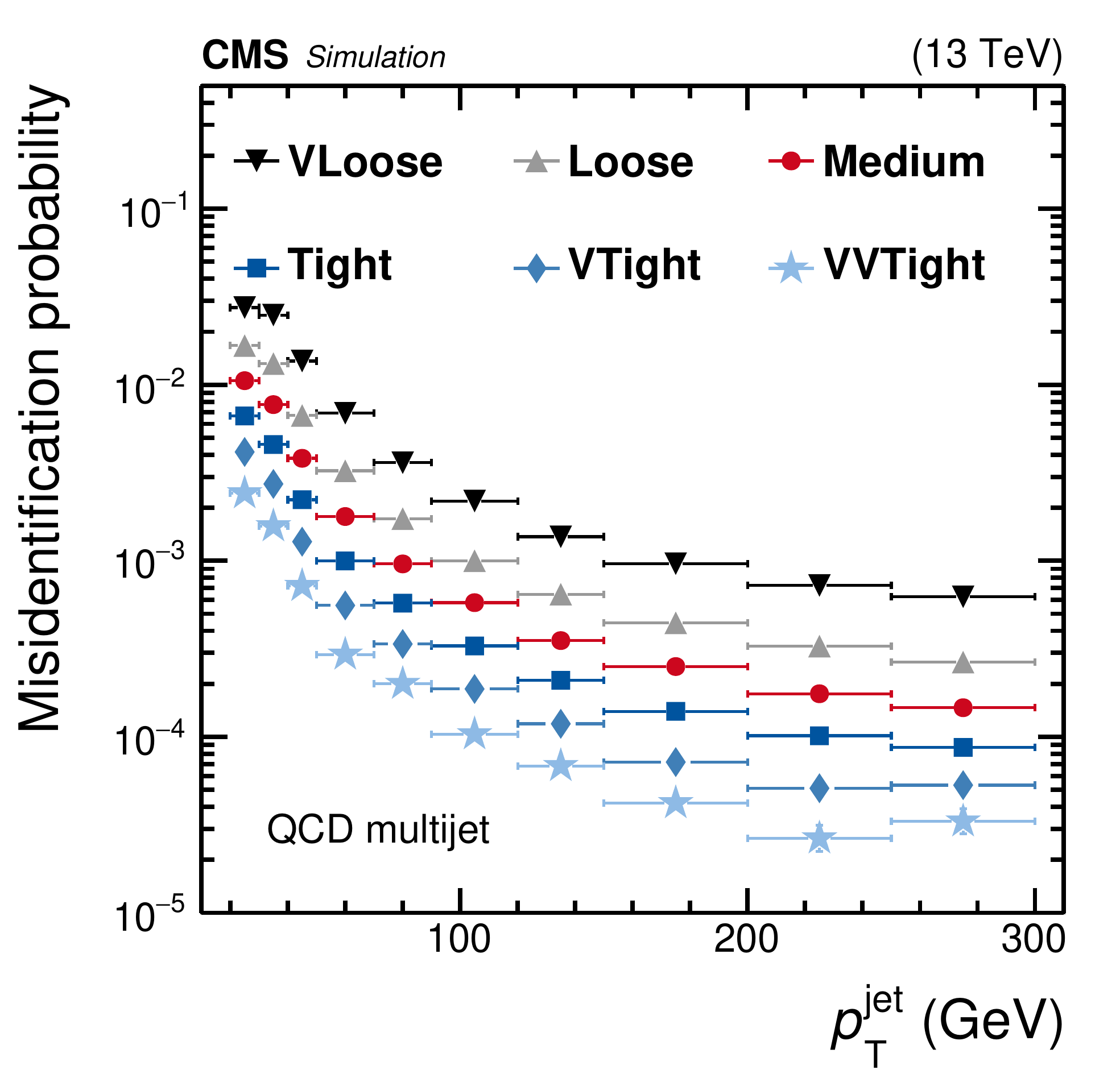}
\caption[dummy text]{Efficiency of $\tauh$ identification, estimated using simulated $\cPZ/\Pggx\to\Pgt\Pgt$ events (left), and the misidentification probability estimated using simulated QCD multijet events (right) are given, for the very loose, loose, medium, tight, very tight, and very-very tight WPs of the MVA-based $\tauh$ isolation algorithm. The efficiency and misidentification probabilities are shown as a function of \pt of the generated $\tauh$ and of the reconstructed jet, respectively. Vertical bars (often smaller than the symbol size) correspond to the statistical uncertainties (the 68\% Clopper-Pearson intervals~\cite{ClopperPearson1934}), while horizontal bars indicate the bin widths.}
\label{fig:tauId_perf_ptbin}

\end{figure}

Figure~\ref{fig:tauId_perf_ptbin} shows the respective expected $\tauh$ identification efficiency (left) and the misidentification probability (right), as a function of \pt of the generated $\tauh$ and of the reconstructed jet.
The efficiency is computed from $\cPZ\to\Pgt\Pgt$ events, while the expected jet~$\mapsto\tauh$ misidentification probability is computed for QCD multijet events with jet $\pt<300\GeV$.

\subsection{Discrimination of \texorpdfstring{$\Pgt$}{tau} leptons against electrons}
\label{sec:antiElectronDiscrMVABased}

Isolated electrons have a high probability to be misidentified as $\tauh$ candidate that decay to either $\oneProngZeroPizero$ or $\oneProngOnePizero$. In particular, electrons crossing the tracker material often emit bremsstrahlung photons mimicking neutral pions in their reconstruction.
An improved version of the MVA electron discriminant used previously~\cite{TAU-14-001} is developed further to reduce the $\Pe\mapsto\tauh$ misidentification probability, while maintaining a high selection efficiency for genuine $\tauh$ decays over a wide $\pt$ range.
The variables used as input for the BDT are identical to the ones described in Ref.~\cite{TAU-14-001}, with the addition of the following photon-related variables:

\begin{enumerate}
 \item the number of photons in any of the strips associated with the $\tauh$ candidate;
 \item the \pt-weighted root-mean-square of the distances in $\eta$ and $\phi$ between all photons included in any strip and the leading track of the $\tauh$ candidate; and
 \item the fraction of $\tauh$ energy carried away by photons.
\end{enumerate}
These variables are computed separately for the photons inside and outside of the $\tauh$ signal cone to improve separation.
The most sensitive variables are the fraction of energy carried by the photon candidates, the ratio of the energy deposited in the ECAL to the sum of energies deposited in the ECAL and HCAL, the ratio of the deposited energy in the ECAL relative to the momentum of the leading charged hadron, the $m_{\tauh}$, and the \pt of the leading charged hadron.

\begin{figure}[!htbp]
\centering
\includegraphics[width=0.48\textwidth]{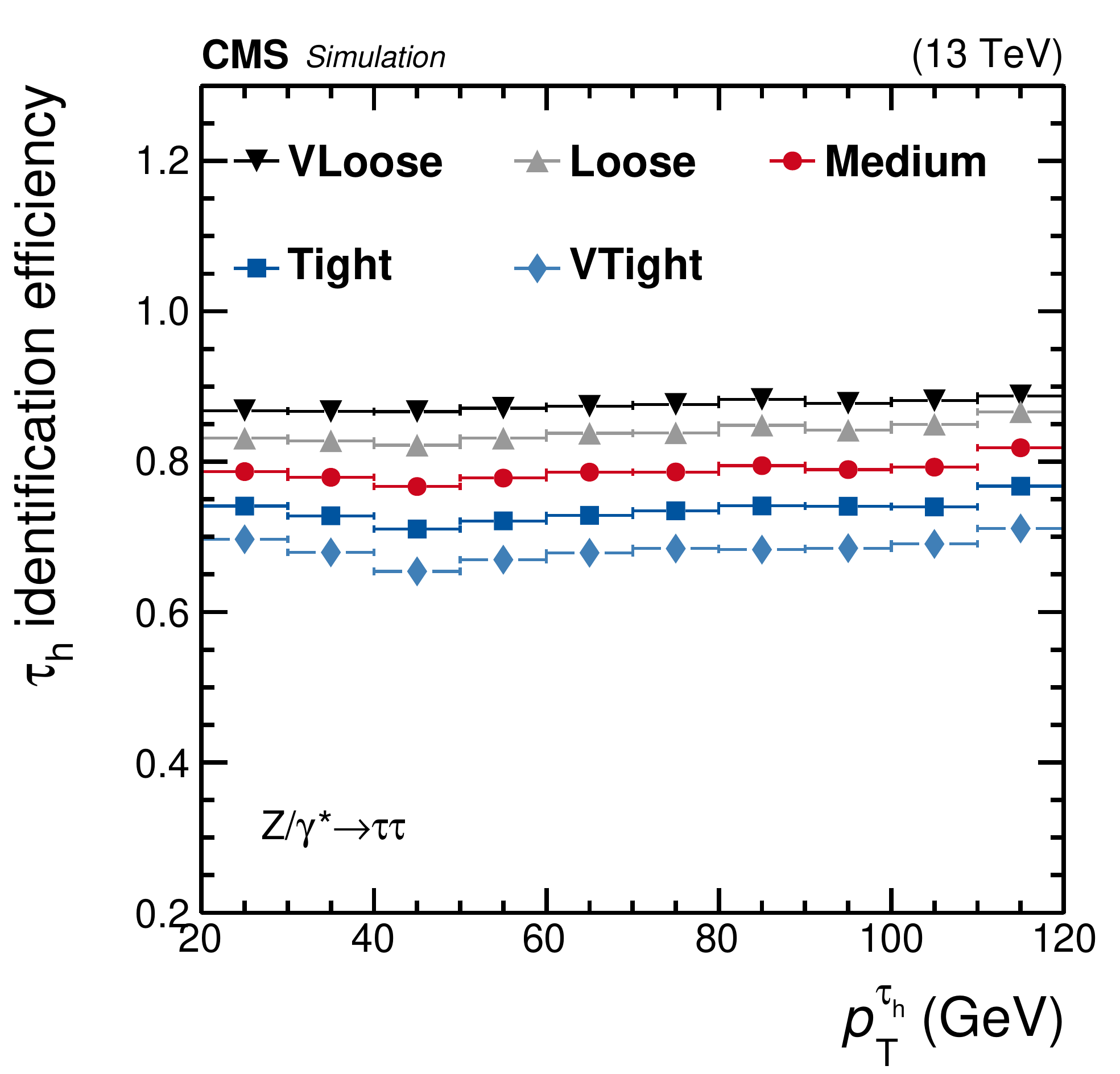}
\includegraphics[width=0.48\textwidth]{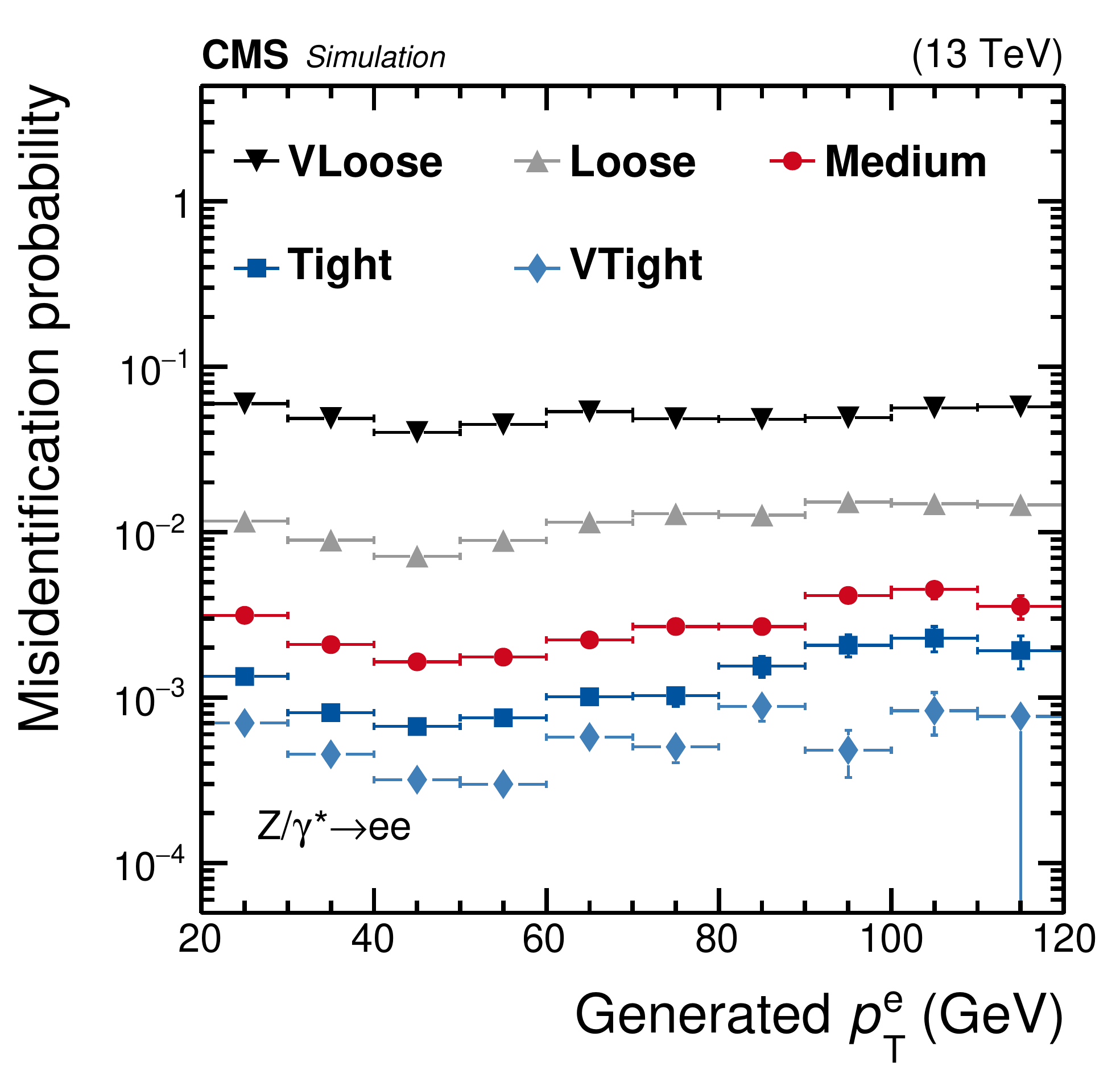}
\caption[dummy text]{
Efficiencies of $\tauh$ identification estimated via simulated $\cPZ/\Pggx\to\Pgt\Pgt$ events (left), and the $\Pe\mapsto\tauh$ misidentification probability estimated using simulated $\cPZ/\Pggx\to\Pe\Pe$ events (right) for the very loose, loose, medium, tight, and very tight WPs of the MVA-based electron discrimination algorithm. The efficiency is shown as a function of \pt of the reconstructed $\tauh$ candidate, while the misidentification probability is shown as a function of the generated electron \pt. The efficiency is calculated for $\tauh$ candidates with a reconstructed decay mode that pass the loose WP of the isolation-sum discriminant, while the misidentification probability is calculated for generated electrons of $\pt>20\GeV$ and $\abs{\eta}<2.3$, excluding the less sensitive detector region of $1.46<\abs{\eta}<1.56$ between the barrel and endcap ECAL regions. Vertical bars (often smaller than the symbol size) indicate the statistical uncertainties (the 68\% Clopper-Pearson intervals), while horizontal bars indicate the bin widths.}
\label{fig:antiEDiscr_eff}

\end{figure}

The BDT is trained using the simulated events listed in Section~\ref{sec:datasamples_and_MonteCarloSimulation}, which contain genuine $\Pgt$ leptons and electrons.
Reconstructed $\tauh$ candidates can be considered as signal or background, depending on whether they are matched to a $\tauh$ decay or to an electron at the generator level.
Different working points are defined according to the requirements on their BDT output and the efficiency for a genuine $\tauh$ candidate to pass the working points of the discriminants.
The expected efficiency of $\tauh$ reconstruction and the $\Pe\mapsto\tauh$ misidentification probability are presented in Fig.~\ref{fig:antiEDiscr_eff}.
Both are found to be approximately uniform over \pt, except for a dip at $\approx$45\GeV, whose depth increases with the tightening of the selection criteria.
This is because the MC events used to model the $\Pe\mapsto\tauh$ misidentification in the training of the MVA discriminant have electron \pt distributions that peak at $\approx$45\GeV, since the sample is dominated by $\cPZ/\Pggx\to\Pe\Pe$ and $\PW\to\Pe\Pgn$ events.

\subsection{Discrimination of \texorpdfstring{$\Pgt$}{tau} leptons against muons}
\label{sec:antiMuonDiscrCutBased}
Muons have a high probability to be misreconstructed as $\tauh$ objects in the $\oneProngZeroPizero$ decay mode.
The discriminant against muons, developed previously~\cite{TAU-14-001}, is based on vetoing $\tauh$ candidates
when signals in the muon detector are found near the $\tauh$ direction.
The two working points corresponding to different $\tauh$ identification efficiencies and $\Pgm \mapsto \tauh$ misidentification rates are:

\begin{enumerate}
\item ``against-$\Pgm$ loose'':
  $\tauh$ candidates fail this working point when
  track segments in at least two muon detector planes are found to lie within a cone of size $\Delta R = 0.3$ centered on the $\tauh$ direction,
  or when the energy deposited in the calorimeters, associated through the PF algorithm to the ``leading'' charged hadron of the $\tauh$ candidate,
  is $<$20\% of its track momentum.
\item ``against-$\Pgm$ tight'':
  $\tauh$ candidates fail this working point when
  they fail condition (i), or
  when a hit is present in the CSC, DT, or RPC detectors located in the two outermost muon stations
  within a cone of size $\Delta R = 0.3$ around the $\tauh$ direction.
\end{enumerate}

The efficiency for $\tauh$ candidates from $\cPZ/\Pggx\to\Pgt\Pgt$ events to pass the against-$\Pgm$ discriminant selection requirements exceeds 99\%.
The $\Pgm \mapsto \tauh$ misidentification probability, for muons in $\cPZ/\Pggx\to\Pgm\Pgm$ events, is ${\approx}3.5\times10^{-3}$ and ${\approx}1.4\times10^{-3}$ for loose and tight WPs, respectively.

\section{Reconstruction of highly boosted \texorpdfstring{$\Pgt$}{tau} lepton pairs}
\label{sec:boostedTauReco}

In events containing a (hypothetical) massive boson with large \pt, \eg, a radion (R) decaying to a pair of Higgs bosons~\cite{Giudice:2000av,Oliveira:2014kla}, with at least one of these decaying to a pair of $\Pgt$ leptons, the jets from the two $\Pgt$ leptons would be emitted very close to each other, thereby forming a single jet.
The performance of the HPS algorithm in such topologies is poor, as it was designed to reconstruct only one $\tauh$ per jet.
A dedicated version of the HPS algorithm was therefore recently developed to reconstruct two $\Pgt$ leptons with large momenta that typically originate from decays of large-momentum \PZ or Higgs bosons.
This algorithm takes advantage of jet substructure techniques, as follows. A collection of ``large-radius jets'' is assembled from the PF candidates using the Cambridge--Aachen algorithm~\cite{Wobisch:1998wt} with a distance parameter of 0.8 (CA8).
Due to the large boosts, the emitted $\Pgt$ lepton decay products are expected to be contained within the same CA8 jet, when its \pt exceeds 100\GeV.
The algorithm proceeds by reversing the final step of the clustering algorithm for each given CA8 jet, to find two subjets $sj_{1}$ and $sj_{2}$ that can be expected to coincide with the two $\Pgt$ leptons from the decay of the boosted massive boson. To reduce the misidentification of jets arising from QCD multijet events, $\mathrm{sj_{1}}$ and $\mathrm{sj_{2}}$ must satisfy the following additional restrictions:
\begin{enumerate}
\item{the \pt of each subjet must be greater than 10\GeV, and}
\item{the mass of the heavier subjet must be less than 2/3 of the large-radius jet mass, where mass refers to the invariant mass of all jet constituents.}
\end{enumerate}
These requirements are obtained from an optimization of the reconstruction efficiency, while maintaining a reasonable misidentification probability.
When these requirements cannot be met, the pair of subjets is discarded, and the procedure is repeated, treating the subjet with largest mass as the initial jet that is then split into two new subjets. If the algorithm is unable to find two subjets satisfying the above criteria within a given CA8 jet, no $\tauh$ reconstruction is performed from this CA8 jet, and the algorithm moves on to the next such jet.
When two subjets satisfying the requirements are found, they are passed to the HPS algorithm as seeds.
At this stage, the algorithm does not differentiate between subjets arising from hadronic or leptonic $\Pgt$ decays.
After reconstruction, the decay-mode criteria (Section~\ref{sec:tauIdAlgorithm_dmreco}) and the MVA-based isolation discriminants (Section~\ref{sec:tauIdDiscrMVABased}) are applied to the reconstructed $\tauh$ candidate, taking into account just the PF candidates belonging to the subjet that seeds the $\tauh$ in the reconstruction and the isolation calculations.
The decay-mode criteria are relaxed relative to those used in the standard HPS algorithm by accepting $\tauh$ candidates with two charged hadrons, and therefore an absolute charge different from unity.
This relaxation recovers $\Pgt$~leptons decaying into three charged hadrons when one of the tracks is not reconstructed in the dense environment of a high-\pt jet.
If an electron or muon, reconstructed and identified through the usual techniques available for these leptons~\cite{Khachatryan:2015hwa,Sirunyan:2018fpa}, is found to be near ($\Delta R<0.1$) to a $\tauh$ candidate reconstructed from a subjet, the corresponding CA8 jet is considered to originate from a semileptonic $\Pgt$ lepton pair decay. Cases in which both $\Pgt$ leptons decay leptonically are not considered.

\begin{figure}[!htbp]
\centering
\subfigure{\includegraphics[width=0.48\textwidth]{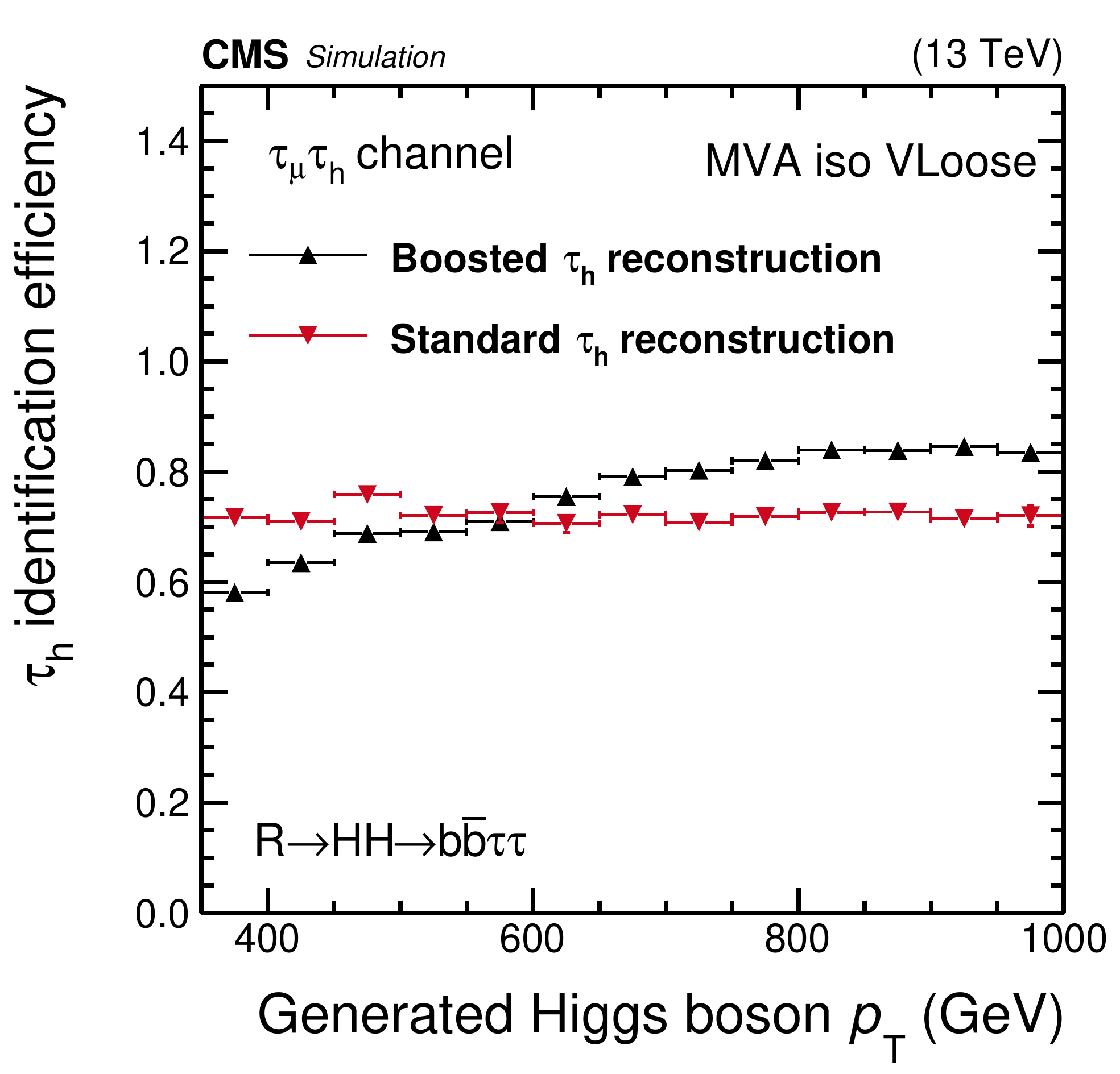}}
\subfigure{\includegraphics[width=0.48\textwidth]{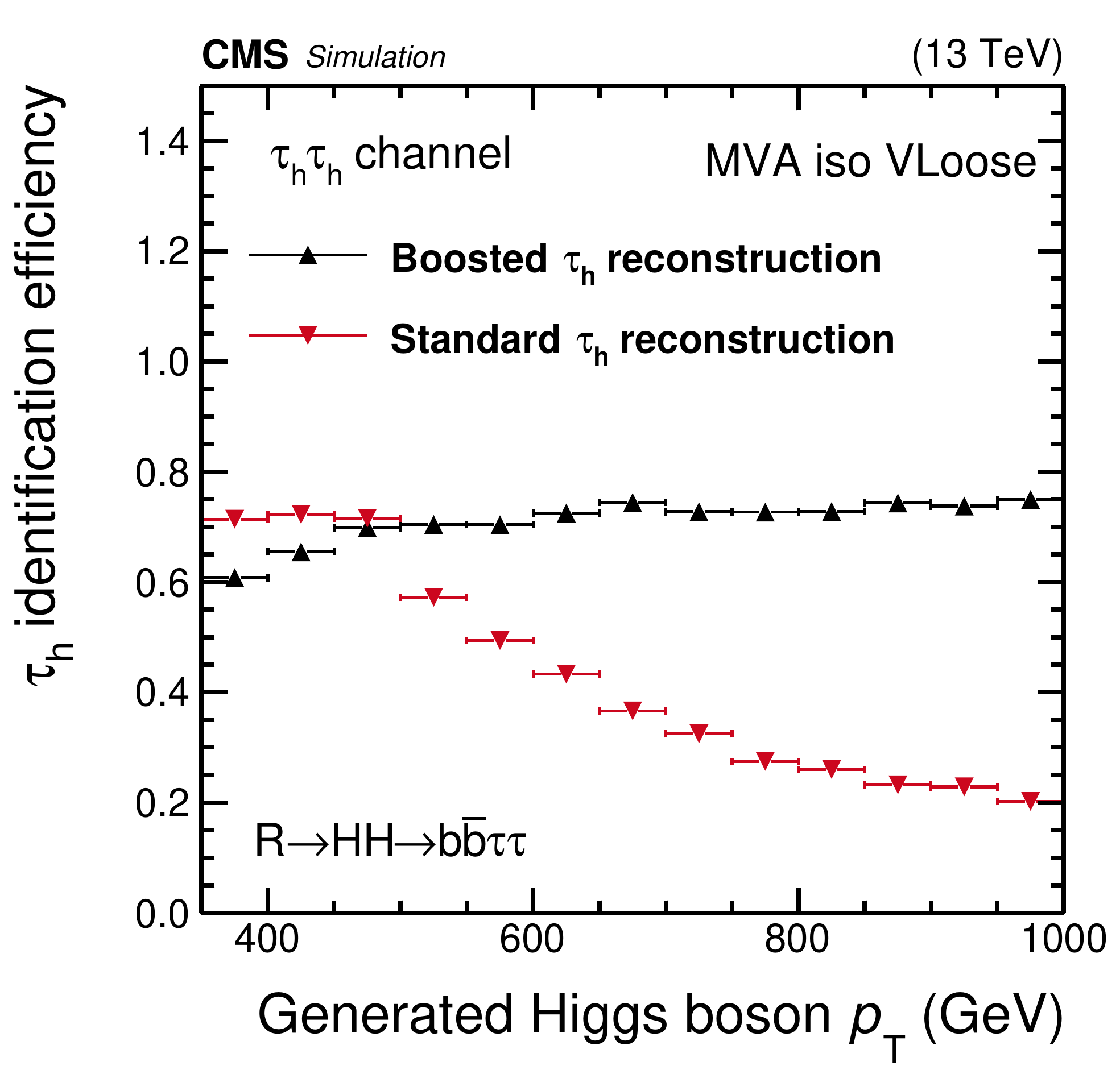}}\\
\subfigure{\includegraphics[width=0.48\textwidth]{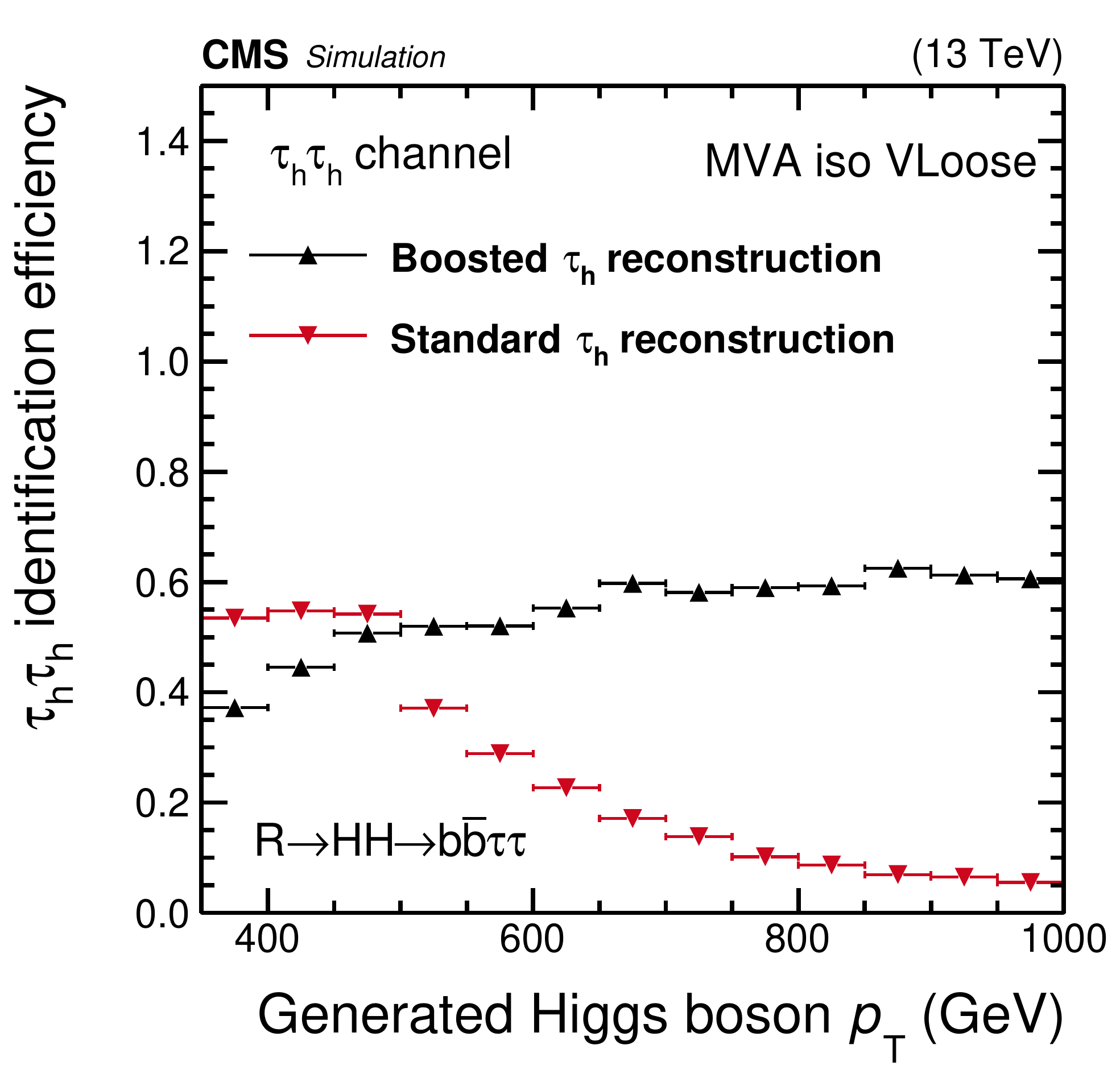}}
\subfigure{\includegraphics[width=0.48\textwidth]{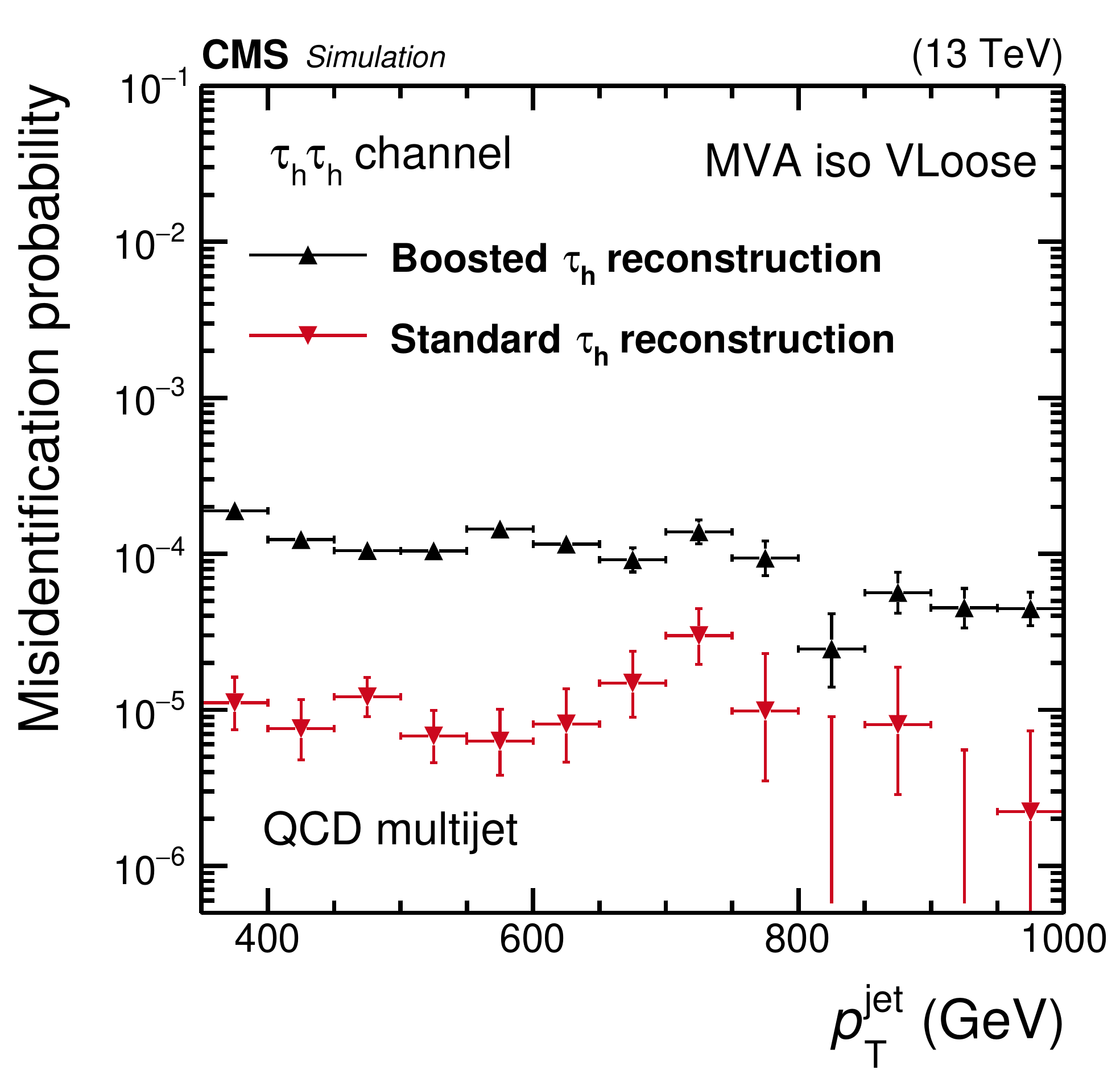}}
\caption{Reconstruction and identification efficiencies for the $\tauh$ in the $\Pgm\tauh$ (upper left) and $\tauh\tauh$ (upper right) final states, and for the $\tauh\tauh$ pair (lower left), as a function of the generated \pt of the Higgs boson, and the probability for large-radius jets in QCD multijet events to be misidentified as $\tauh\tauh$ final states (lower right), as a function of the large-radius~jet~\pt. Vertical bars correspond to the statistical uncertainties (the 68\% Clopper-Pearson intervals), while horizontal bars indicate the bin widths.}\label{fig:efficiency_sig_boosted}
\end{figure}

Figure~\ref{fig:efficiency_sig_boosted} compares the efficiencies in standard reconstruction with that for highly boosted $\Pgt$ lepton pairs in simulated events of R~$\to\PH\PH\to\bbbar\Pgt\Pgt$ decays in the $\tauh\tauh$ and $\Pgt_{\Pgm}\tauh$ final states. In addition, the expected probability for large-radius jets to be misidentified as $\tauh$ pairs is shown for simulated QCD multijet events.
While the efficiency in $\Pgt_{\Pgm}\tauh$ events is computed just for the $\tauh$ candidate, it is computed once relative to one $\tauh$ candidate and once relative to both $\tauh$ candidates in $\tauh\tauh$ events.
The misidentification probability is calculated in $\tauh\tauh$ final states for both $\tauh$ candidates.
The $\tauh$ candidates are selected requiring $\pt>20\GeV$ and $\abs{\eta}<2.3$, using the very loose WP of the MVA-based isolation.

The algorithm used for highly boosted events provides a considerably higher efficiency than the standard HPS algorithm for $\Pgt$ lepton pairs with \pt greater than $\approx$0.5\TeV, with an expected increase in misidentification probability.
Since at such high \pt, the contributions from background are highly suppressed, and the misidentification rate remains of the order of $10^{-4}$, this algorithm can be used for searches in this kinematic regime.

\section{Identification of \texorpdfstring{$\tauh$}{tau[h]} candidates in the high-level trigger}
\label{sec:Tau_Trigger}

Several analyses are based on experimental signatures that include $\tauh$ signals, and therefore, along with the offline reconstruction discussed in Sections~\ref{sec:tauId} and \ref{sec:boostedTauReco}, we also employ dedicated $\tauh$ identification algorithms in the trigger system, at both L1 and HLT.

The L1 system went through a series of upgrades~\cite{Tapper:1556311} in 2015 and 2016, and it is now based on more powerful, fully-programmable FPGA processors and $\mu$TCA logic boards.
This allows more sophisticated $\tauh$ reconstruction and isolation algorithms at L1, the performance of which can be found in Ref.~\cite{CMS-DP-2017-022}.

The HLT system uses the full-granularity information of all CMS subdetectors, and runs a version of the CMS reconstruction that is slightly different than that used offline, as the HLT decision is made within 150 ms, on average, a factor of 100 faster than offline reconstruction.
This is achieved using specialized, fast, or regional versions of reconstruction algorithms, and through implementation of multistep selection logic, designed to reduce the number of events processed by more complex, and therefore more time consuming subsequent steps. Both methods are exploited in the $\tauh$ reconstruction at the HLT.

The $\tauh$ HLT algorithm has three steps.
The first step, referred to as Level~2 (L2), uses only the energy depositions in the calorimeter towers in regions around the L1 $\tauh$ objects with $\Delta R<0.8$. The depositions are clustered into narrow L2 $\tauh$ jets using the anti-\kt algorithm with a distance parameter of 0.2. The only selection criterion required at L2 is a \pt threshold.

In the second step, known as Level 2.5 (L2.5), a simple form of charged-particle isolation is implemented, using just the information from the pixel detector.
Tracks are reconstructed from hits in the pixel detector around the L2 $\tauh$ jets (rectangular regions of $\Delta\eta{\times}\Delta\phi = 0.5{\times}0.5$), and used to form vertices.
If no vertex is found, the $\tauh$ jet is passed to the following step for more detailed scrutiny.
If, on the other hand, at least one vertex is found, the one with highest $\sum\pt^2$ of its tracks is assumed to be the primary hard-scattering vertex in the event. Tracks originating from within $d_z<0.1\unit{cm}$ of the hard-scattering vertex, in an annulus of $0.15<\Delta R<0.4$ centered on the $\tauh$ jet direction, and with at least three hits in the pixel detector, are used in the computation of the $\tauh$ jet isolation.
An L2 $\tauh$ jet is considered isolated if the scalar sum of the \pt of the associated pixel tracks $\sum\pt^\text{track}$ is less than 1.85\GeV.

Finally, at Level 3 (L3), full track reconstruction, using both pixel and strip detectors, is executed using rectangular regions of size $\Delta\eta{\times} \Delta\phi = 0.5{\times}0.5$ around the L2 $\tauh$ jets, followed by the PF reconstruction. Both components are tuned specifically for the fast processing at HLT, as discussed in Ref.~\cite{CMS-PRF-14-001}.

The L3 $\tauh$ algorithm starts with jets clustered from PF particles by the anti-\kt algorithm using a distance parameter of 0.4.
First, photons, contained in a jet, within a fixed
$\Delta\eta{\times}\Delta\phi$ area of $0.05{\times}0.2$ are clustered into the strips, and assigned the $\Pgpz$ mass.
A variable signal-cone size of $\Delta R_\text{sig}^\text{L3} = (3.6\GeV)/\pt^\text{jet}$, with $\Delta R_\text{sig}^\text{L3}$ limited to the range of 0.08--0.12, and an isolation cone of $\Delta R = 0.4$, are defined around the direction of the charged hadron in the jet with highest \pt.
The L3 $\tauh$ candidate is then constructed from the following constituents found within the signal cone: up to three charged hadrons that are ordered in decreasing \pt, and assumed to be charged pions, and all the available $\Pgpz$ candidates.
To recover possible tracking inefficiencies, neutral hadrons within a distance of $\Delta R = 0.1$ from the leading charged hadron are also considered as being part of the $\tauh$ candidate.
The vertex with smallest $d_z$ relative to the track of the leading charged hadron is considered as the vertex of the $\tauh$ production.
To maximize the HLT reconstruction efficiency, these identification criteria are chosen to be fairly inclusive, not requiring strict consistency with the $\tauh$ decay modes, with the respective sizes of the signal and isolation cones chosen to be larger and smaller than the sizes of the corresponding cones in the offline algorithm.

Two types of isolations were defined for L3 $\tauh$ candidates in 2016.
First is the charged isolation ($\sum\pt^{\text{charged}}$), computed by summing the
scalar \pt of charged hadrons (other than those constituting the L3 $\tauh$ candidate) with $d_z<0.2\unit{cm}$
relative to the $\tauh$ vertex, located within the isolation cone; defining the
loose, medium, and tight WPs through $\sum\pt^{\text{charged}}$ being smaller than
3.0, 2.0, and 1.5\GeV, respectively.

The second type is the combined isolation, $I_{\Pgt}^\text{L3}$, defined as
\begin{linenomath}
\begin{equation}
I_{\Pgt}^\text{L3} = \sum\pt^{\text{charged}} + 0.3\,\max\left(0,\sum\pt^{\Pgg}-\pt^{\text{PU}}\right),
\label{eq:l3tauCombIsolation}
\end{equation}
\end{linenomath}
where $\sum\pt^{\Pgg}$ is the sum of the scalar \pt of photons within
an annulus between the signal and isolation cones that do not belong to the signal strips,
and $\pt^{\text{PU}}$ is the neutral contribution to the isolation from PU,
estimated using the jet area method~\cite{Cacciari:2007fd}.
The respective loose, medium, and tight WPs of the combined isolation require $I_{\Pgt}^\text{L3}$ to be smaller than 3.0, 2.3, and 2.0\GeV.

The absolute isolation cutoff values (for both isolation types) are often relaxed by a few percent, depending on the trigger, as a function of $\pt^{\tauh}$, starting at values of about twice the trigger threshold.
This relaxation increases the reconstruction efficiency for genuine \tauh candidates, and is possible because of the number of misidentified \tauh candidates decreases with \pt, providing thereby a control of the trigger rates.

Finally, the scalar \pt sum of photons that are included in the strips of the
L3 $\tauh$ candidate, but are located outside of its signal cone ($R_\text{sig}^\text{L3}$), is defined as for offline $\tauh$ candidates in Eq.~(\ref{eq:pTouterCut}).
This variable was not used for $\tauh$ triggers in 2016, but is included in triggers during data taking in 2017.

The $\tauh$ reconstruction and identification algorithms described in this section
are employed to define a set of triggers for data taking during 2016. The
triggers and their performance are discussed in Section~\ref{sec:Tau_Trigger_Performance}.

\section{Event selection and systematic uncertainties}
\label{sec:validation}

This section describes the selection requirements employed to define event samples used in the following measurements of the performance of $\tauh$ reconstruction and identification in data and simulation, as well as their related systematic uncertainties.
Differences between data and simulated events in trigger, identification, and isolation efficiencies are taken into account through the reweighting of simulated events.
In addition, the number of PU interactions in simulation is reweighted to match that measured in data.

\subsection{The \texorpdfstring{$\cPZ/\Pggx \to \Pgt\Pgt$}{Z/gamma* to tau tau} events}
\label{sec:validation_eventSelection_ZTT}

A sample of $\cPZ/\Pggx$ events decaying into $\Pe\tauh$ or $\Pgm\tauh$ final states is selected by requiring at least one well-identified and isolated electron or muon, referred to as the ``tag'', and one $\tauh$ candidate that passes loose preselection criteria, which corresponds to the ``probe''.

The events in the $\Pe\tauh$ final state are required to pass an isolated single-electron trigger with  $\pt>25\GeV$.
Offline, the electron candidate is required to have $\pt>26\GeV$ and $\abs{\eta}<2.1$, pass the tight WP of the MVA-based electron identification (with an average efficiency of 80\%)~\cite{Khachatryan:2015hwa,CMS-DP-2016-049}, and have $I_\text{rel}^{\Pe} < 0.1$, as defined in Eq.~(\ref{eq:relIso}).
In the $\Pgm\tauh$ final state, events are required to pass an isolated single-muon trigger with $\pt>22\GeV$. Offline, the muon candidate is required to have $\pt>23\GeV$ and $\abs{\eta}<2.1$, pass the medium identification WP~\cite{Sirunyan:2018fpa}, and have $I_\text{rel}^{\Pgm}<0.15$.
The $\tauh$ candidate is preselected to have $\pt>20\GeV$, $\abs{\eta}<2.3$, no overlap with any global muon~\cite{Sirunyan:2018fpa} with $\pt>5\GeV$, to pass the against-lepton discriminant selection requirements defined in Sections~\ref{sec:antiElectronDiscrMVABased} and~\ref{sec:antiMuonDiscrCutBased}, and to have at least one charged hadron with $\pt>5\GeV$.
The $\tauh$ and electron or muon are required to be separated by at least $\Delta R=0.5$,
and to carry opposite electric charges. If several $\Pe\tauh$ or $\Pgm\tauh$ pairs in one event
pass this set of selection criteria, the pair formed from the most isolated $\tauh$ and the most isolated electron or muon is selected.
The events are rejected if they contain an additional electron or muon passing relaxed selection criteria. The relaxed selection requires that an electron satisfies the very loose WP of the MVA-based identification (with an average efficiency of 95\%), a muon has to be reconstructed as a global muon, and both the electron or muon must have $\pt>10\GeV$ and $I_\text{rel}^{\Pe/\Pgm}<0.3$.
To reduce the W+jets background contribution, the transverse mass of the electron or muon and \ptvecmiss, $\mT \equiv \sqrt{\smash[b]{2 \pt^{\Pe/\Pgm} \ptmiss (1-\cos\Delta\phi)}}$, is required to be less than 40\GeV, where $\Delta\phi$ is the difference in azimuthal angle between the electron or muon \ptvec and \ptvecmiss.
In addition, a linear combination of the variables $P_{\zeta}^{\,\ptmiss}$ and $P_\zeta^{\,\text{vis}}$, originally developed by the CDF experiment~\cite{CDFrefPzeta}, namely $D_\zeta=P_\zeta^{\,\ptmiss}\!-0.85\,P_{\zeta}^{\,\text{vis}}$, is used to benefit from the fact that in $\cPZ/\Pggx\to \Pgt\Pgt$ events the \ptvecmiss from the neutrinos produced in $\Pgt$ decays typically forms a small angle with the visible $\tauh$ decay products.
The $D_\zeta$ is required to be greater than $-25\GeV$.

\subsection{The \texorpdfstring{$\Pgm\tauh$ final states in $\ttbar$}{gamma tau[h] final states in ttbar} events}
\label{sec:validation_eventSelection_ttbar}
The $\ttbar\to\Pgm\tauh$+jets events are selected in the same way as the $\cPZ/\Pggx~\to\Pgt\Pgt\to\Pgm\tauh$ events, except for the requirements on $\mT$ and $D_{\zeta}$, which are not applied.
The events are also required to have at least one \cPqb-tagged jet to enrich the content in $\ttbar$ events.

\subsection{The \texorpdfstring{$\cPZ/\Pggx \to \Pgm\Pgm$ events to constrain the $\cPZ/\Pggx \to \Plepton\Plepton$}{Z/gamma* to muon muon events to contstrain the Z/gamma* to lepton lepton} normalization}
\label{sec:validation_eventSelection_Zmm_TnP}
A high purity sample of $\cPZ/\Pggx \to \Pgm\Pgm$ events is selected to constrain the normalization of the Drell--Yan (DY, $\cPq\cPaq \to \cPZ/\Pggx \to \Plepton^{+}\Plepton^{-}$) events in the measurement of $\tauh$ efficiency through the tag-and-probe method~\cite{Khachatryan:2010xn}, described in detail in Section~\ref{sec:tauIdEfficiency_TnP_Ztautau}.
The events are required to have a pair of well-separated ($\Delta R>0.5$), oppositely-charged muons.
The leading (in \pt) muon is required to pass the same selection as used in the $\Pgm\tauh$ final states of $\cPZ/\Pggx$ events.
The subleading muon is required to pass the same selection as the leading muon, except for the $\eta$ requirement, which is relaxed to $\abs{\eta}<2.4$.
The invariant mass of the dimuon pair is required to be within 60--120\GeV.
Events are rejected if they contain an additional electron or muon passing the relaxed selection criteria.

\subsection{Off-shell \texorpdfstring{$\PW \to \Pgt\Pgn$}{W to tau nu} events}
\label{sec:validation_eventSelection_high_Wjets}

Here, we use events in which a virtual $\PW$ boson that decays into a $\tauh$ and a $\Pgn$ is produced with small \pt (and no accompanying hard jet).
The $\ptvec$ of the $\tauh$ and the \ptvecmiss are expected to be well balanced in such events.

Events are required to pass a trigger where $\ensuremath{p_{\text{T, no}\Pgm}^{\text{miss}}}$ and $\ensuremath{H_{\text{T, no}\Pgm}^{\text{miss}}}$ are both greater than 110\GeV, with $p_{\text{T, no}\Pgm}^{\text{miss}}$ being the magnitude of \ptvecmiss computed using all particles in an event except muons, and $H_{\text{T, no}\Pgm}^{\text{miss}}$ being the magnitude of \ptvecmiss computed using jets with $\pt>20\GeV$, reconstructed from all particles except muons.
Offline, events are required to have one $\tauh$ candidate with $\pt>100\GeV$, and
$\ptmiss>120\GeV$.
To ensure back-to-back topologies between the $\tauh$ candidate and $\ptmiss$,
we require $\Delta\phi(\tauh,\ptmiss)>2.8\unit{rad}$.
The event is discarded if it has at least one jet with $\pt>30\GeV$ and $\abs{\eta}<4.7$, except the one corresponding to the $\tauh$, or an additional electron or muon passing the relaxed selection criteria.

\subsection{Off-shell \texorpdfstring{$\PW \to \Pgm\Pgn$}{W to mu nu} events to constrain the \texorpdfstring{$\PW \to \Pgt\Pgn$}{W to tau nu} normalization}
\label{sec:validation_eventSelection_high_Wjets_munu}

This event sample is used to constrain the normalization of off-shell $\PW$ boson production for $m_{\PW}>200\GeV$, used in the $\tauh$ efficiency measurement, as described in Section~\ref{sec:tauIdEfficiency_Wtaunu}.
Events are selected with an isolated single-muon trigger with $\pt>22\GeV$ and $\abs{\eta}<2.1$.
Offline, the muon candidate must have $\pt>120\GeV$ and
$\abs{\eta}<2.1$; it must also pass the medium identification WP, and have a relative isolation of less than 0.15.
The event must also have $\ptmiss>120\GeV$ and $\Delta\phi(\Pgm,\ptmiss)>2.8\unit{rad}$.
The event is discarded if it has at least one jet with $\pt>30\GeV$ and $\abs{\eta}<4.7$, or an additional electron or muon passing the relaxed selection criteria.

\subsection{Events from \texorpdfstring{$\PW \to \Pgm\Pgn$+jet}{W to mu nu + jet} production}
\label{sec:validation_eventSelection_Wjets}

These events are triggered using a single isolated-muon trigger with $\pt>24\GeV$ and $\abs{\eta}<2.1$.
Offline, we require one well-identified and isolated muon with $\pt>25\GeV$.
Events with additional electrons or muons passing the relaxed selection criteria are rejected.
In addition, the transverse mass of the muon and $\ptvecmiss$ is required to be greater than 60\GeV, to suppress events with genuine $\tauh$ candidates, in particular from $\cPZ/\Pggx$~bosons.
Events should contain exactly one jet with $\pt>20\GeV$ and $\abs{\eta}<2.4$, and there should be no
additional jets (in $\abs{\eta}>2.4$) with $\pt>20\GeV$.
To ensure that the $\PW$ boson is balanced in \pt with the jet, the following selections are applied: $\Delta\phi(\PW,\text{jet})>2.4\unit{rad}$, and the ratio of jet \pt and $\PW$ boson \pt must be between 0.7 and 1.3, where the \pt of the $\PW$ boson is reconstructed from the vector sum of muon $\ptvec$ and $\ptvecmiss$.

\subsection{The \texorpdfstring{$\Pe\Pgm$}{e mu} final states in \texorpdfstring{$\ttbar$}{ttbar} events}
\label{sec:validation_eventSelection_ttbar_emu}
These events are triggered using a single isolated-muon trigger with $\pt>24\GeV$,
and are required to have one well-identified and isolated electron and one well-identified and isolated muon both of $\pt>26\GeV$ and $\abs{\eta}<2.4$.
Events with additional electrons or muons passing the relaxed selection criteria are rejected.

\subsection{The \texorpdfstring{$\cPZ/\Pggx \to \Pe\Pe,\,\Pgm\Pgm$}{Z/gamma* to ee mu mu} events for measuring the \texorpdfstring{$\Pe/\Pgm\mapsto\tauh$}{e/mu maps to tau[h]} misidentification probability}
\label{sec:validation_eventSelection_Zll}

High-purity samples of $\cPZ/\Pggx \to \Pe\Pe$ and $\cPZ/\Pggx \to \Pgm\Pgm$ events are selected for measuring their respective $\Pe\mapsto\tauh$ and $\Pgm\mapsto\tauh$ misidentification probabilities. Consequently, again, we require at least one well-identified, isolated electron or muon (tag) and one isolated $\tauh$ candidate (probe).

The $\cPZ/\Pggx \to \Pe\Pe$ events are selected by requiring a single-electron trigger to have fired. Offline, the electron candidate must match the trigger object (within $\Delta R<0.5$), have $\pt>26\GeV$ and $\abs{\eta}<2.1$, pass the most-restrictive electron-identification criteria, and have an $I_\text{rel}^{\Pe}<0.1$. The $\cPZ/\Pggx \to \Pgm\Pgm$ events are collected using a single isolated-muon trigger with $\pt>24\GeV$. Offline, the muon candidate must match the trigger object (within $\Delta R<0.5$), be selected with $\pt>26\GeV$ and $\abs{\eta}< 2.1$, after passing medium muon-identification criteria, and $I_\text{rel}^{\Pgm}<0.15$.

The $\tauh$ candidate is required to satisfy $\pt>20\GeV$ and $\abs{\eta}<2.3$, be reconstructed in one of the decay modes $\oneProngZeroPizero$, $\oneProngOnePizero$, $\oneProngTwoPizero$, or $\threeProngZeroPizero$, and pass the tight WP of the MVA-based isolation discriminant described in Section~\ref{sec:tauIdDiscrMVABased}. It must also be separated from the electron or muon by $\Delta R>0.5$, and have an electric charge opposite to that of the electron or muon.
The $\tauh$ candidate must pass the loose WP of the against-$\Pgm$ discriminant described in Section~\ref{sec:antiMuonDiscrCutBased} when selecting $\cPZ/\Pggx \to \Pe\Pe$ events. The purity of the sample is increased by requiring the invariant mass of the tag-and-probe pair to be between 60--120 or 70--120\GeV for $\cPZ/\Pggx \to \Pe\Pe$ and $\cPZ/\Pggx \to \Pgm\Pgm$ events, respectively.

The $\PW$+jets and $\ttbar$ backgrounds are reduced by requiring the selected events to have $\mT$ (of the tag electron or muon and \ptvecmiss) not exceeding 30\GeV.

\subsection{Systematic uncertainties affecting all studied final states}
\label{sec:systematics}

The generic systematic uncertainties affecting most of the measurements presented in Sections~\ref{sec:Tau_identification_efficiency}--\ref{sec:TauEnergyScale} are discussed in this section.
Uncertainties concerning particular analyses are not covered here, but are discussed in their corresponding sections.
The same is true for deviations in the values of the systematic uncertainties.

The uncertainty in the measured integrated luminosity is 2.5\%~\cite{CMS-PAS-LUM-17-001}, and affects the normalization of all processes modelled via MC simulation. The combination of trigger, identification, and isolation efficiencies for electrons and muons, measured using the tag-and-probe technique, result in normalization uncertainties of 2\% that also affect the normalization of processes modelled in simulation.
Uncertainties in the normalization of production cross sections~\cite{Campbell:1999ah,Campbell:2011bn,Campbell:2015qma,Gavin:2010az,Gavin:2012sy,Li:2012wna,Czakon:2011xx} or in the method used to extract the normalization of $\ttbar$ (3--10\%), diboson (5--15\%), and DY (2--4\%) production, are also taken into account. Uncertainties in the $\tauh$ energy scale, affecting the distributions in simulated events that depend on $E_{\tauh}$, and range between $1.2\%$ (as determined in Section~\ref{sec:TauEnergyScale}) and $3\%$ for high-$\pt$ $\tauh$ candidates.
Furthermore, to account for statistical fluctuations caused by the limited number of simulated events, we use the ``Barlow-Beeston light'' approach~\cite{Barlow:1993, Conway:2011in}, which assigns a single nuisance parameter per bin that rescales the total bin yield.
Most of the analyses discussed in the following sections correct the simulated $\pt$ distributions of the $\cPZ/\Pggx$ boson in DY events and of the top quark in $\ttbar$ events to the spectra observed in data through measured weights.
This reweighting corrects only the differential distributions without changing their normalization.
Uncertainties in these weights are propagated through the analyses, where the downward changes by one standard deviation are computed as a difference between the weighted distribution and the one without weight, while the upward changes by one standard deviation are computed as a difference between weighted distributions with nominal weight and with the square of that weight. Finally, the uncertainty related to the PU distribution is estimated by changing the minimum-bias $\Pp\Pp$ cross section by $\pm$5\%.

A comprehensive overview of these uncertainties is given in Table~\ref{tab:systematics}.

\begin{table}[!hbtp]
\topcaption{Systematic uncertainties affecting the measurements described in Sections~\ref{sec:Tau_identification_efficiency}--\ref{sec:TauEnergyScale}. Given are the source of the uncertainty and whether the distribution in question is affected.}
\label{tab:systematics}
\centering
\resizebox{\textwidth}{!}{
\begin{tabular}{llc}
Uncertainty & Value & Affecting distribution? \\
\hline
Integrated luminosity & 2.5\% & No \\
$\Pe$ trigger, identification, and isolation efficiency & 2\% & No \\
$\Pgm$ trigger, identification, and isolation efficiency & 2\% & No \\
DY normalization & 2--4\% & No \\
$\ttbar$ normalization & 3--10\% & No \\
Diboson normalization & 5--15\% & No \\
$\tauh$ energy scale & 1.2--3\% & Yes \\
\multirow{2}{*}{Limited number of events} & Statistical uncertainty & \multirow{2}{*}{Yes} \\
 & in individual bin & \\
DY $\pt$ & (Weight)$^{2}$ -- no weight  & Yes \\
$\ttbar$ $\pt$ & (Weight)$^{2}$ -- no weight & Yes \\
Number of PU events & 5\% & Yes \\
\hline
\end{tabular}
}

\end{table}

\section{Measurement of the \texorpdfstring{$\tauh$}{tau[h]} identification efficiency}
\label{sec:Tau_identification_efficiency}
The measurements of $\tauh$ reconstruction and identification efficiencies in data use approaches similar to those of Ref.~\cite{TAU-14-001},
and provide data-to-simulation scale factors and their uncertainties that can be used to correct the simulated predictions in analyses.
The efficiency is measured in different $\pt^{\tauh}$ regions: small $\pt^{\tauh}$ between 20 and $\approx$60\GeV, using the $\Pgm\tauh$ final state of $\cPZ/\Pggx\to\Pgt\Pgt$ events, as discussed in Section~\ref{sec:tauIdEfficiency_TnP_Ztautau};
intermediate $\pt^{\tauh}$ of up to $\approx$100\GeV, using the $\Pgm\tauh$ final states in $\ttbar$ events, as discussed in Section~\ref{sec:tauIdEfficiency_ttbar};
and high $\pt^{\tauh}$ of $>$100\GeV, using a selection of highly virtual $\PW$ bosons~($m_{\PW}>200\GeV$) decaying into $\Pgt$ leptons, as presented in Section~\ref{sec:tauIdEfficiency_Wtaunu}.
The data-to-simulation scale factors obtained through these measurements are combined, as described in Section~\ref{sec:tauIdEfficiency_extrapolation}, to extrapolate to higher-$\pt^{\tauh}$ regions not covered by these measurements. Finally, the identification efficiency for $\tauh$ candidates reconstructed using the algorithm dedicated to highly boosted $\Pgt$ lepton pairs is measured using the tag-and-probe method, as described in Section~\ref{sec:tauIdEfficiency_TnP_BoostedTaus}.

\subsection{Using the tag-and-probe method in \texorpdfstring{$\cPZ/\Pggx$}{Z/gamma*} events}
\label{sec:tauIdEfficiency_TnP_Ztautau}
The $\tauh$ identification efficiency for $\pt^{\tauh}$ up to $\approx$60\GeV is estimated in $\Pgm\tauh$ final states of $\cPZ/\Pggx$ events, selected as described in Section~\ref{sec:validation_eventSelection_ZTT}.
The events are subdivided into passing (``pass'' region) and failing (``fail'' region) categories, depending on whether the $\tauh$ candidate passes or fails the appropriate working point of the $\tauh$ isolation discriminant.
The data-to-simulation scale factor for the $\tauh$ identification efficiency is extracted from a maximum likelihood fit of the invariant mass distribution of the reconstructed (visible) $\Pgm\tauh$ system, referred to as $\mvis$.
The expected SM contributions are fitted to the observed data simultaneously in both categories.

The predictions for SM processes contributing to the distribution in $\mvis$ consist of a signal sample of $\cPZ/\Pggx\to\Pgt\Pgt\to\Pgm\tauh$ events, where the reconstructed $\tauh$ candidate is required to be matched to the generated one, and a set of backgrounds.
All background events, except for QCD multijet production, rely on simulated $\mvis$ distributions. Diboson, single top quark, and $\ttbar$ samples are normalized to their theoretical cross sections. A sample of dimuon events, as described in Section~\ref{sec:validation_eventSelection_Zmm_TnP}, is used to constrain the normalization of the DY process, by using them simultaneously in the fit, along with the events in the passing and failing categories.
The DY processes, other than the $\cPZ/\Pggx\to\Pgt\Pgt\to\Pgm\tauh$ signal, where $\tauh$ candidates from the misidentification of $\Pe$, $\Pgm$, or jets, contribute to the background, and are denoted as ``other DY''.

The normalization of the contribution from $\PW$+jets events is estimated using control samples in data.
A data-to-simulation scale factor is estimated in a sample enriched in $\PW$+jets events, defined in a way similar to the signal sample, but without the $D_\zeta$ requirement having been applied, and with $\mT>80\GeV$, where small contributions from other processes are subtracted from data, based on their estimated cross sections. The scale factor is then applied to the simulation of the $\PW$+jets events in the low-$\mT$ signal sample.

The distribution and normalization of the QCD multijet background is estimated from control samples in data.
The distribution is extracted from a sample selected using the nominal selection criteria discussed previously, but requiring the $\Pgm$ and $\tauh$ candidates to have the same-sign (SS) electric charge. All other processes contributing to this sample are estimated using the procedures detailed above, and are subtracted from the data. The normalization is controlled using the ratio of events found in two separate control samples requiring same- and opposite-sign (OS) charge for the $\Pgm$ and $\tauh$ candidates, respectively.
Otherwise, both samples are defined in ways similar to that of the signal sample, but with an inverted muon isolation criterion.

The following uncertainties are considered in addition to the ones outlined in Section~\ref{sec:systematics}, that is, uncertainties in the $\PW$+jets background normalization that arise from a possible difference between the low- and high-$\mT$ regions and from the uncertainties in \ptmiss, which are used in computing $\mT$. The uncertainty in the yield of $\PW$+jets events is estimated to be about 10\%.
The uncertainty in the OS/SS scale factor, used in the estimation of the QCD multijet background is $\approx$5\%, which is mostly due to the limited number of events in the OS and SS control regions. The normalization of the DY process is extracted from the dimuon control region. An extrapolation uncertainty of 2\% is used for the $\Pgm\tauh$ sample to account for the differences in lepton kinematics (mostly in \pt).

\begin{figure}[!htbp]
\centering
\includegraphics[width=0.48\textwidth]{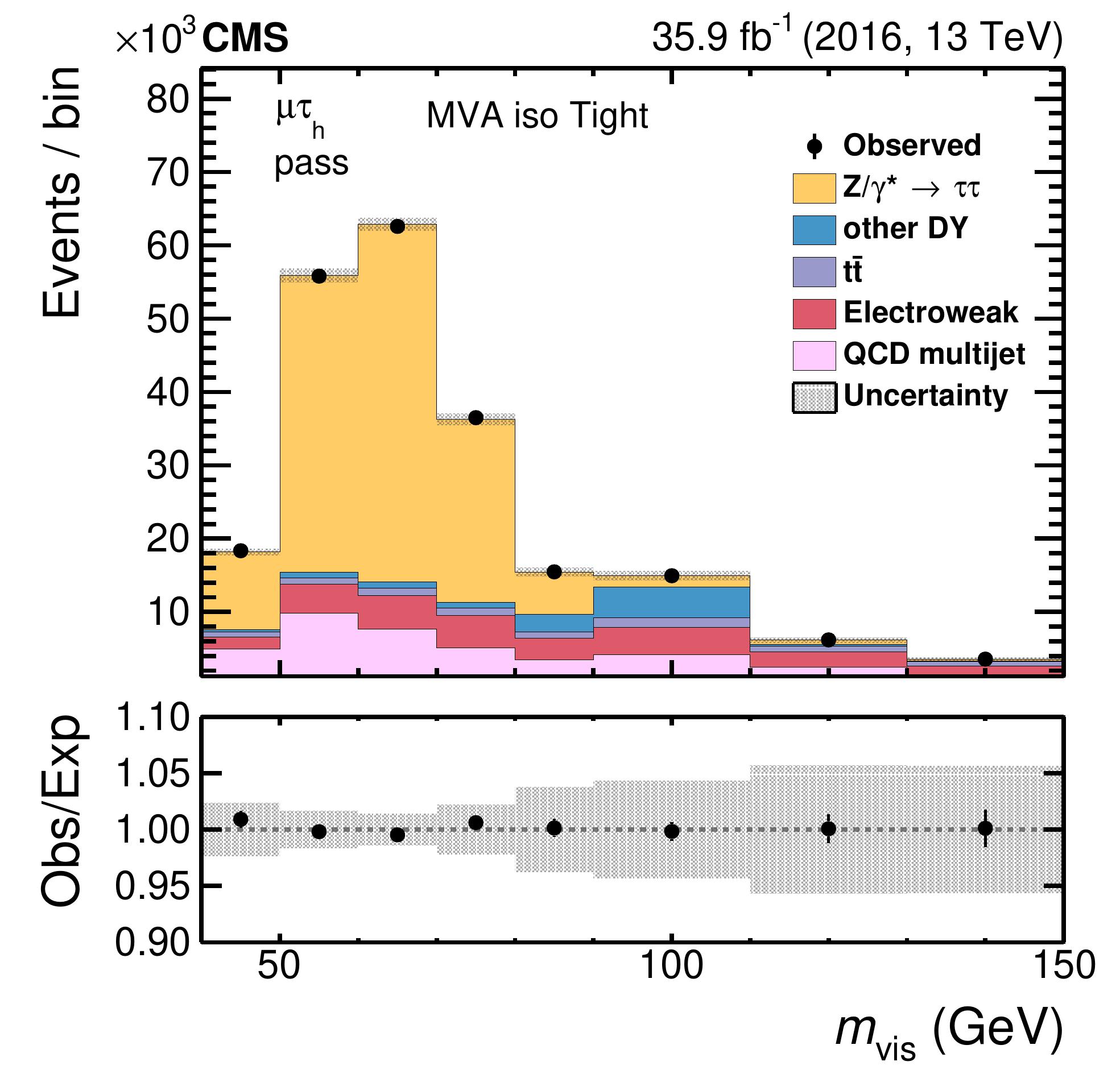}
\includegraphics[width=0.48\textwidth]{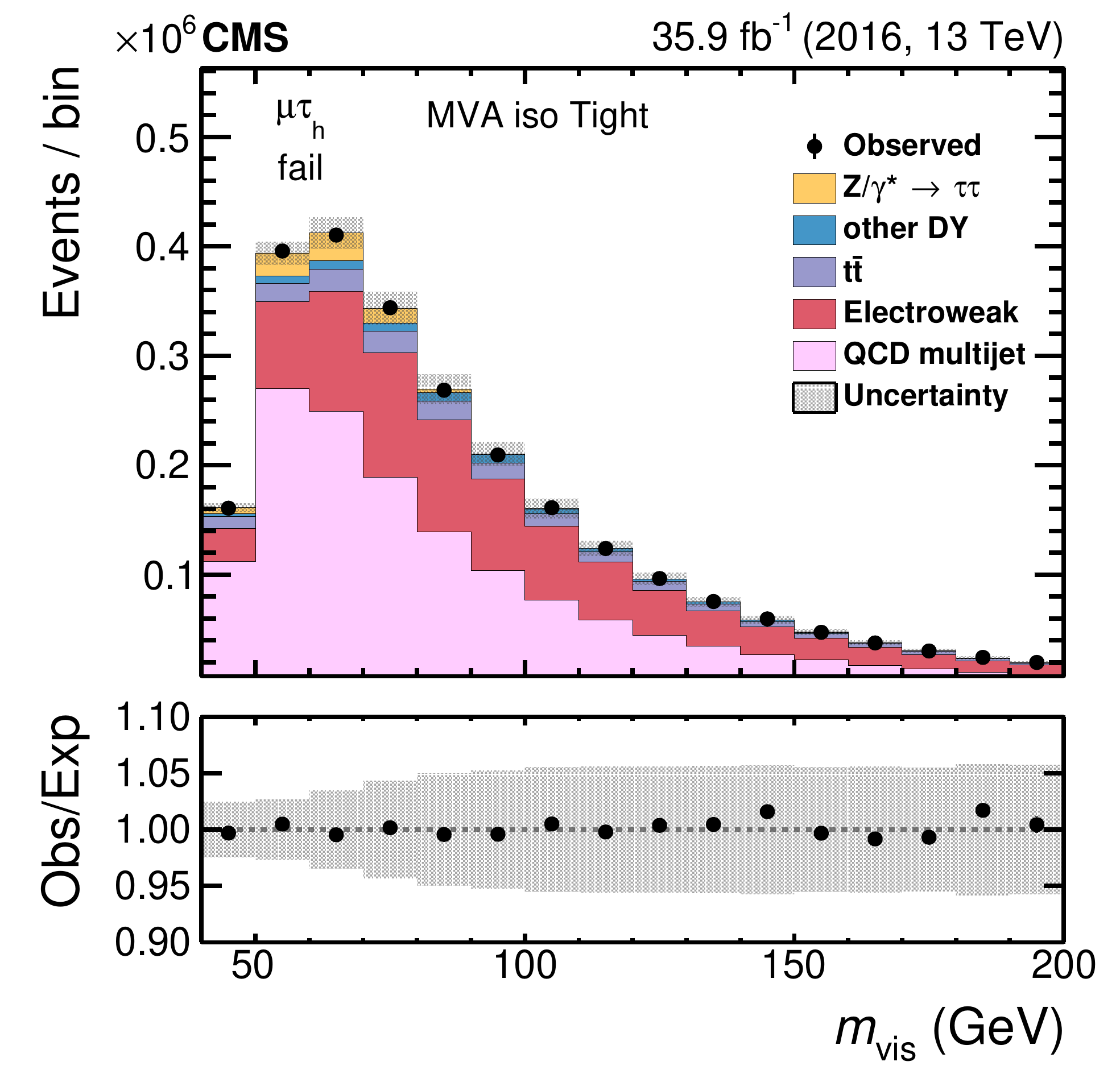}
\caption[dummy text]{The fitted distribution in $\mvis$ in the passing (left) and failing (right) categories for the tight WP of the MVA-based isolation. The electroweak background includes contributions from $\PW$+jets (dominating), diboson, and single top quark events. Vertical bars correspond to the statistical uncertainties in the data points (68\% frequentist confidence intervals), while shaded bands to the quadratic sum of the fitted statistical and systematic uncertainties.
}
\label{fig:tauID_MVA3T}

\end{figure}

The results obtained for different working points of the MVA-based discriminant with $\Delta R=0.5$ are shown in Table~\ref{tab:tauID_res1}.
An uncertainty of 3.9\% is added in quadrature to the one returned by the fit, to account for the uncertainty associated with the track reconstruction efficiency~\cite{Chatrchyan:2014fea}.
The scale factors obtained for different working points of the isolation sum discriminants are found to be close to 90\%, with uncertainties of 5\%, and the scale factors obtained for the MVA-based discriminants trained using 2016 simulations as well as for $\Delta R=0.3$, are found to be compatible with those presented in Table~\ref{tab:tauID_res1}.  The measured scale factors vary from 0.92 to 0.99, depending on the working point, with uncertainties of about 5\%.
The fitted distributions that maximize the likelihood for the
tight WP of the MVA-based isolation are shown in Fig.~\ref{fig:tauID_MVA3T}.
The scale factors are also measured in different ranges of $\pt^{\tauh}$ for the tight WP of the MVA-based isolation discriminant with $\Delta R=0.5$, and enter the extrapolation to high $\pt^{\tauh}$, as discussed in Section~\ref{sec:tauIdEfficiency_extrapolation}.

\begin{table}[!hbtp]
\topcaption{Data-to-simulation scale factors for different MVA-based isolation working points with $\Delta R=0.5$, measured using $\cPZ/\Pggx$ events. An uncertainty of 3.9\% is added in quadrature to the uncertainty returned by the fit to account for the uncertainty in track reconstruction efficiency.}
\label{tab:tauID_res1}
\centering
\begin{tabular}{lc}
Working point & Scale factor \\
\hline
Very loose & 0.99 $\pm$ 0.05 \\
Loose  & 0.98 $\pm$ 0.05 \\
Medium  & 0.97 $\pm$ 0.05 \\
Tight & 0.95 $\pm$ 0.05 \\
Very tight & 0.92 $\pm$ 0.05 \\
Very-very tight & 0.93 $\pm$ 0.05 \\
\hline
\end{tabular}

\end{table}

The efficiency for $\tauh$ candidates to pass the working points of the discriminants used to reject electrons and muons, described in Sections~\ref{sec:antiElectronDiscrMVABased} and~\ref{sec:antiMuonDiscrCutBased}, respectively, are also measured in the $\Pgm\tauh$ final states of $\cPZ/\Pggx\to\Pgt\Pgt$ events, which are selected as described above. The $\tauh$ candidates are required to have $\pt>20\GeV$, $\abs{\eta}<2.3$, and to pass the tight WP of the MVA-based $\tauh$ isolation discriminant.
The events are again subdivided into passing and failing categories, depending on whether the $\tauh$ candidate passes or fails the appropriate working points of the discriminants used against electrons or muons.
The data-to-simulation scale factor is obtained from a maximum likelihood fit to the $\mvis$ distribution. The scale factors are compatible with unity to within the uncertainty in the measurements that range between 1 and 3\%.

\subsection{Using \texorpdfstring{$\ttbar$}{ttbar} events}
\label{sec:tauIdEfficiency_ttbar}

A sample of $\ttbar$ events with a muon and a $\tauh$ in the final state is used to measure the $\tauh$ identification efficiency for $\pt^{\tauh}$ up to 100\GeV. The selection requirements are described in Section~\ref{sec:validation_eventSelection_ttbar}. The selected $\tauh$ candidate must be accepted using the appropriate working point of the $\tauh$ isolation discriminant.
The distribution in $\mT$ of the muon and \ptvecmiss is used to determine the data-to-simulation scale factors.

Contributions to $\mT$ distribution from $\cPZ/\Pggx\to\Pgt\Pgt$, single top quark, diboson, and $\PW$+jets events are modelled using simulations normalized to theoretical cross sections. Background from QCD multijet production is determined as described in Section~\ref{sec:tauIdEfficiency_TnP_Ztautau}. The major background contribution is from $\ttbar$ events where a jet is misidentified as a $\tauh$ candidate. The distribution is taken from simulation and a dedicated sample of events is selected to constrain the normalization of this background, as well as the probability of a jet to be misidentified as a $\tauh$ candidate. Events have to pass the same criteria as discussed in Section~\ref{sec:validation_eventSelection_ttbar}, but must also contain an additional isolated electron of electric charge opposite to that of the selected muon. This selects the $\Pe\Pgm$ final state of $\ttbar$ events with an additional jet which can be misidentified as a $\tauh$ candidate. These $\Pe\Pgm$ events are then subdivided into passing and failing categories, based on whether the requirements imposed on the $\tauh$ candidate are met in the $\tauh$ isolation discriminant.
A simultaneous likelihood fit is performed to the $\mT$ distribution in all three samples, constraining thereby the $\ttbar$ contribution and the probability for jets to be identified as $\tauh$ candidates, as well as measuring the efficiency of the $\tauh$ identification relative to that expected in simulation.

\begin{figure}[!htbp]
\centering
\begin{tabular}{cc}
\multicolumn{2}{c}{\includegraphics[width=0.48\textwidth]{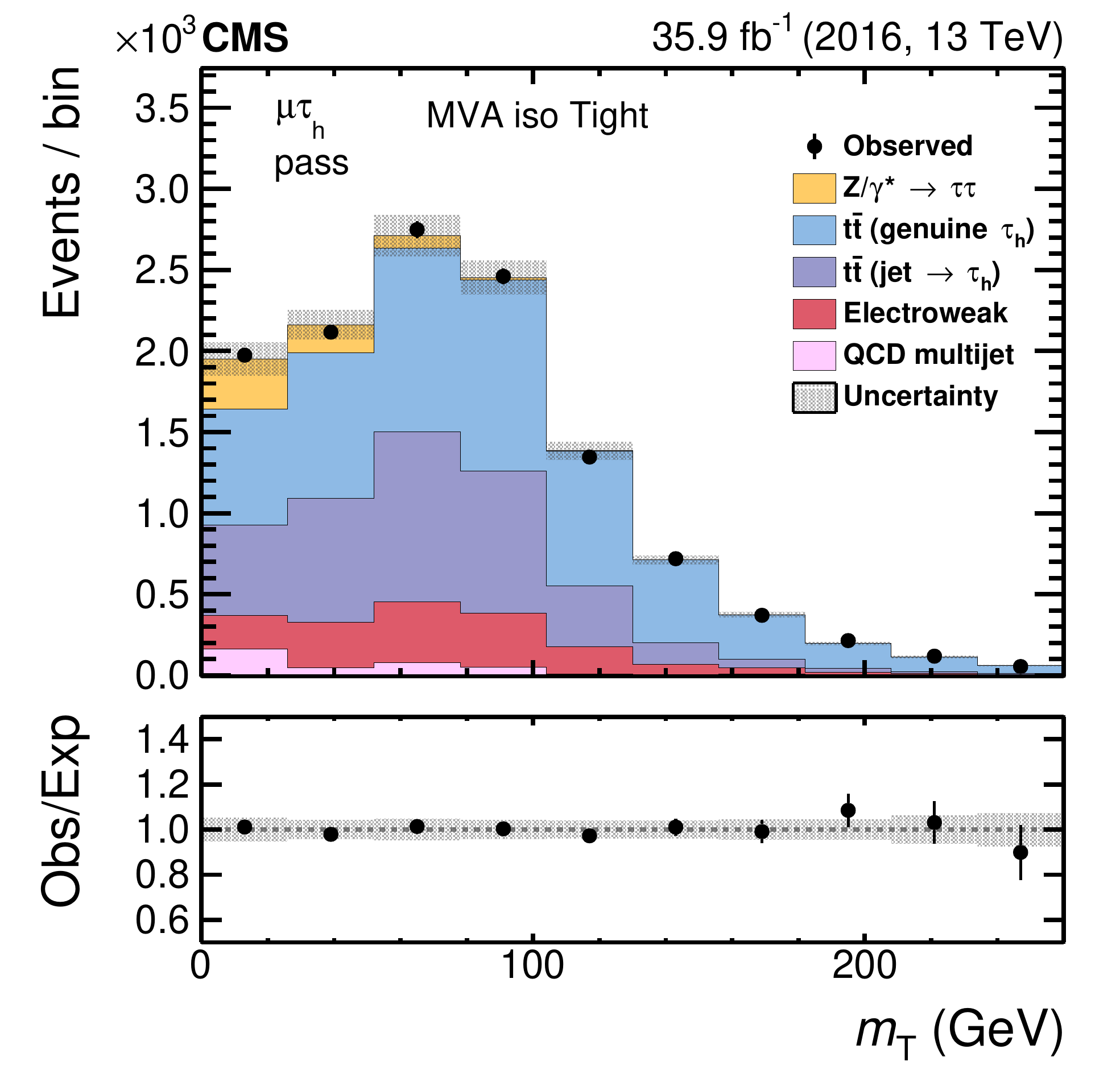}}\\
\includegraphics[width=0.48\textwidth]{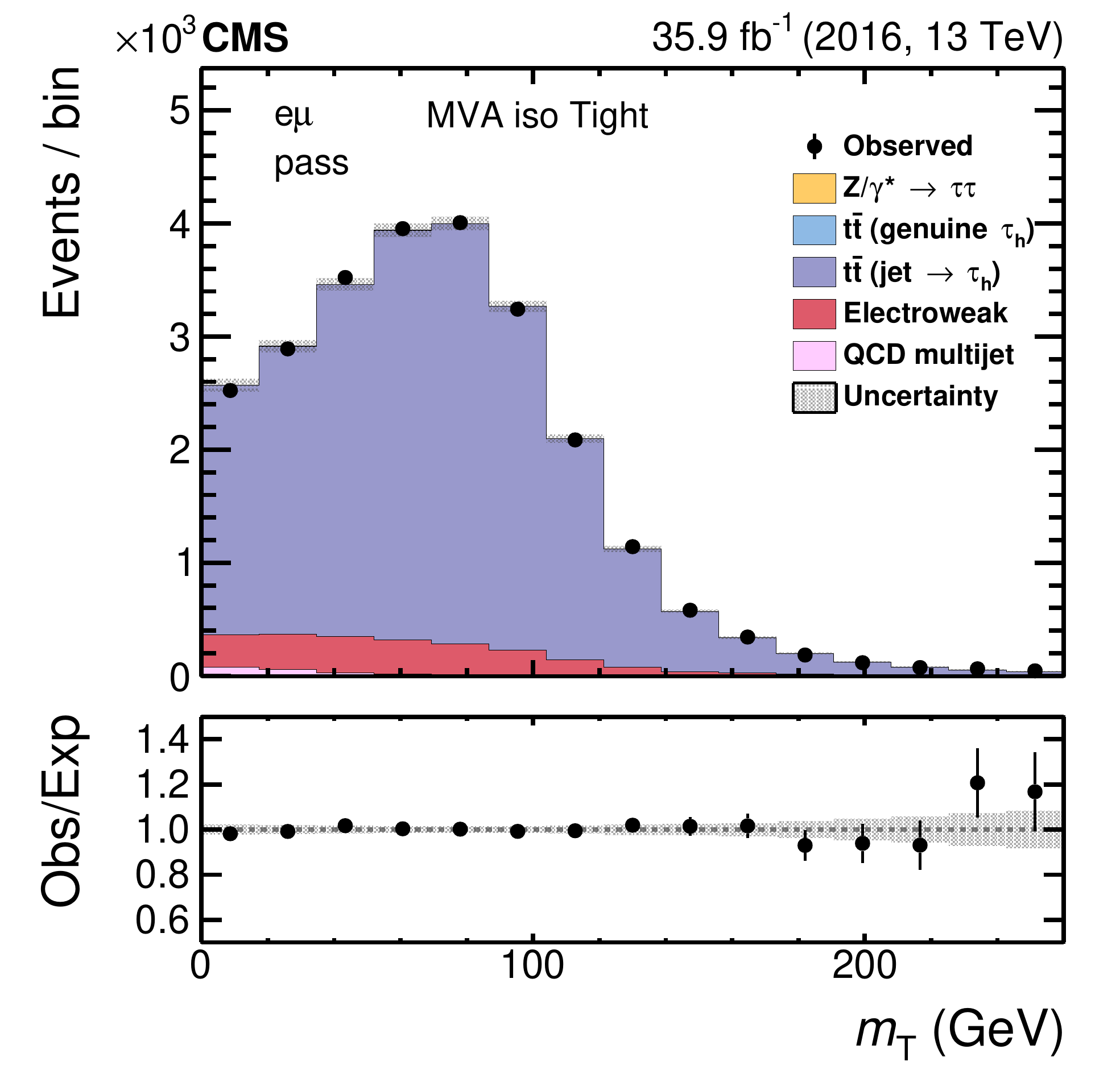} &
\includegraphics[width=0.48\textwidth]{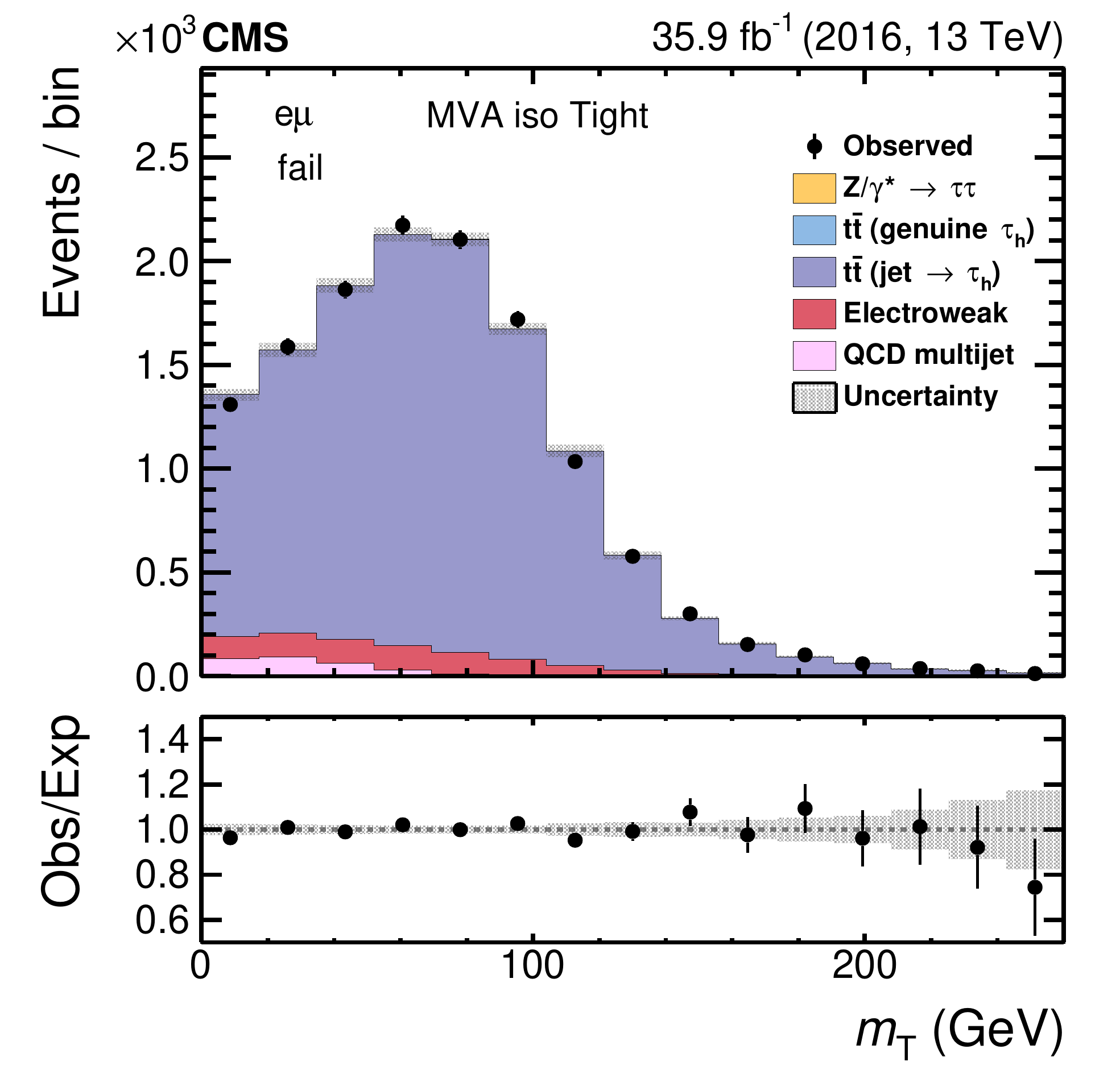} \\
\end{tabular}
\caption[dummy text]{Fitted distributions for the signal (upper), $\Pe\Pgm$ passing (lower left), and the $\Pe\Pgm$ failing (lower right) categories, using the $\mT$ for the $\ptvec^{\mu}$ and \ptvecmiss vectors as observables for the tight WP of the MVA-based isolation with $\pt^{\tauh}$ between 30 and 40\GeV. The electroweak background includes contributions from $\PW$+jets (dominating), diboson, and single top quark events. Vertical bars correspond to the statistical uncertainties in the data points (68\% frequentist confidence intervals), while the shaded bands reflect the quadratic sum of the statistical and systematic uncertainties after the fit.}
\label{fig:tauID_ttbar_tight}

\end{figure}

The systematic uncertainties are similar to those listed in Section~\ref{sec:systematics}, except for additional uncertainties related to the $\cPqb$ tagging performance (3\% effect on the normalization), and the cross section for $\cPZ/\Pggx$+jet process (30\%), given that the $\cPZ/\Pggx$+$\cPqb$~jet cross section is not well measured. A 3.9\% uncertainty in the track reconstruction efficiency is added to the signal processes. The uncertainty in the jet $\mapsto\tauh$ misidentification probability is correlated between the signal and the control sample, where the $\tauh$ candidate passes the identification requirement. The $\Pe\Pgm$ failing category is used to further constrain both the normalization for $\ttbar$ production as well as the uncertainty in $\cPqb$ tagging performance. Figure~\ref{fig:tauID_ttbar_tight} shows the fitted distributions in $\mT$ for the tight WP of the MVA-based isolation.

The measurement is repeated for different isolation working points of the MVA-based discriminant, as well as for the tight WP in different regions of $\pt^{\tauh}$, and individually for each reconstructed decay mode.
Although the mean value of the scale factor in the $\threeProngZeroPizero$ decay mode is slightly below those of the other decay modes, no significant differences are observed between the three decay modes. The measured scale factors in different $\pt^{\tauh}$ regions enter the extrapolation as outlined in Section~\ref{sec:tauIdEfficiency_extrapolation}, and Table~\ref{tab:tauID_res_ttbar} summarizes the results for the working points of the MVA-based isolation discriminants.
The scale factors measured from the inclusive $\ttbar$ events are slightly lower than those from $\cPZ/\Pggx\to\Pgt\Pgt$.
This is because the jet~$\mapsto\tauh$ misidentification probability is slightly higher in simulation than in data, causing the $\tauh$ identification efficiency scale factor to be pulled down towards lower values, where the distributions of the $\ttbar$ events with genuine $\tauh$ and the misidentified jet~$\mapsto\tauh$ candidates become similar.
However, this is mitigated for the measurement in bins of $\pt^{\tauh}$, by constraining the normalization of the $\ttbar$ background with a jet misidentified as a $\tauh$ candidate, using the $\Pe\Pgm$ passing sample, as discussed above.

\begin{table}[!hbtp]
\topcaption{Data-to-simulation scale factors for different MVA-based isolation working points obtained from $\ttbar$ events.}
\label{tab:tauID_res_ttbar}
\centering
\begin{tabular}{lc}
Working point & Scale factor \\
\hline
Very loose & 0.99 $\pm$ 0.07 \\
Loose & 0.94 $\pm$ 0.07 \\
Medium & 0.91 $\pm$ 0.07 \\
Tight & 0.92 $\pm$ 0.06 \\
Very tight & 0.89 $\pm$ 0.06 \\
Very-very tight & 0.86 $\pm$ 0.06 \\
\hline
\end{tabular}

\end{table}

\subsection{Using off-shell \texorpdfstring{$\PW\to\Pgt\Pgn$}{W to tau nu} events}
\label{sec:tauIdEfficiency_Wtaunu}

The identification efficiency
for \tauh leptons with $\pt>100\GeV$ is measured using
a sample of events in which a highly virtual W boson ($m_{\PW}>200\GeV$) is produced at small \pt (and often without an accompanying hard jet), and decays into a $\Pgt$ lepton and $\Pnut$.
The signature for such events consists of a single \tauh decay and \ptvecmiss balanced by the $\ptvec^{\tauh}$.

The selection requirements for the $\PW\to\Pgt\Pgn$ sample are described in Section~\ref{sec:validation_eventSelection_high_Wjets}.
A large fraction of events selected in this channel
originate from processes where a jet is misidentified as a $\tauh$ candidate.
The main processes contributing to this kind of background are
QCD multijet,  $\cPZ/\Pggx\to\Pgn{\Pagn}$+jets, and  $\PW\to\ell\Pgn$+jets events.

The background from events where a jet is misidentified as a $\tauh$ candidate is estimated using a control sample obtained by applying the same set of requirements as used in the selection of the $\PW\to\Pgt\Pgn$ events, except for the $\tauh$ isolation criterion, which is inverted.
Events in this control sample are then extrapolated to the signal region using the ratio of probabilities for a jet to pass to that to fail the $\tauh$ isolation.
The $\PW\to\Pgm\Pgn$+1\,jet and QCD dijet events are utilized to estimate the extrapolation factor.
The method is verified with simulated samples of $\PW\to\ell\Pgn$+jets and $\cPZ/\Pggx\to\Pgn\bar{\Pgn}$+jets events.

The study shows that the set of requirements outlined in Section~\ref{sec:validation_eventSelection_high_Wjets}, selects $\PW\to\Pgt\Pgn$ events with an invariant mass of
the $\Pgt\Pgn$ pair $m_{\Pgt\Pgn}\equiv m_{\PW}>200\GeV$.
A dedicated auxiliary sample of $\PW\to\Pgm\Pgn$ events is used to constrain the normalization of virtual $\PW$ boson production with $m_{\PW}>200\GeV$.
The $\PW\to\Pgm\Pgn$ events are selected as described in Section~\ref{sec:validation_eventSelection_high_Wjets_munu}, and verified using MC simulation that the phase space covered by the $\PW\to\Pgm\Pgn$ and $\PW\to\Pgt\Pgn$ samples tend to largely overlap.

The signal is extracted using a simultaneous maximum likelihood fit to the $\mT$ (of the $\ptvec^{\tauh/\Pgm}$ and \ptvecmiss) distribution for both the $\PW\to\Pgt\Pgn$ signal and $\PW\to\Pgm\Pgn$ control samples. This procedure minimizes the uncertainties related to the normalization of $\PW$ boson events.
The fit is performed using two freely floating parameters:

\begin{enumerate}
\item{the scale factor in the $\tauh$ identification efficiency, \ie, the ratio of the measured value of the $\tauh$ identification efficiency to the value predicted by simulation, and}
\item{the normalization for $\PW$ production with $m_{\PW}>200\GeV$, relative to the theoretical prediction ($r_{\PW}$).}
\end{enumerate}

In addition to the uncertainties listed in Section~\ref{sec:systematics}, the following systematic uncertainties are also taken into account in the fit.
An uncertainty of 1\% in the momentum scale of the muon that also alters the differential distributions.
The energy scale of the \ptmiss is taken into account in propagating the uncertainty in the jet energy scale, as well as in the scale of the unclustered energy depositions.
Uncertainties in the extrapolation factor used in the estimation of background from jets misidentified as \tauh is also taken into account.
The backgrounds with genuine $\Pgt$ leptons in $\PW\to\Pgt\Pgn$ events are dominated by diboson events, which are estimated via MC simulation.
The normalization of the diboson background is verified in dedicated control regions, indicating discrepancies of up to 30\%.
An uncertainty of 30\% is therefore used in the normalization of backgrounds containing genuine $\Pgt$ leptons.

\begin{figure*}[!htbp]
\centering
\includegraphics[width=0.48\textwidth]{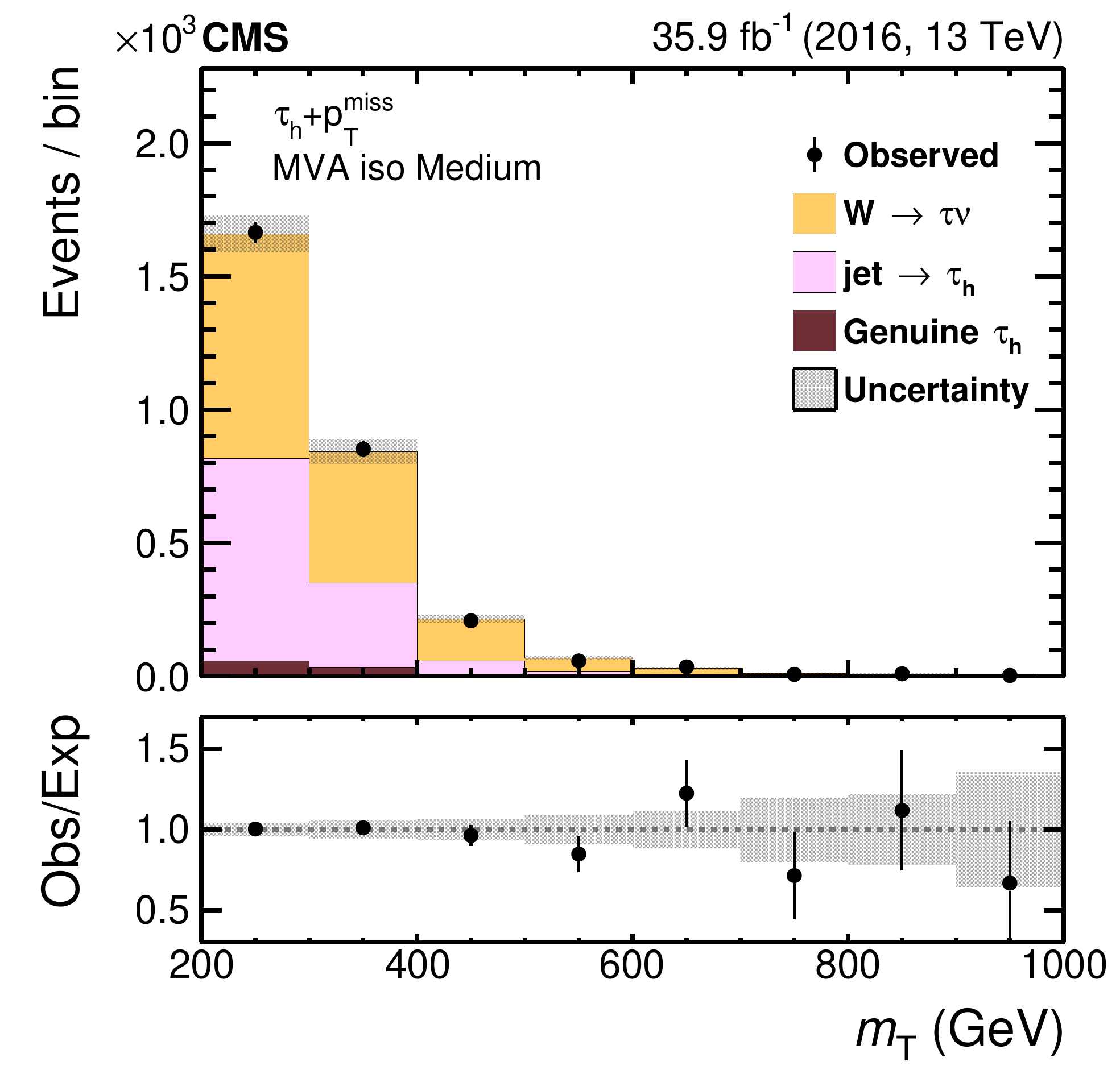}
\includegraphics[width=0.48\textwidth]{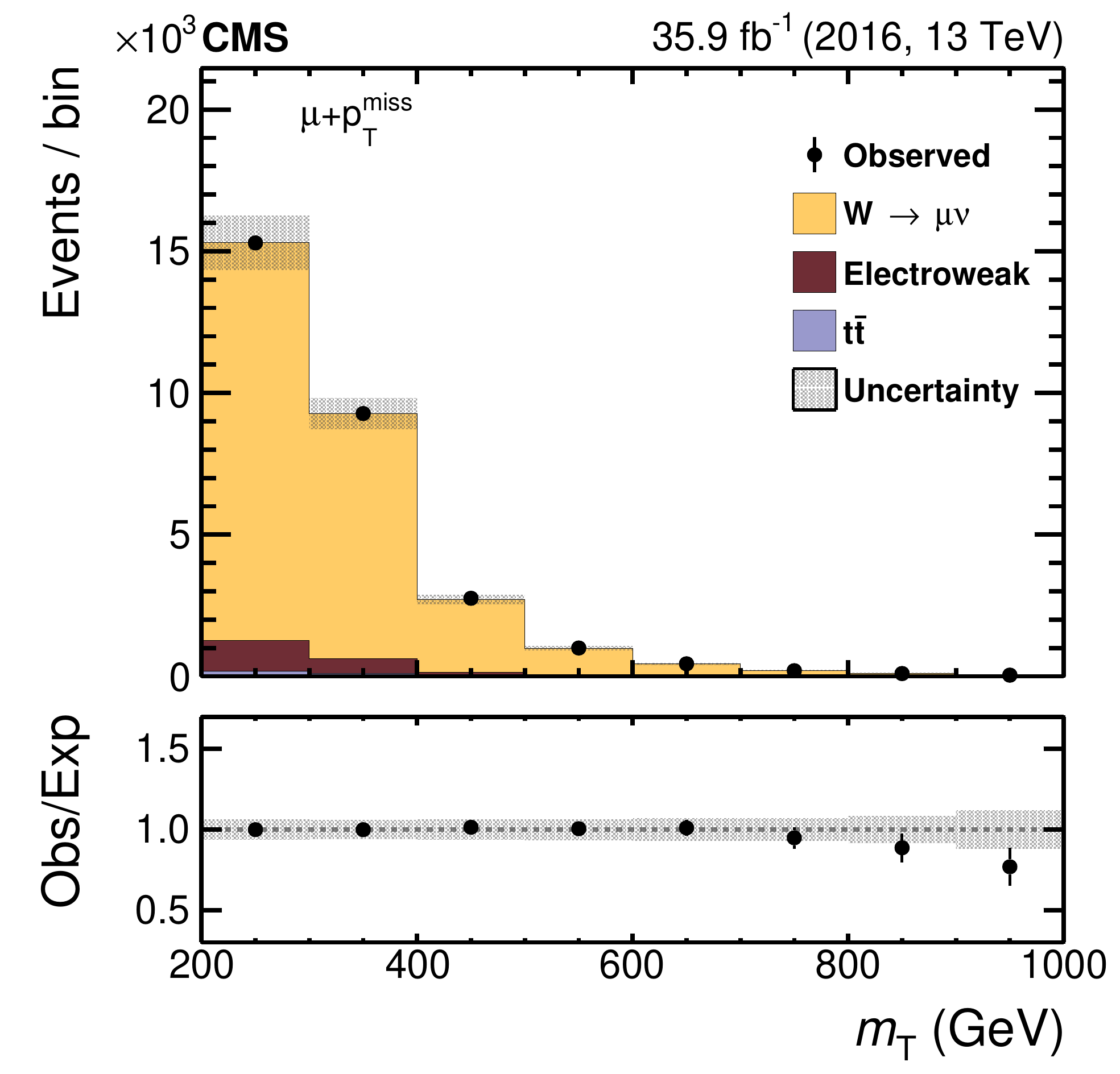}
\caption{The $\mT$ distribution for selected $\PW\to\Pgt\Pgn$ (left) and $\PW\to\Pgm\Pgn$ (right) events after the maximum likelihood fit. The medium WP of the MVA-based isolation discriminant is applied to select $\PW\to\Pgt\Pgn$ events. The electroweak background contribution includes diboson and single top quark events. Vertical bars correspond to the statistical uncertainties in the data points (68\% frequentist confidence intervals), while the shaded bands to the quadratic sum of the statistical and systematic uncertainties after the fit.
}
\label{fig:wmunu_taunu}
\end{figure*}

Figure~\ref{fig:wmunu_taunu} shows the fitted $\mT$ distributions for the $\PW\to\Pgt\Pgn$ signal and $\PW\to\Pgm\Pgn$ control samples.
The scale factor in the $\tauh$ identification efficiency, the parameter $r_{\PW}$, and the correlation coefficient between the two quantities
obtained from the fits, are detailed in Table~\ref{tab:wtaunu_tauid} for different working points of the MVA-based $\tauh$ isolation discriminants.
The data-to-simulation scale factors range between 0.89 for the very tight WP and 0.96 for the loose WP. The fitted value of the $\PW$ boson production cross section for $m_{\PW}>200\GeV$ is consistent with theoretical predictions. The $\PW$ boson sample normalization factor is anticorrelated with the scale factor for \tauh identification efficiency, as an increase in the $\PW$ boson yield is compensated in the fit by a reduction in the scale factor.
The correlation between the scale factor and $r_{\PW}$ increases with tighter $\tauh$ isolation, as expected, due to an increase in the purity of the signal region.

\begin{table*}[!hbtp]
\topcaption{The scale factor in the $\tauh$ identification efficiency, the normalization of $\PW$ boson production with $m_{\PW}>200\GeV$, $r_{\PW}$, and the correlation
coefficient between the two quantities obtained from the fit, measured for MVA-based discriminants using $\Delta R =0.5$ in $\PW\to\Pgt\Pgn$ events.}
\label{tab:wtaunu_tauid}
\centering
\begin{tabular}{lccc}
Working point & Scale factor & $r_{\PW}$ & Correlation \\
\hline
Loose & $0.96\pm0.08$ & $1.03\pm0.06$ & $-0.34$ \\
Medium  & $0.93\pm0.07$ & $1.02\pm0.07$ & $-0.44$ \\
Tight & $0.91\pm0.07$ & $1.02\pm0.07$ & $-0.46$ \\
Very tight & $0.89\pm0.07$ & $1.02\pm0.06$ & $-0.47$ \\
\hline
\end{tabular}
\end{table*}

We also measure the $\tauh$ identification efficiencies in bins of $\pt^{\tauh}$, with the
data-to-simulation scale factors extracted in a simultaneous fit
to the $\mT$ distribution in four signal samples, corresponding to four bins of $\pt^{\tauh}$, and of $\pt^{\Pgm}$ in the
$\PW\to\Pgm\Pgn$ control sample.
The results enter in the extrapolation of the scale factor to high $\pt^{\tauh}$, as discussed in Section~\ref{sec:tauIdEfficiency_extrapolation}.

\subsection{Extrapolation of the \texorpdfstring{$\tauh$}{tau[h]} identification efficiency to large \texorpdfstring{$\pt^{\tauh}$}{pt(tau[h])}}
\label{sec:tauIdEfficiency_extrapolation}

To extrapolate the scale factors for the $\tauh$ identification efficiency to high $\pt^{\tauh}$, a fit is performed to the values obtained in Sections~\ref{sec:tauIdEfficiency_TnP_Ztautau}, \ref{sec:tauIdEfficiency_ttbar}, and \ref{sec:tauIdEfficiency_Wtaunu}, as a function of $\pt^{\tauh}$. These measurements cover a $\pt^{\tauh}$ range between 20 and $\approx$300\GeV, with the mean value in each $\pt^{\tauh}$ bin used as a representative number for that bin. Fits to a zero- (constant) and first-order polynomial are performed, without considering
the uncertainty in track reconstruction efficiency, as it is correlated among the individual measurements.
Nevertheless, it is found to contribute very little to the overall uncertainty, with the exception of measurements at low $\pt^{\tauh}$, where other uncertainties are small because of the large number of events and the high purity of the event samples.
Despite having other possible correlations between $\pt^{\tauh}$ bins in a single measurement, or between different measurements, all measurements entering the fit are assumed to be uncorrelated.

\begin{figure}[!htbp]
\centering
\includegraphics[width=0.48\textwidth]{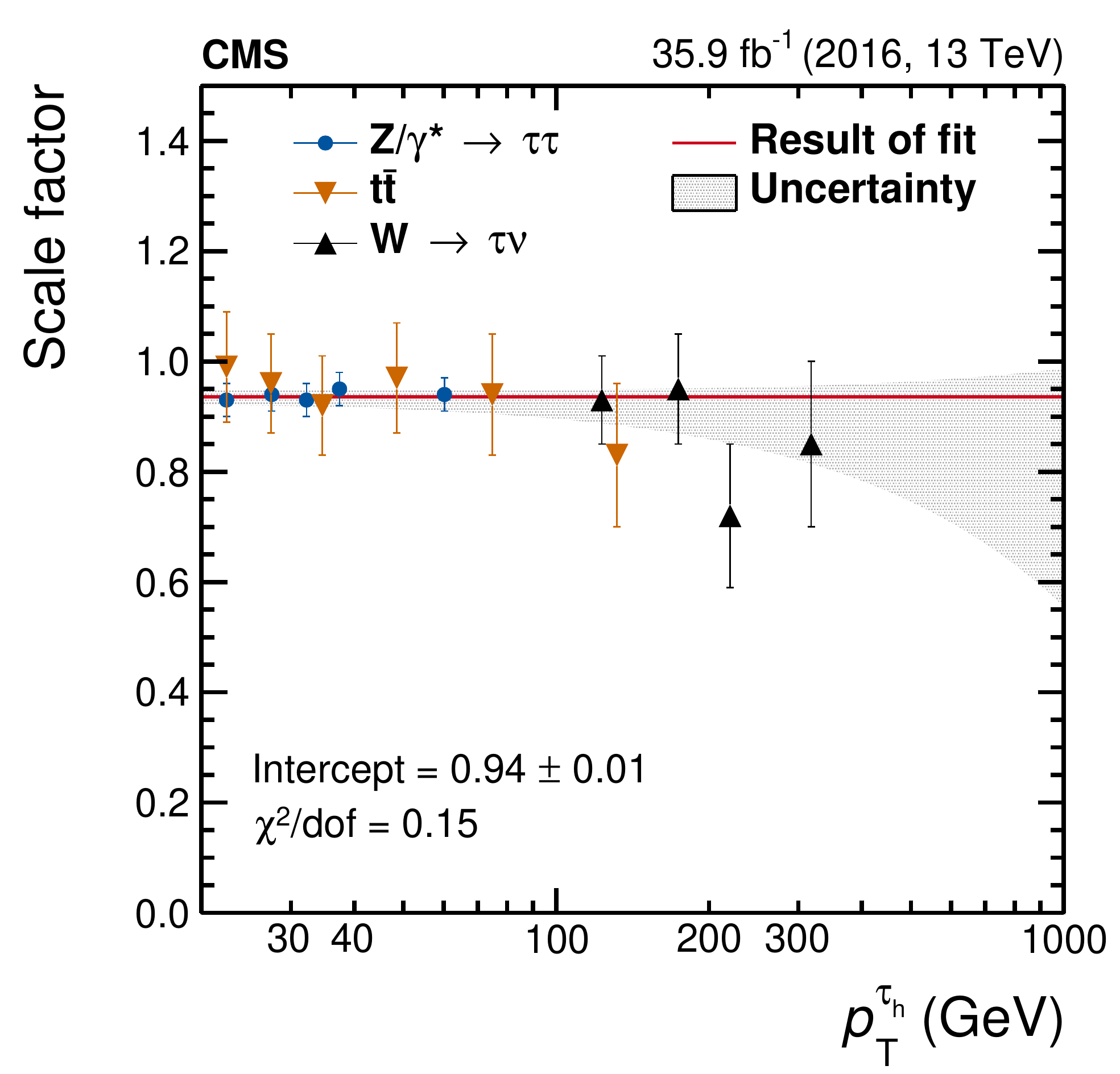}
\caption[dummy text]{Fit of the measured scale factors to a constant value in the $\tauh$ identification efficiency, for the tight WP of the MVA-based isolation discriminant in $\cPZ/\Pggx$, $\ttbar$, and $\PW$ events, as a function of $\pt^{\tauh}$. The shaded band represents the uncertainties in the fit, where the result is combined with the difference obtained using a first-order polynomial instead of a constant for the downward deviations, which also contain an additional contribution from the uncertainty in track-reconstruction efficiency.}
\label{fig:tau_id_extrapolation}

\end{figure}

The fit to a first-order polynomial provides a smaller goodness-of-fit per degree of freedom, $\chi^{2}$/dof, than that to a constant, indicating that the scale factor for $\tauh$ identification efficiency may decrease with $\pt^{\tauh}$; but, given that the slope of the fitted first-order polynomial barely deviates from zero (by only about one standard deviation), the scale factor is compatible with being constant. As there are no indications that components of $\tauh$ reconstruction or identification behave abnormally at high $\pt^{\tauh}$, a constant scale factor with an asymmetric uncertainty that increases with $\pt^{\tauh}$ is defined by adding in quadrature the uncertainty in the fit to a constant, and the difference between the fit to a first-order polynomial and to a constant for the downward deviation.
In addition, this also takes into account the uncertainty in the efficiency of track reconstruction, yielding the total (asymmetric) uncertainty of $+5\% \times \pt^{\tauh}$~(\TeVns) and $-35\% \times \pt^{\tauh}\,(\TeVns)$. The fit to a constant using the combined uncertainty is shown in Fig.~\ref{fig:tau_id_extrapolation}.

\subsection{Using the tag-and-probe method in \texorpdfstring{$\cPZ/\Pggx$}{Z/gamma*} events for highly boosted \texorpdfstring{$\Pgt$}{tau} lepton pairs}
\label{sec:tauIdEfficiency_TnP_BoostedTaus}

The identification efficiency for highly boosted $\Pgt$ lepton pairs in $\tauh$ final states is measured using the same tag-and-probe method as described in Section~\ref{sec:tauIdEfficiency_TnP_Ztautau}. The selection is optimized to have a pure sample of $\Pgt$ leptons from the decay of high-\pt $\cPZ$ bosons, where one $\Pgt$ lepton decays leptonically and the other one into hadrons and a neutrino.
As the trigger thresholds for nonisolated leptons are very high, too few events are available to reliably measure the identification efficiency for very high \pt $\Pgt$ lepton pairs.
Single isolated-lepton triggers with lower thresholds are used therefore to select $\Pe\tauh$ and $\Pgm\tauh$ events.
However, events in which a \tauh is within the isolation area around a triggering lepton ($\Delta R<0.4$) are not accessible in this measurement.

The selection requires one isolated electron or muon fulfilling tight identification criteria, and satisfying $\pt>40$ or $>$26\GeV, respectively.
Furthermore, as discussed in Section~\ref{sec:boostedTauReco}, at least one $\tauh$ candidate must be reconstructed with $\pt>20\GeV$ and $\abs{\eta}<2.3$, in compliance with relaxed decay mode criteria. The $\Delta R$ between the selected lepton and $\tauh$ candidate must be between 0.4 and 0.8, and the $m_{\mathrm{T}}$ of the $\ptvec^{\,\Plepton}$ and \ptvecmiss system must be $<$40\GeV.
Moreover, \ptmiss must exceed 75\GeV, the scalar \pt sum of all measured particles has to be greater than 200\GeV, and there cannot be any identified $\cPqb$~jets in the event.
If more than one $\Pe\tauh$ or $\Pgm\tauh$ pair is present, the one with the largest \pt is chosen for further analysis.

The contribution from DY events is modelled using MC simulation. It is split into the signal contribution by matching the reconstructed leptons to those generated and those contributing via misidentified $\cPZ$ boson decays.
The distributions of the backgrounds from $\PW$+jets and $\ttbar$ production are also modelled using simulation, but their normalizations are obtained from dedicated control data samples. The control sample for $\PW$+jets production is defined by inverting the requirement on $\mT$. The control sample for $\ttbar$ production is established by demanding at least one $\cPqb$-tagged jet.

The background from QCD multijet production is estimated from a sample selected in the same way as the signal, except for the requirement on \ptmiss, which is inverted to $\ptmiss<75\GeV$. Contributions from other processes are subtracted based on simulation. The extrapolation factor from the sample with an inverted \ptmiss requirement to the signal region is obtained from the ratio of events in two other control samples, where the $\Delta R$ between the lepton and the $\tauh$ candidate is between 0.8 and 2.0, one which uses the nominal and the other an inverted \ptmiss requirement, respectively.
Contributions from other processes are also subtracted from data using MC simulation in these two control regions.

The systematic uncertainties discussed in Section~\ref{sec:systematics} are taken into account in the procedure, as are the additional uncertainties in the estimation of the QCD multijet background, which are dominated by the limited number of events in the control samples. Finally, the uncertainties in the normalization of background from $\ttbar$ and $\PW$+jets production are determined from their respective control samples, and amount to 3 and 13\%, respectively.

\begin{figure}[!htbp]
\centering
\includegraphics[width=0.48\textwidth]{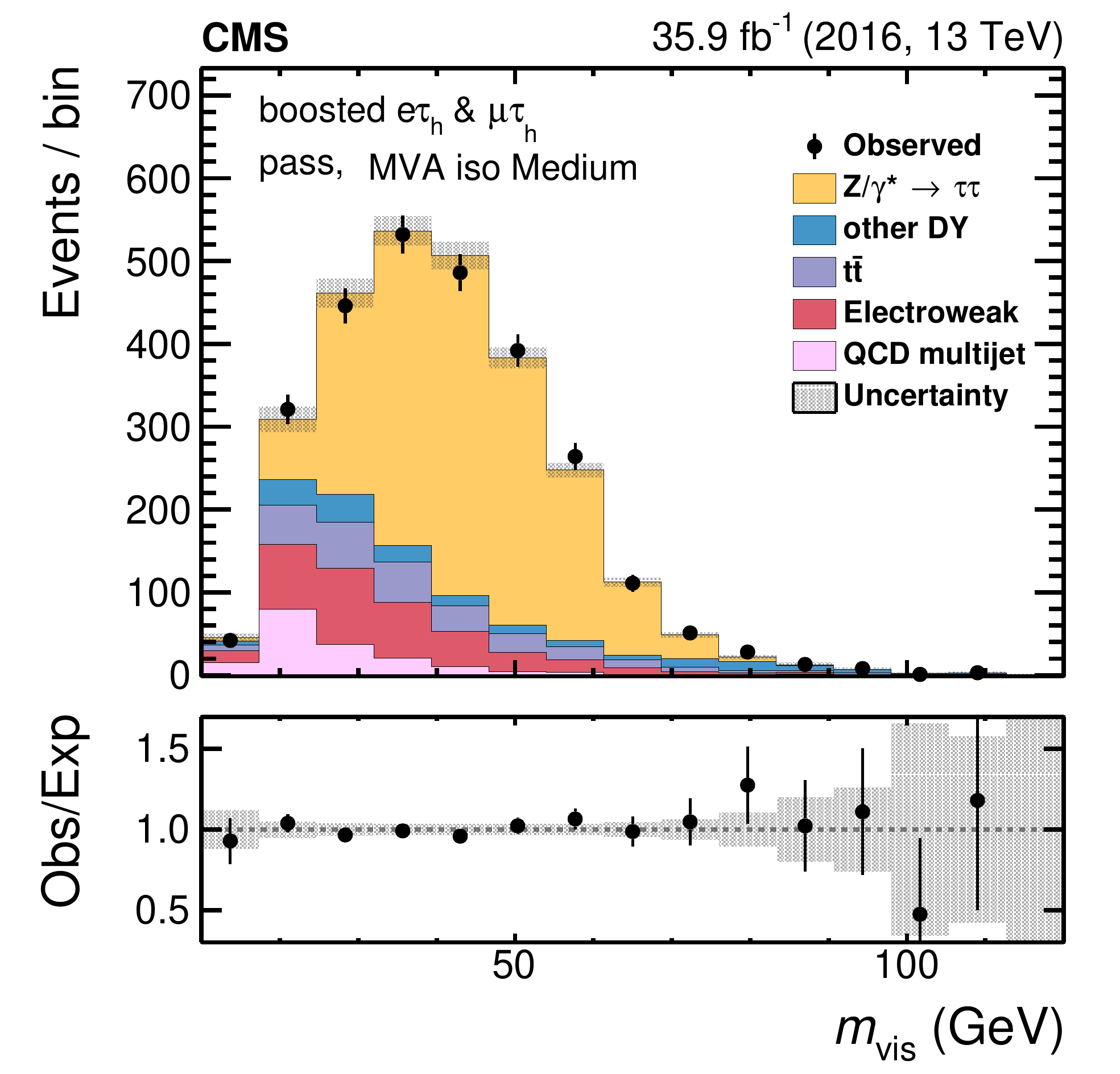}
\includegraphics[width=0.48\textwidth]{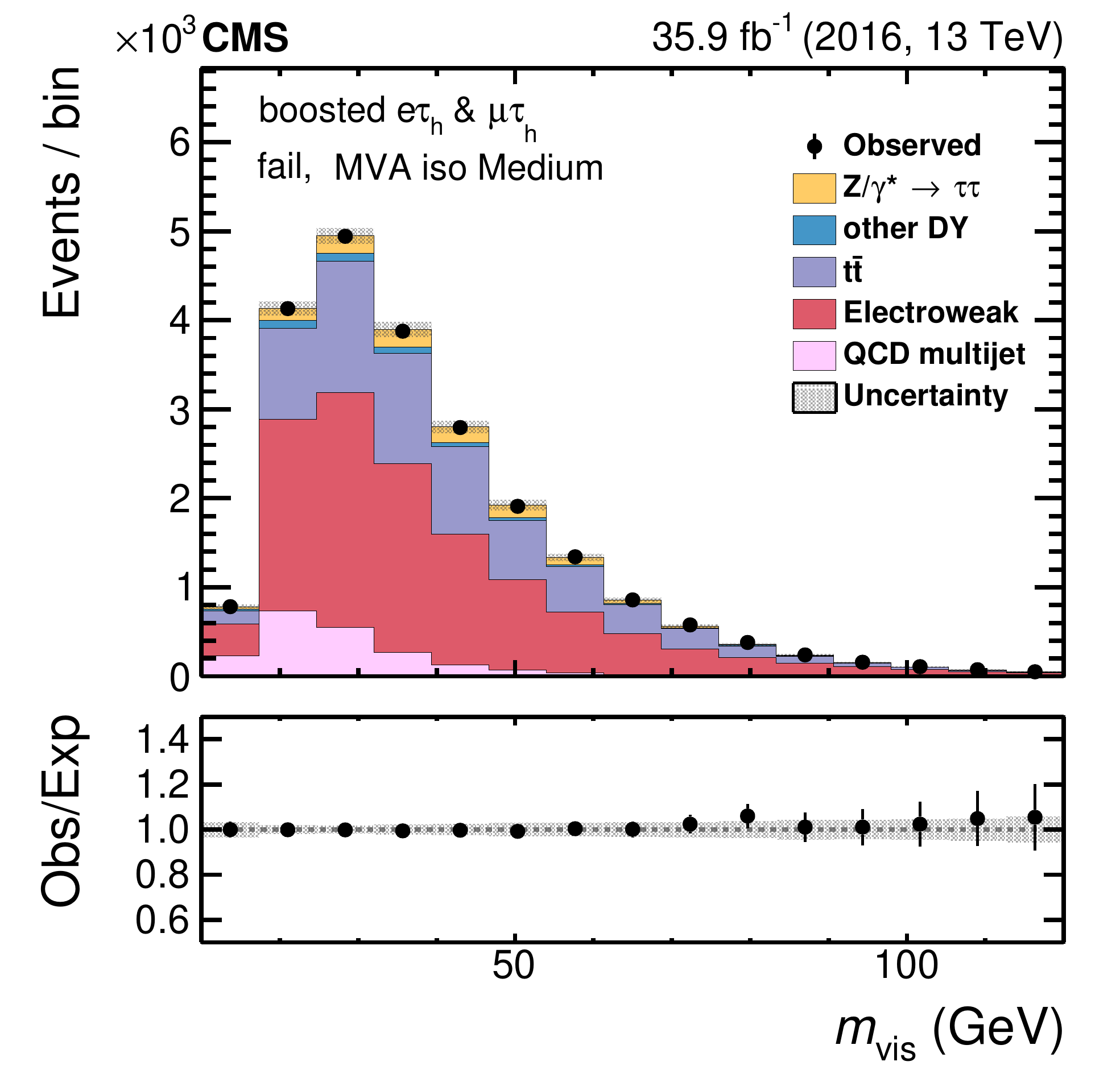}
\caption{Fitted distributions to the passing (left) and failing (right) events for $\tauh$ from highly boosted $\Pgt$ lepton pairs that pass the medium WP of the MVA-based isolation discriminant. The electroweak background includes contributions from $\PW$+jets (dominating), diboson, and single top quark events.
Vertical bars correspond to the statistical uncertainties in the data points (68\% frequentist confidence intervals), while the shaded bands provide the quadratic sum of the statistical and systematic uncertainties after the fit.}
\label{fig:BoostedTaus_MediumID}

\end{figure}

The data-to-simulation scale factors are evaluated in the same way as outlined in Section~\ref{sec:tauIdEfficiency_TnP_Ztautau}.
The passing and failing events are defined by requiring the $\tauh$ to pass or fail a given working point of the MVA-based isolation discriminant. The scale factors for the six MVA-based working points are shown in Table~\ref{tab:BoostedTauID}. The values are compatible with unity, as well as with the scale factors obtained through the measurements described in Sections~\ref{sec:tauIdEfficiency_TnP_Ztautau}--\ref{sec:tauIdEfficiency_Wtaunu}.
The dependence of the scale factor on the $\Delta R$ between $\tauh$ and the lepton, is studied without revealing a significant effect.
The fitted distributions corresponding to the medium isolation WP are shown in Fig.~\ref{fig:BoostedTaus_MediumID}.

\begin{table}[!hbtp]
\topcaption{Data-to-simulation scale factors for different working points of the MVA-based isolation discriminant, using highly boosted $\cPZ/\Pggx$ events decaying to $\Pgt$ lepton pairs.}\label{tab:BoostedTauID}
\centering
\begin{tabular}{lc}
Working point & Scale factor \\
\hline
Very loose & 0.97$\pm$0.09 \\
Loose      & 0.99$\pm$0.09 \\
Medium     & 0.98$\pm$0.09 \\
Tight      & 0.96$\pm$0.08 \\
Very tight & 0.95$\pm$0.09 \\
Very-very tight & 0.90$\pm$0.08 \\
\hline
\end{tabular}

\end{table}

\section[\texorpdfstring{$\text{Jet} \mapsto \tauh$}{Jet maps to tau[h]} misidentification probability]{Measurement of the \texorpdfstring{$\text{jet} \mapsto \tauh$}{Jet maps to tau[h]} misidentification probability}
\label{sec:jetToTauFakeRate}
\subsection{Using \texorpdfstring{$\PW\to\Pgm\Pgn$+jet}{W to mu nu + jet} events}
\label{sec:jetToTauFakeRateW}

The probability to misidentify a quark or gluon jet as a $\tauh$ candidate is measured as a function of jet $\pt$ and $\eta$ in a sample of $\PW\to\Pgm\Pgn$+jet events, selected as described in Section~\ref{sec:validation_eventSelection_Wjets}.
In addition to $\pt^\text{jet}$ and $\eta^\text{jet}$, the misidentification probability also depends on parton flavour, as well as whether the parton initiating the jet and the reconstructed $\tauh$ have the same or opposite charge.
These factors cause differences of up to a factor of four between misidentification probabilities for c~quark and gluon jets, and up to a factor of two for whether the initiating parton has the same or opposite charge as the $\tauh$ candidate.
This means that the misidentification probabilities given in this section are indicative, in that they are mainly valid for $\PW\to\Pgm\Pgn$+jet events,
which contain a large fraction of light-quark jets, and therefore have a relatively high misidentification probability.

The misidentification probability is given by the ratio of the number of jets
that are identified as $\tauh$ candidates with $\pt>20\GeV$, $\abs{\eta}<2.3$, and passing any one of the working points of the discriminants described in Section~\ref{sec:tauIdDiscriminators_jets}, to the total number of jets with $\pt>20\GeV$ and $\abs{\eta}<2.3$.
It should be recognized that $\pt^\text{jet}$ differs from $\pt^{\tauh}$ because the four-momentum of the jet is computed by summing the momenta of all its constituents, while the $\tauh$ four-momentum is computed only from the charged hadrons and photons used in the reconstruction of the specified decay mode of the $\tauh$ candidate.
For $\pt^\text{jet} < 300\GeV$, the $\pt^{\tauh}$ constitutes on average only 40\% of the jet \pt.
Furthermore, $\pt^\text{jet}$ is subject to additional jet energy corrections, whereas $\pt^{\tauh}$ is not.

In the measurement of the misidentification probability, backgrounds with genuine $\tauh$ are subtracted, based on the expectations from simulated events.
The fraction of events with genuine $\tauh$ candidates in the sample passing the $\tauh$ identification criteria is well below 10\% for $\tauh$ with $\pt<100\GeV$, but reaches up to 50\% for $\pt\approx300\GeV$.
Furthermore, backgrounds with prompt electrons and muons giving rise to $\tauh$ candidates are also subtracted based on expectations from simulated events.
To reject events from $\cPZ/\Pggx\to\Pgm\Pgm$ production, the loose WP of the against-$\Pgm$ discriminant described in Section~\ref{sec:antiMuonDiscrCutBased} is applied to the reconstructed $\tauh$ candidates.

The subtraction of backgrounds containing genuine $\tauh$ is subject to an uncertainty of 30\%,
leading to an uncertainty of up to 15\% in the jet~$\mapsto\tauh$ misidentification probability.
Because of threshold effects, the jet energy scale also leads to a significant uncertainty, especially in the lowest bin of $\pt^\text{jet}$.
Additional uncertainties are considered for probabilities with which electrons are reconstructed as $\tauh$ candidates (with $\approx$100\% relative values), and with which muons are reconstructed as $\tauh$ candidates that pass the loose WP of the against-$\Pgm$ discriminant (at 50\%).
These lead to uncertainties in the measured misidentification probabilities of at most a few percent.

\begin{figure}[!htbp]
\centering
\includegraphics[width=0.4\textwidth]{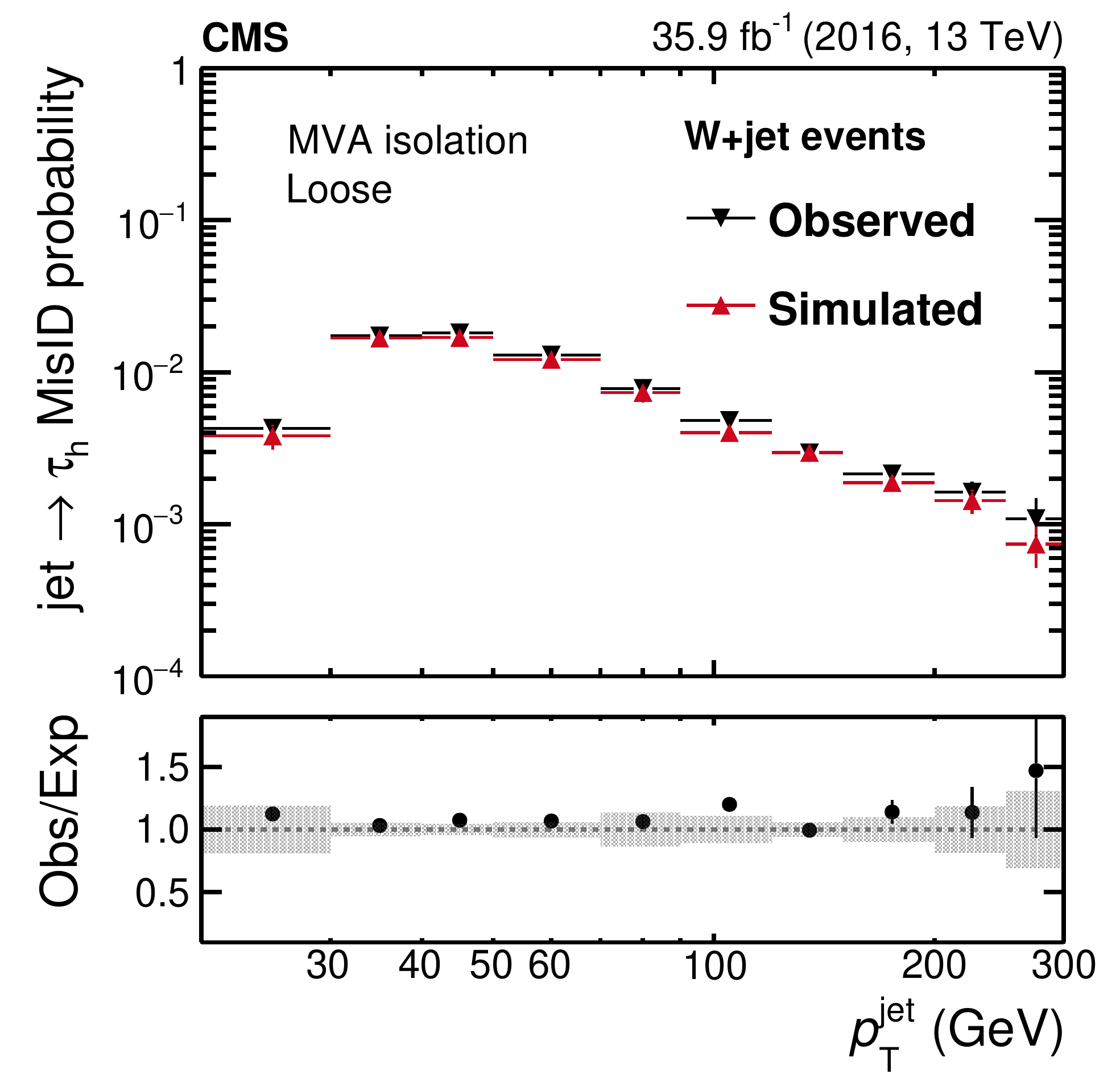}
\includegraphics[width=0.4\textwidth]{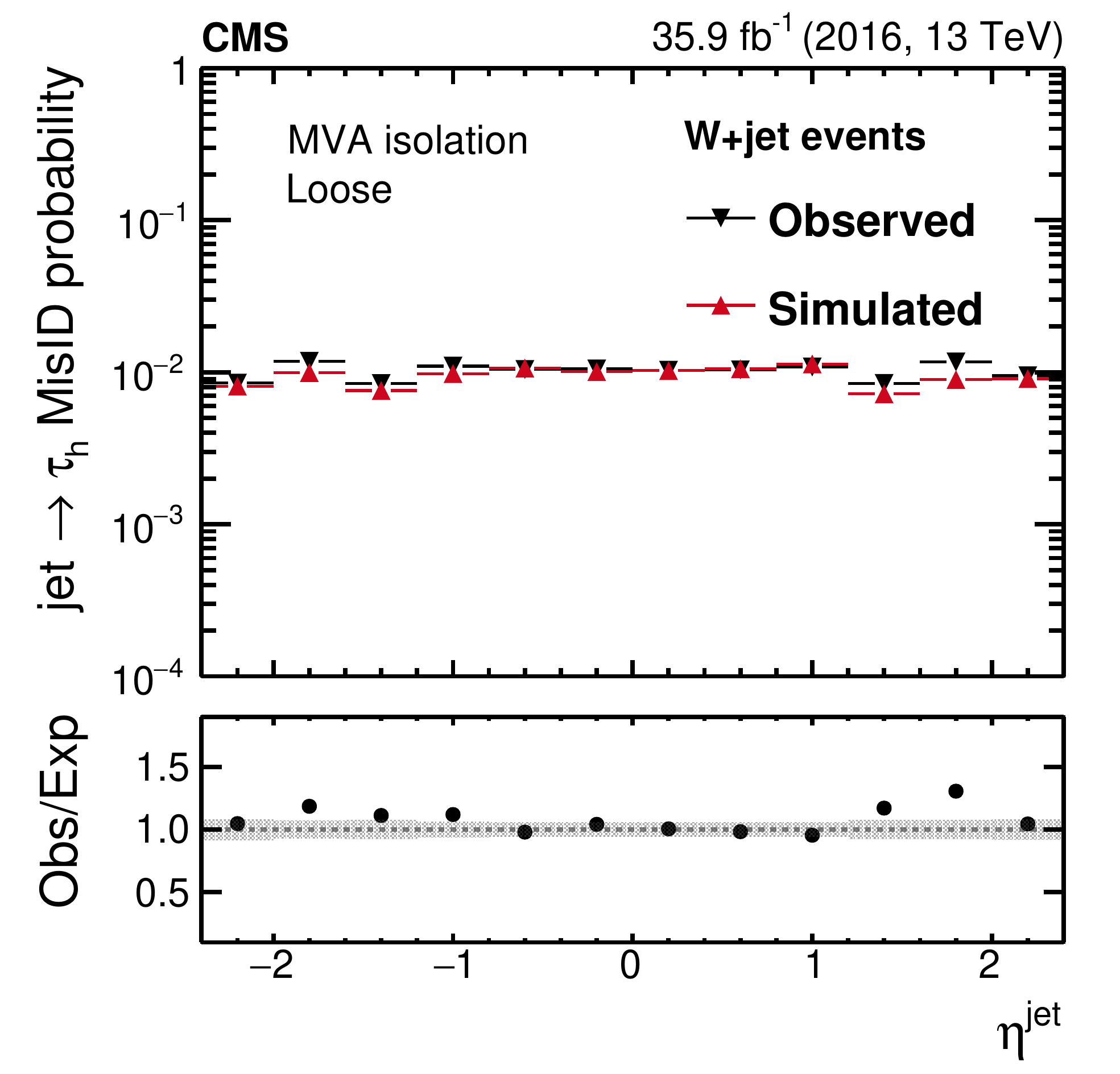}
\includegraphics[width=0.4\textwidth]{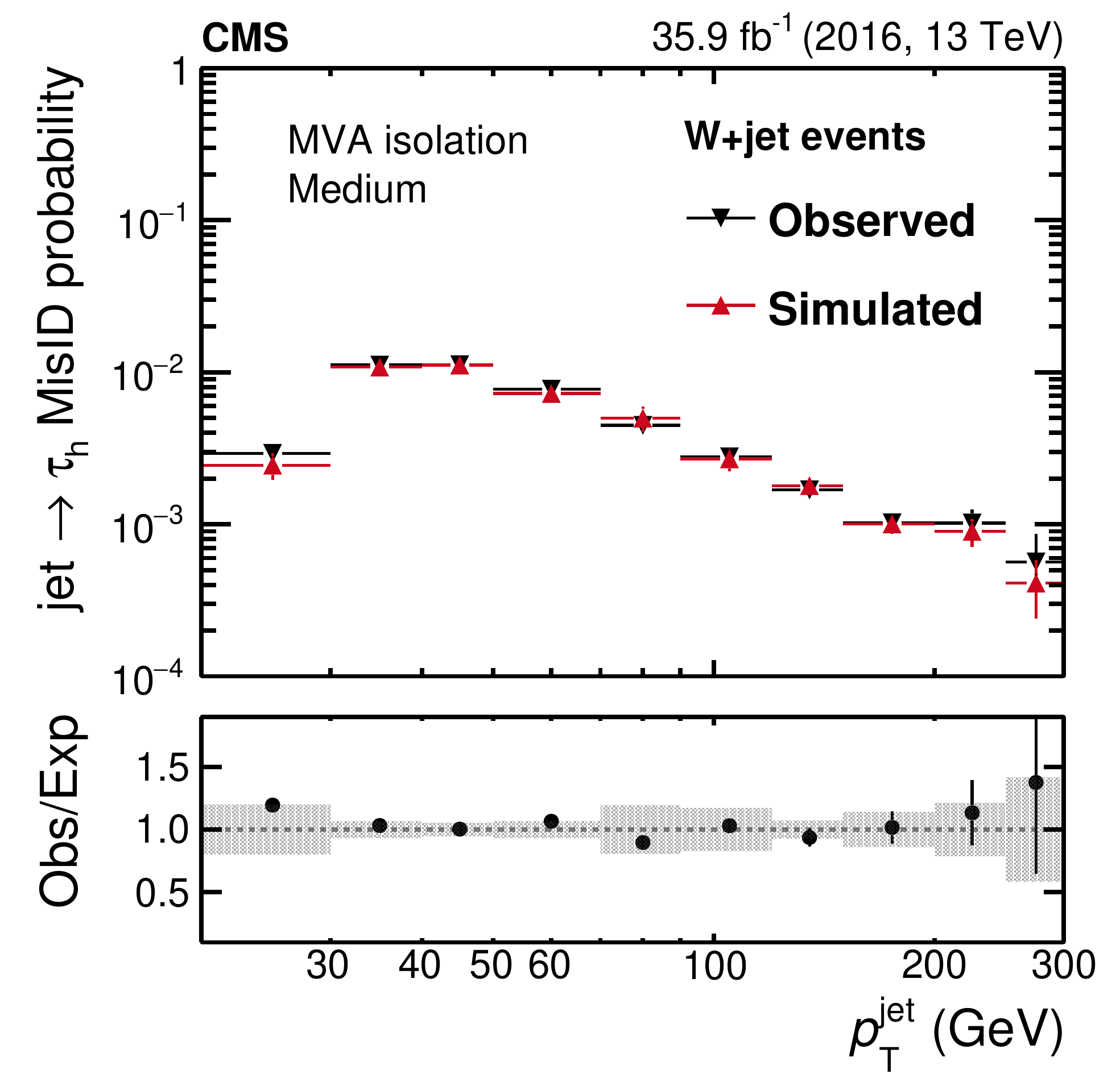}
\includegraphics[width=0.4\textwidth]{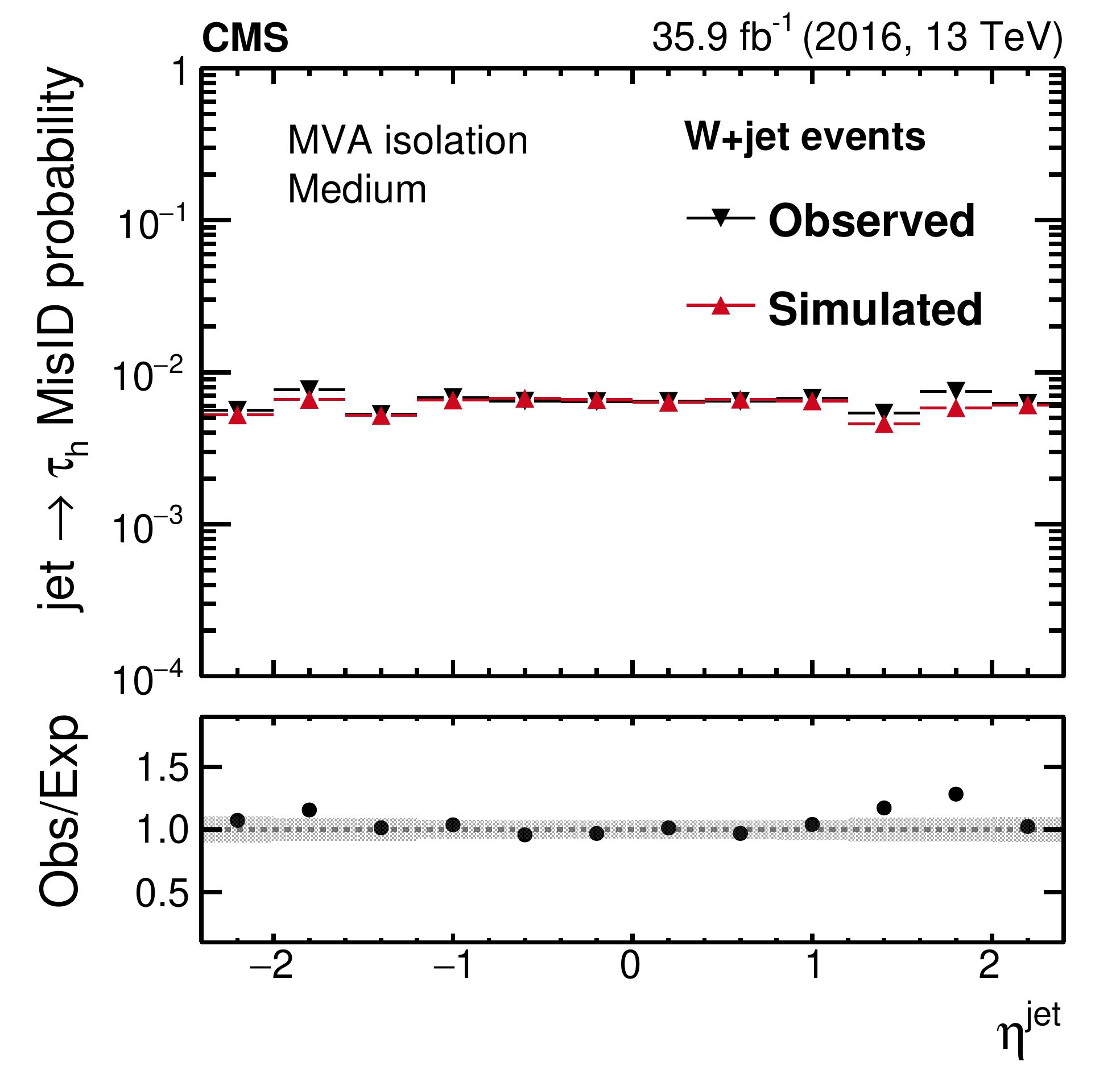}
\includegraphics[width=0.4\textwidth]{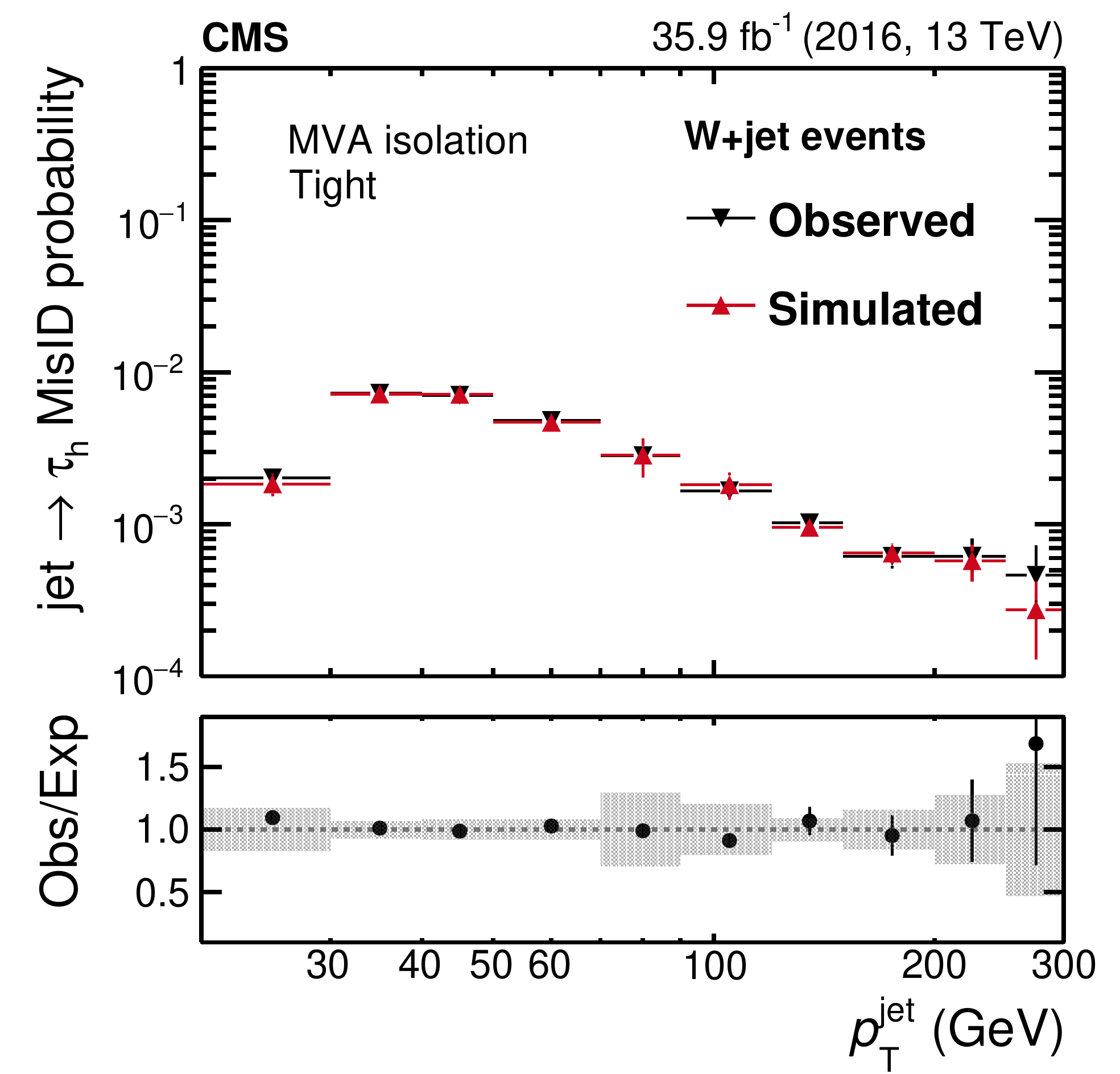}
\includegraphics[width=0.4\textwidth]{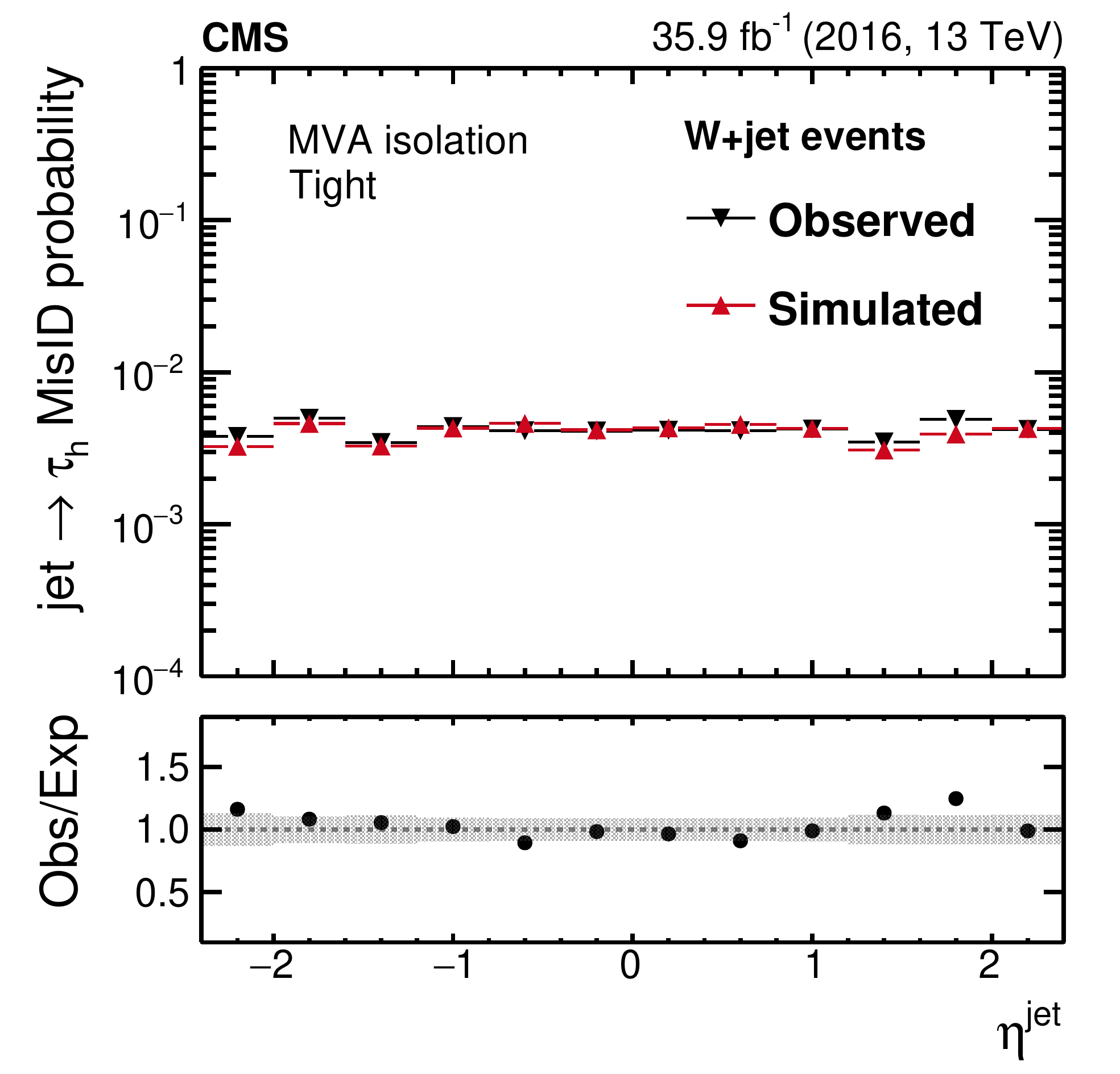}
\caption[]{Probabilities for quark and gluon jets in $\PW$+jet events to pass the loose (uppermost), medium (middle), and tight (lowest) WPs of the MVA-based isolation discriminant as a function of $\pt^\text{jet}$ (left) and $\eta^\text{jet}$ (right).
The misidentification probabilities in data are compared to expectations from simulation.
The vertical bars in the simulated and observed misidentification probabilities include statistical uncertainties from the limited event count in both data and simulated samples, including the background subtraction.
The shaded bands contain the systematic uncertainties related to background subtraction and the jet energy scale.}
\label{fig:jetTauFR_wjet}

\end{figure}

The observed and simulated jet~$\mapsto \tauh$ misidentification probabilities for the loose, medium, and tight WPs of the MVA-based isolation discriminant are shown in Fig.~\ref{fig:jetTauFR_wjet}, as a function of $\pt^\text{jet}$ and $\eta^\text{jet}$.
The probabilities are observed to be almost constant as a function of $\eta^\text{jet}$, while they decrease monotonically with increasing $\pt^\text{jet}$ from $\approx$40\GeV, as the absolute isolation increases for quark- and gluon-initiated jets with increasing jet \pt. The values of the misidentification probability as a function of $\pt^\text{jet}$ range between 2.0 and 0.1\% for the loose WP of the MVA-based isolation discriminant, and between 1.0 and less than 0.1\% for the tight WP.
The observed probabilities show a difference of 10--20\% relative to expectations from MC simulation.
This difference is well within the range of the misidentification probabilities obtained under variations of the parton shower models and underlying-event tunes, and reflects precision of modelling untypical, narrow and low multiplicity, quark and gluon jets being able to pass $\tauh$ identification criteria.

\subsection{Using \texorpdfstring{$\Pe\Pgm$+jets}{e/mu + jets} events}
\label{sec:jetToTauFakeRateTTbar}

\begin{figure}[!htbp]
\centering
\includegraphics[width=0.4\textwidth]{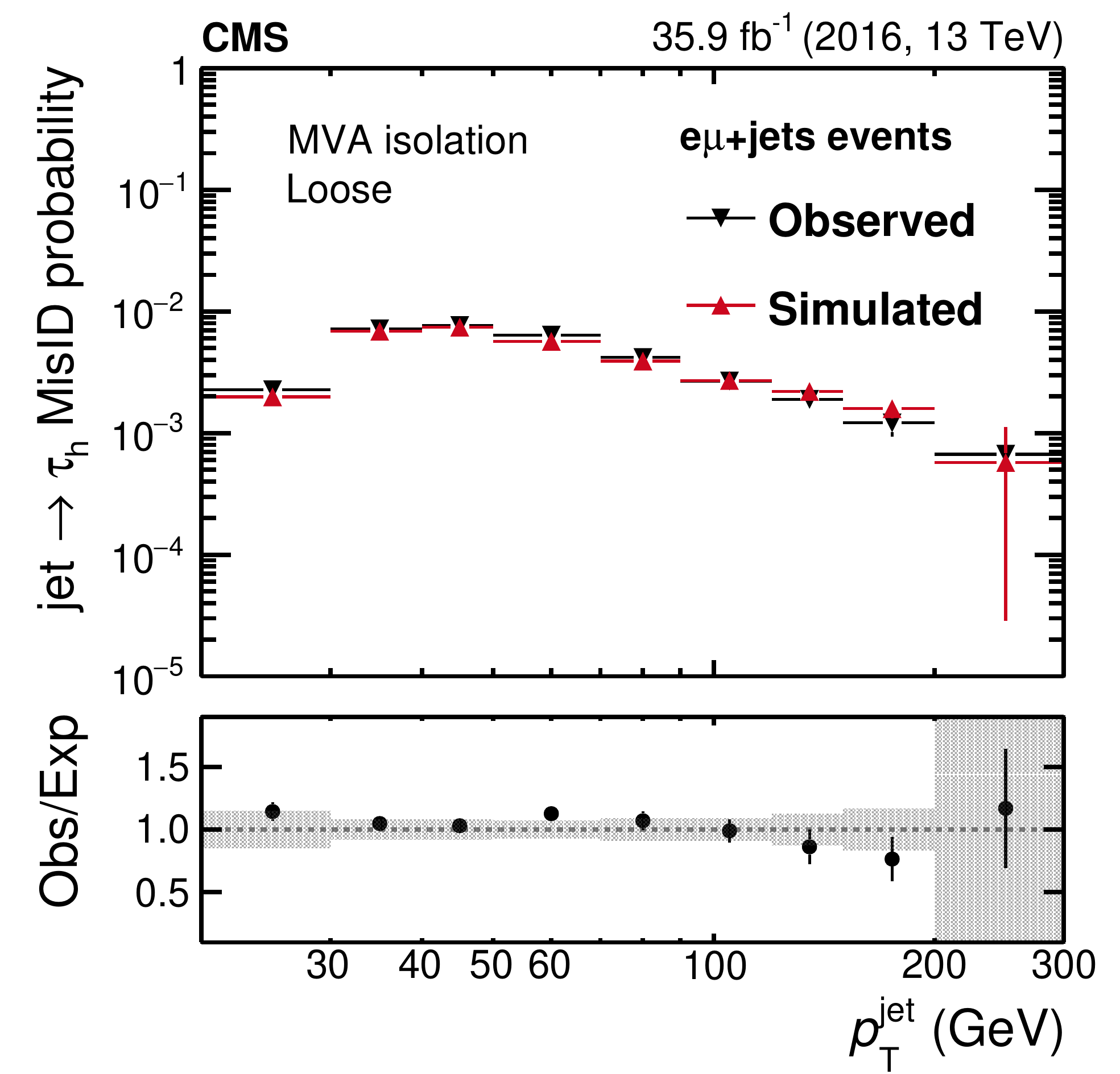}
\includegraphics[width=0.4\textwidth]{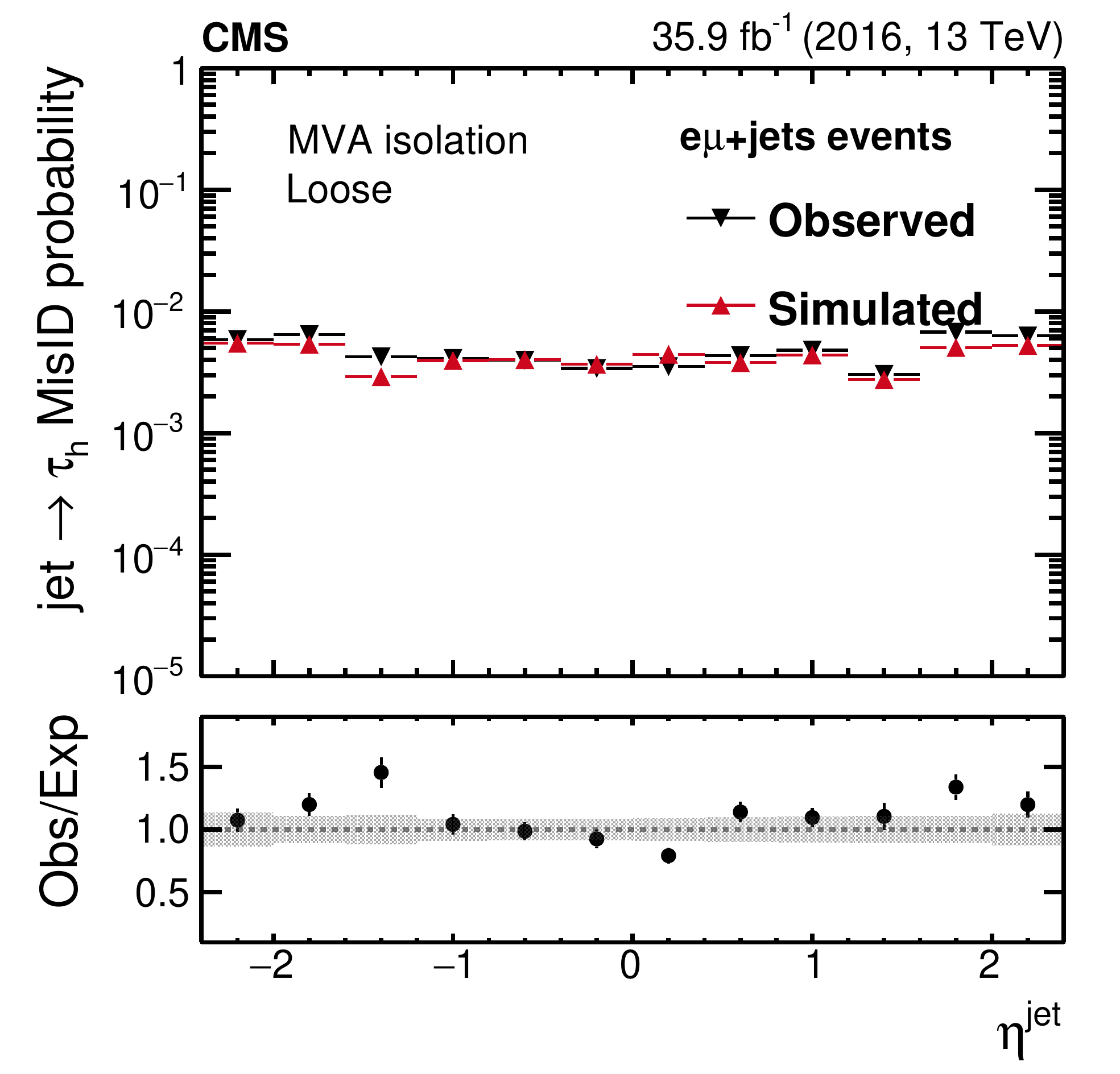}
\includegraphics[width=0.4\textwidth]{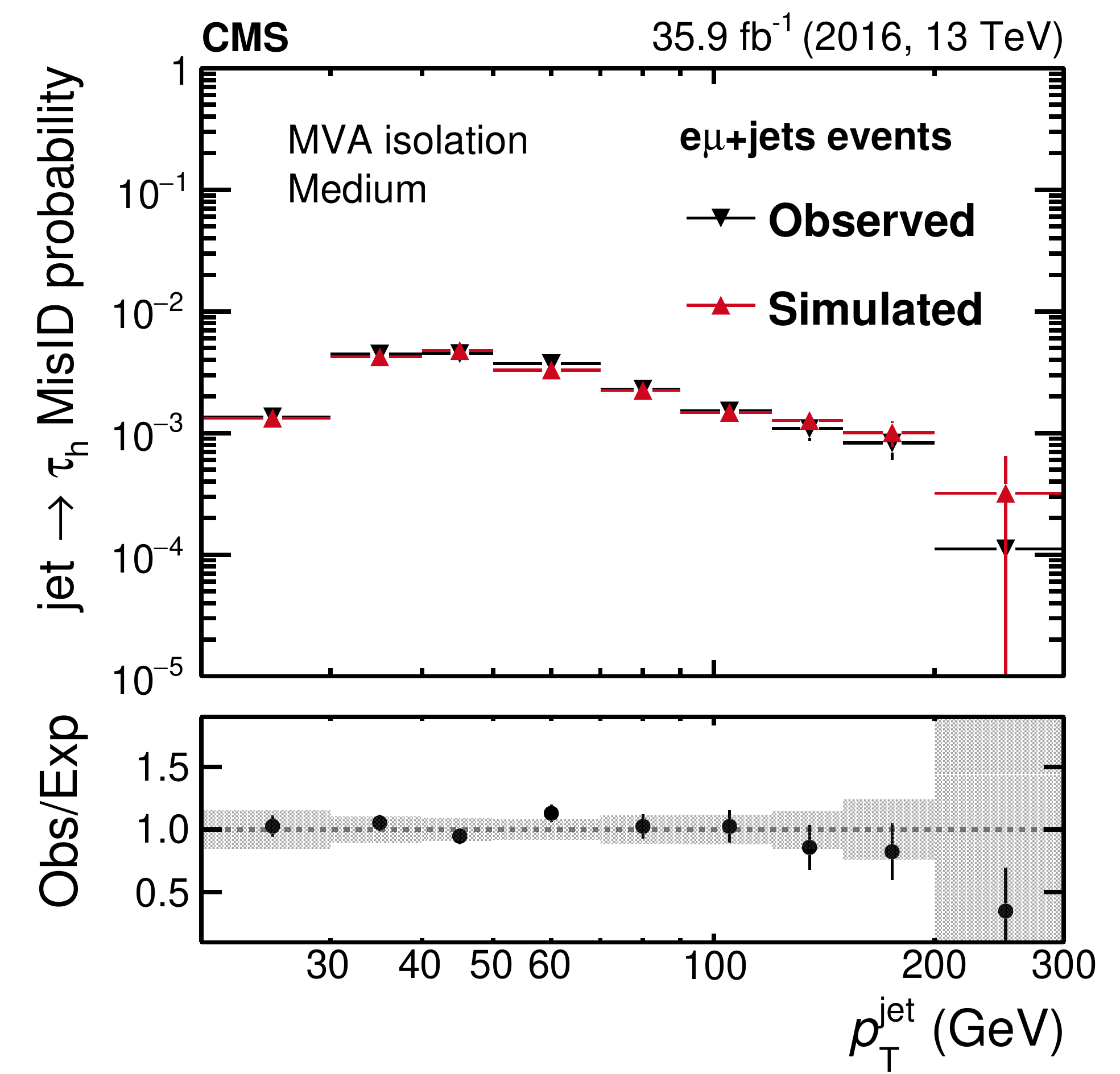}
\includegraphics[width=0.4\textwidth]{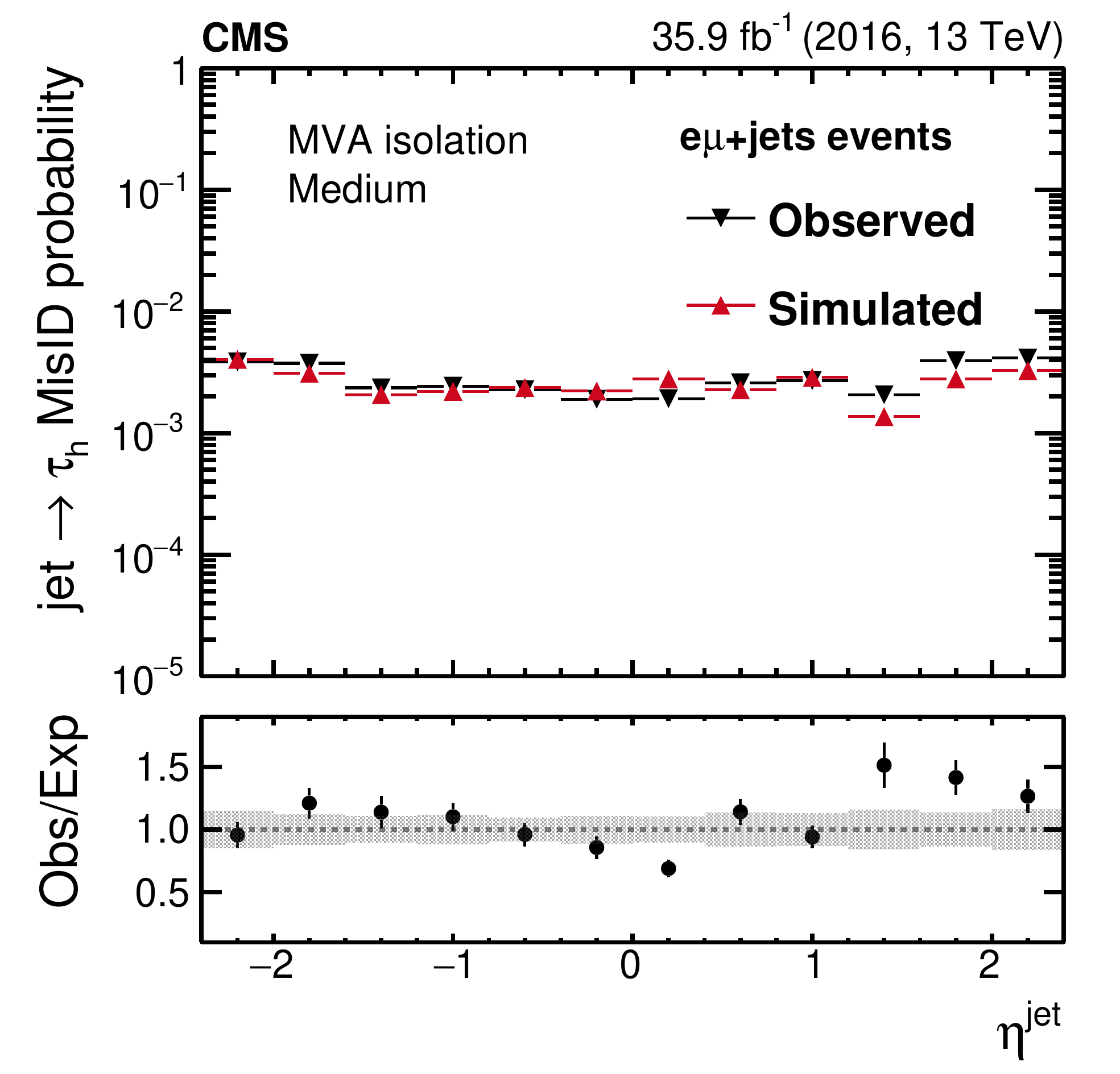}
\includegraphics[width=0.4\textwidth]{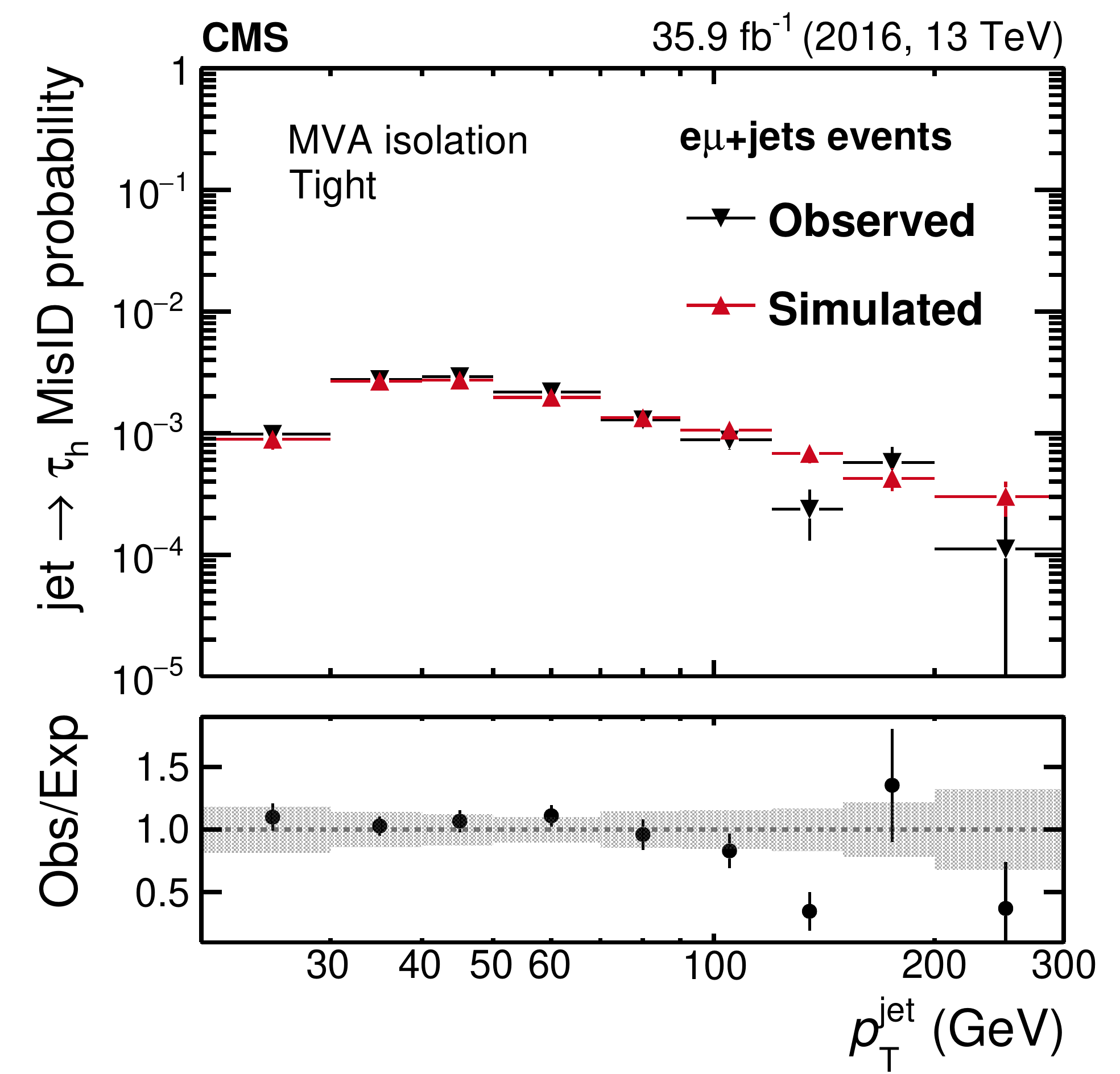}
\includegraphics[width=0.4\textwidth]{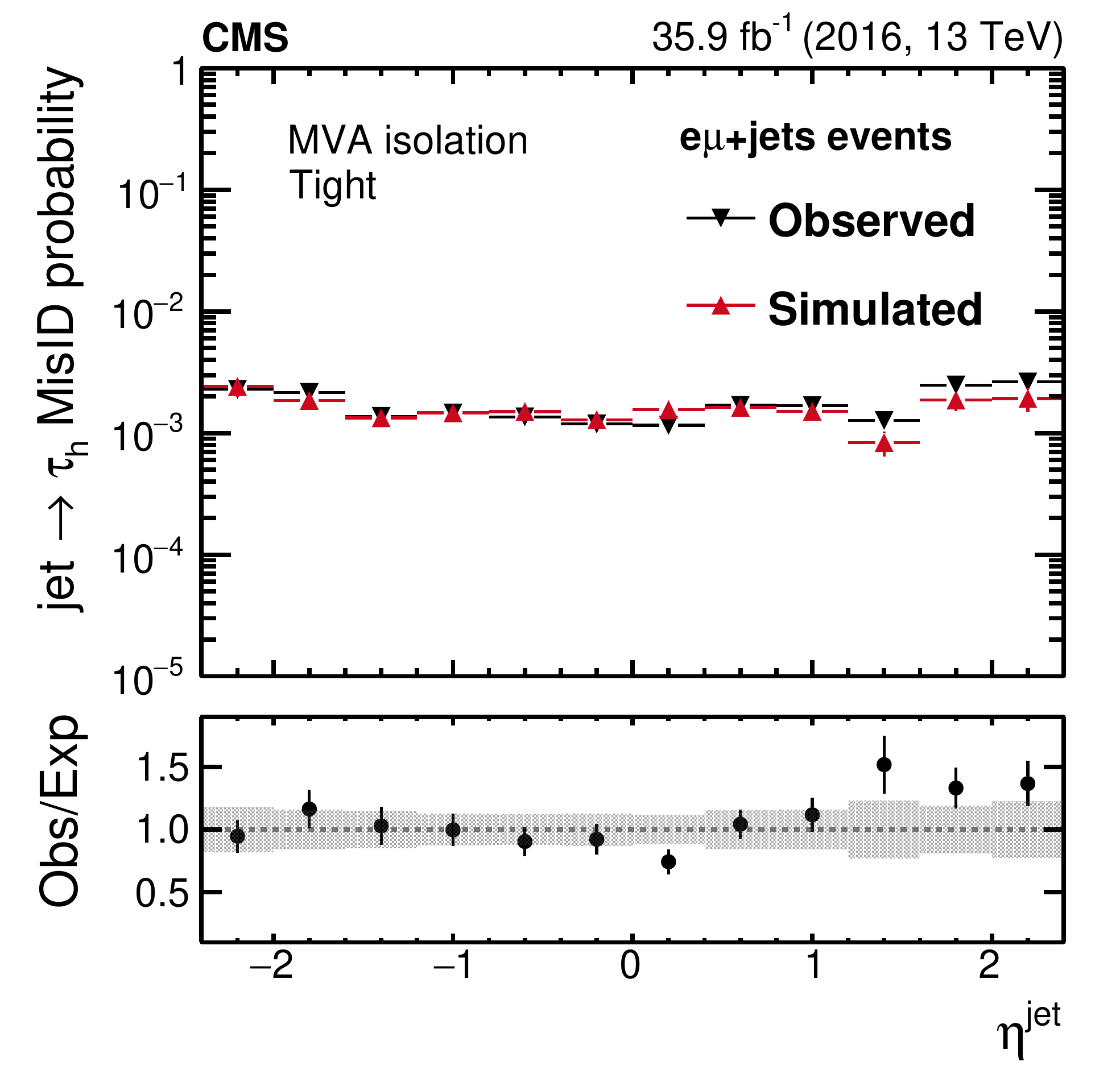}
\caption[]{Probabilities for quark and gluon jets in $\Pe\Pgm$+jets events to pass the loose (uppermost), medium (middle), and tight (lowest) WPs of the MVA-based isolation discriminant as a function of $\pt^\text{jet}$ (left) and $\eta^\text{jet}$ (right).
The misidentification probabilities in data are compared to expectations from simulation.
The vertical bars in the simulated and observed misidentification probabilities include statistical uncertainties from the limited event count in both data and simulated samples, including the background subtraction.
The shaded bands contain the systematic uncertainties related to background subtraction and the jet energy scale.}
\label{fig:jetTauFR_emu}

\end{figure}

The probability to misidentify quark and gluon jets as $\tauh$ candidates is also measured in the $\Pe\Pgm$ final state of $\ttbar$ events using the same methodology and uncertainties outlined in Section~\ref{sec:jetToTauFakeRateW}.
The events are selected as described in Section~\ref{sec:validation_eventSelection_ttbar_emu},
with the largest contributions being from $\ttbar$ and single top quark events, where the misidentified $\tauh$ candidates are dominated by $\cPqb$~quark jets.
The contribution from other processes is $<$10\%.
The observed and simulated jet~$\mapsto \tauh$ misidentification probabilities for the loose, medium, and tight WPs of the MVA-based isolation discriminant are shown in Fig.~\ref{fig:jetTauFR_emu}, as a function of $\pt^\text{jet}$ and $\eta^\text{jet}$.
The observed probabilities show a 10--20\% difference relative to expectations from simulation, except in a few $\eta^\text{jet}$ bins where the differences are as large as 50\%.
The jet~$\mapsto \tauh$ misidentification probabilities in $\Pe\Pgm$+jets events are found to be smaller than those for $\PW$+jet events because of the larger fraction of $\cPqb$~quark jets.
The $\cPqb$ quark jets are typically less collimated than the light-quark jets, providing thereby smaller probabilities to pass the $\tauh$ isolation discriminant selection requirements.

\section{The \texorpdfstring{$\Pe/\Pgm\mapsto\tauh$}{e/mu maps to tau[h]} misidentification probability}
\label{sec:lToTauFakeRate}
\subsection{Measurement of the \texorpdfstring{$\Pe \mapsto \tauh$}{e maps to tau[h]} probability}
\label{sec:eToTauFakeRate}

The $\Pe\mapsto\tauh$ misidentification probability is obtained from data using a tag-and-probe method in $\cPZ/\Pggx\to\Pe\Pe$ events selected as described in Section~\ref{sec:validation_eventSelection_Zll}.

Depending on whether the probe passes or fails a given working point of the against-$\Pe$ discriminant, the event enters the passing or failing category, respectively.
The $\Pe\mapsto\tauh$ misidentification rate is then measured in a simultaneous fit to the number of $\cPZ/\Pggx\to\Pe\Pe$ events in both categories. The $\mvis$ distribution in the range $60 < \mvis < 120\GeV$ is used in the passing category, obtained from the templates for $\cPZ/\Pggx\to\Pe\Pe$ signal and for the $\cPZ/\Pggx\to\Pgt\Pgt$, $\PW$+jets, $\cPqt\cPaqt$, single top quark, diboson ($\PW\PW$, $\PW\cPZ$, $\cPZ\cPZ$), and QCD multijet backgrounds. In the failing category, the total number of events in the same range of $\mvis$ is used to constrain the normalization of the $\cPZ/\Pggx\to\Pe\Pe$ process.

The differential templates for signal and all background distributions, except for QCD multijet, are taken from MC simulation.
The normalization is performed according to the cross section for the specific sample of events, with the exception of the $\PW$+jets background, which is obtained from data, using an enriched sample of $\PW$+jets events with $\mT>70\GeV$.
The scale factor between the sideband and the signal region is extracted from simulation.
The differential distribution and normalization of the QCD multijet background is obtained from data in a control sample where the tag and the probe are of SS.
The contributions from all other backgrounds are estimated using simulation, and are subtracted from the SS control sample in this procedure.

Systematic uncertainties are represented through nuisance parameters in the fit, and account for the effects listed in Section~\ref{sec:systematics}, as well as for the energy scale of tag electrons, which is changed by its uncertainty of $\pm$1\% in the barrel region ($\abs{\eta}<1.46$) and $\pm$2.5\% in the endcap regions ($\abs{\eta}>1.56$), with the difference in the $\mvis$ template considered as an uncertainty in the differential distribution.
Similarly, the energy scale of probe electrons and $\tauh$ are changed by $\pm$1.5 and $\pm$3\%, respectively.
The energy scale of leptons have been measured using the method described in Ref.~\cite{Chatrchyan:2013mxa}.
Uncertainties in the normalization of $\PW$+jets and QCD multijet production are dominated by number of events in the relevant control regions, and each amount to 20\%.
Finally, an additional 3\% uncertainty is associated with the $\cPZ/\Pggx\to\Pe\Pe$ normalization because of the need to disentangle possible differences between the $\cPZ/\Pggx\to\Pe\Pe$ and $\cPZ/\Pggx\to\Pgt\Pgt$ normalizations.
Separate fits are used for probes in the barrel and in the endcap regions.

\begin{figure}[!htbp]
\centering
\includegraphics[width=0.48\textwidth]{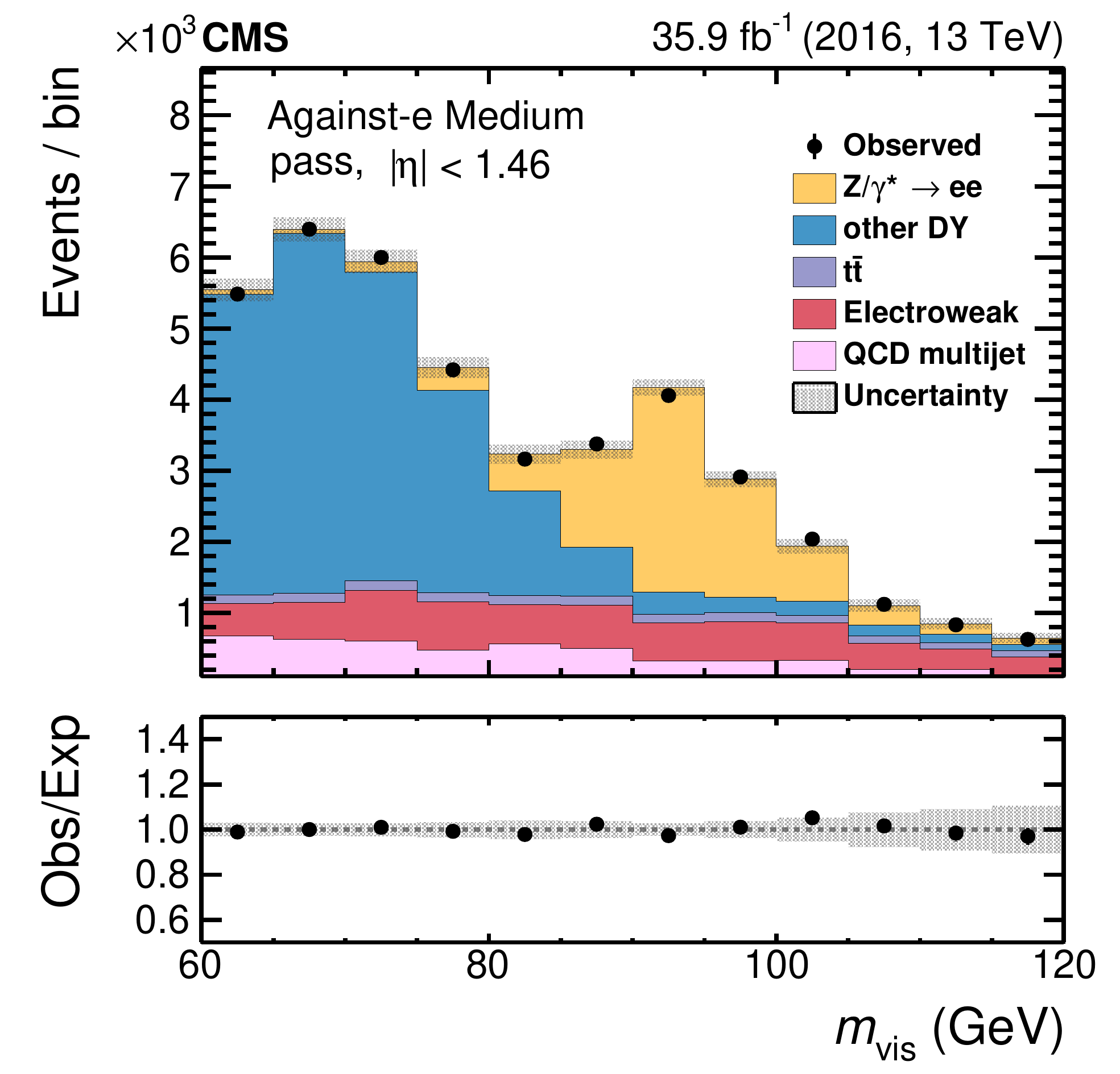}
\includegraphics[width=0.48\textwidth]{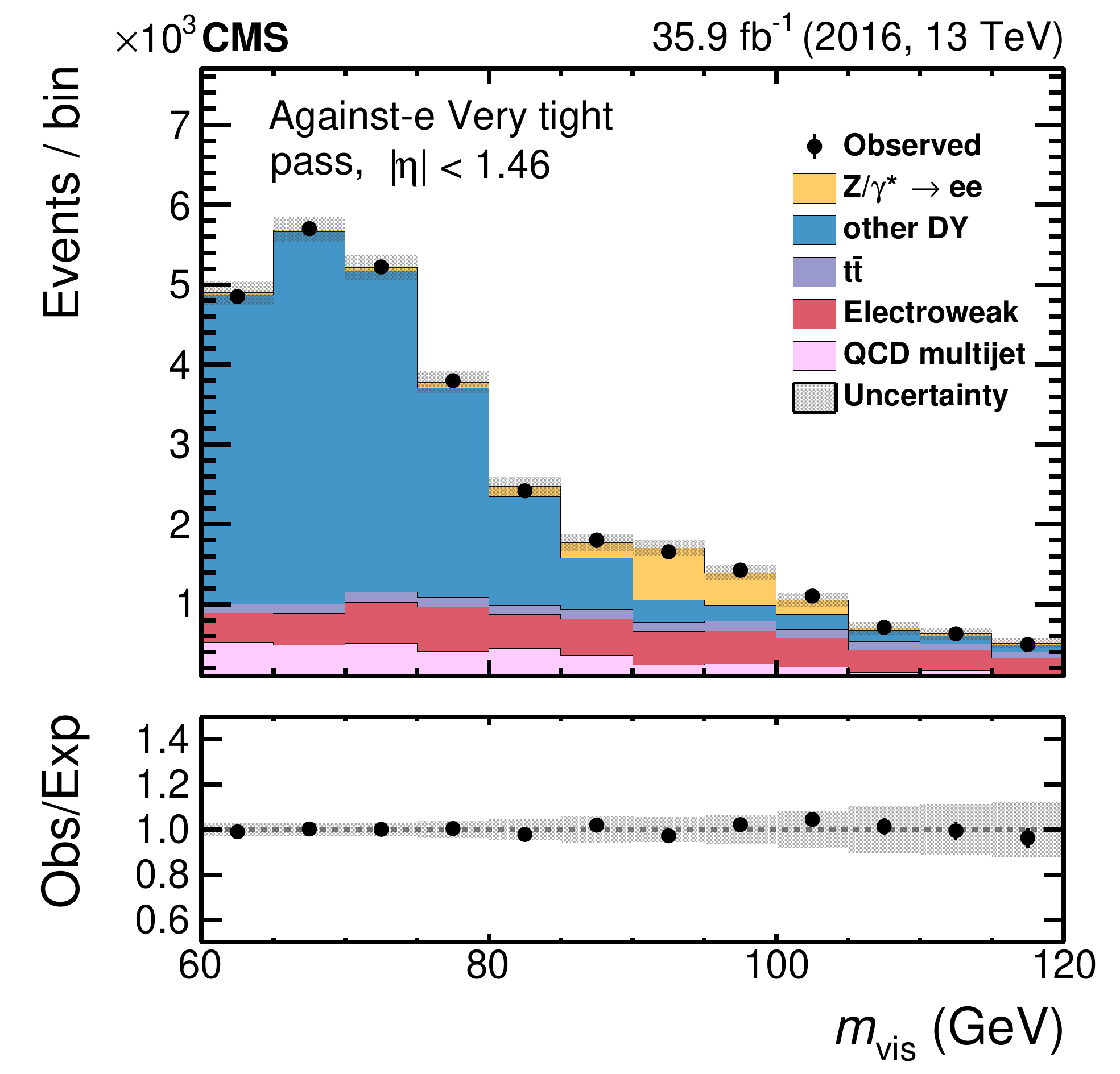}
\caption[dummy text]{Fitted distributions in $\mvis$ in the passing category for the medium (left) and very tight (right) WPs of the against-$\Pe$ discriminant in the barrel region. The electroweak background includes contributions from $\PW$+jets (dominating), diboson, and single top quark events.
Vertical bars correspond to the small (not visible) statistical uncertainties in the data points (68\% frequentist confidence intervals), while the shaded bands provide the quadratic sum of the statistical and systematic uncertainties after the fit.}
\label{fig:eTauFR}

\end{figure}

\begin{figure}[!htbp]
\centering
\includegraphics[width=0.48\textwidth]{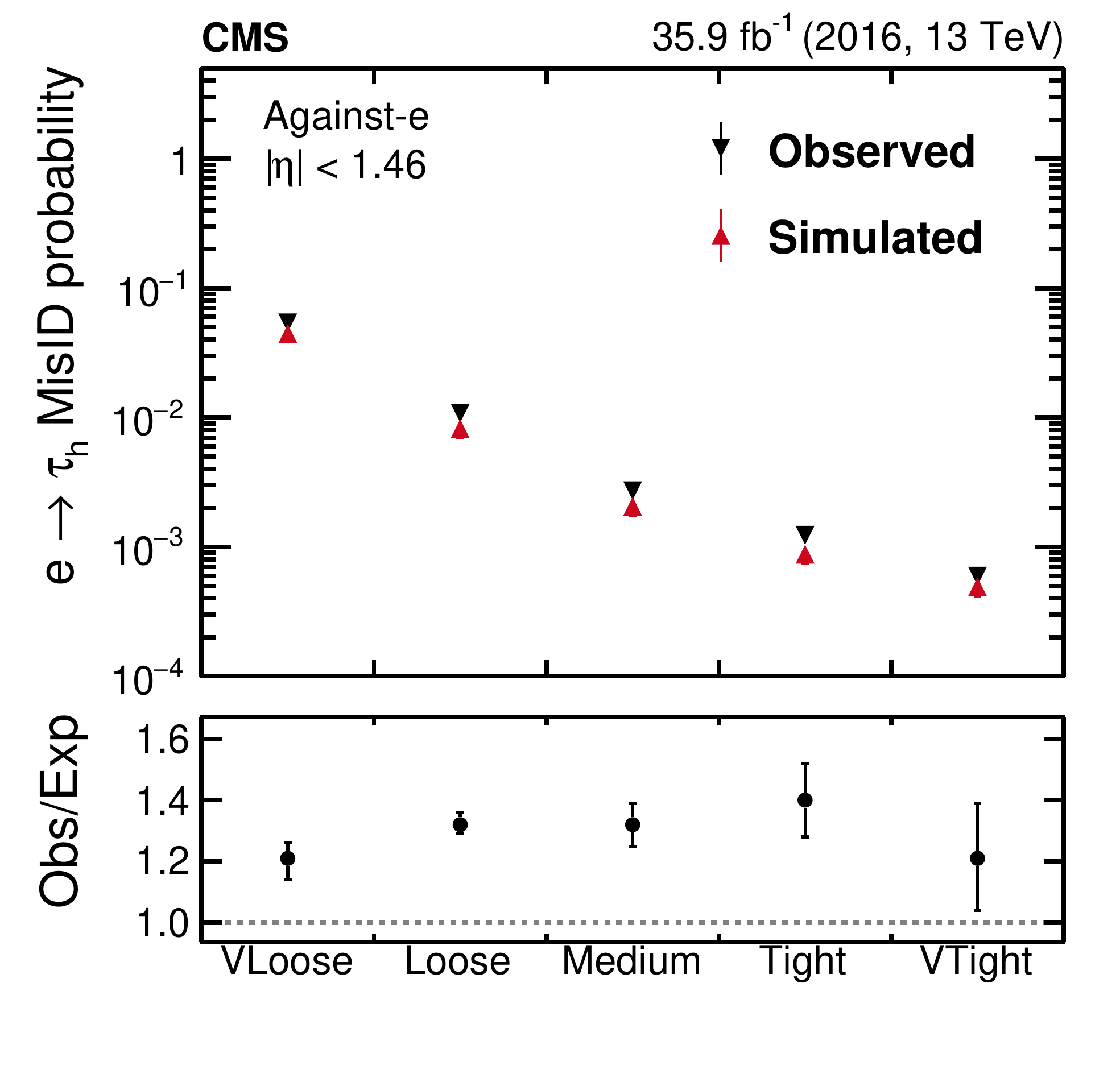}
\includegraphics[width=0.48\textwidth]{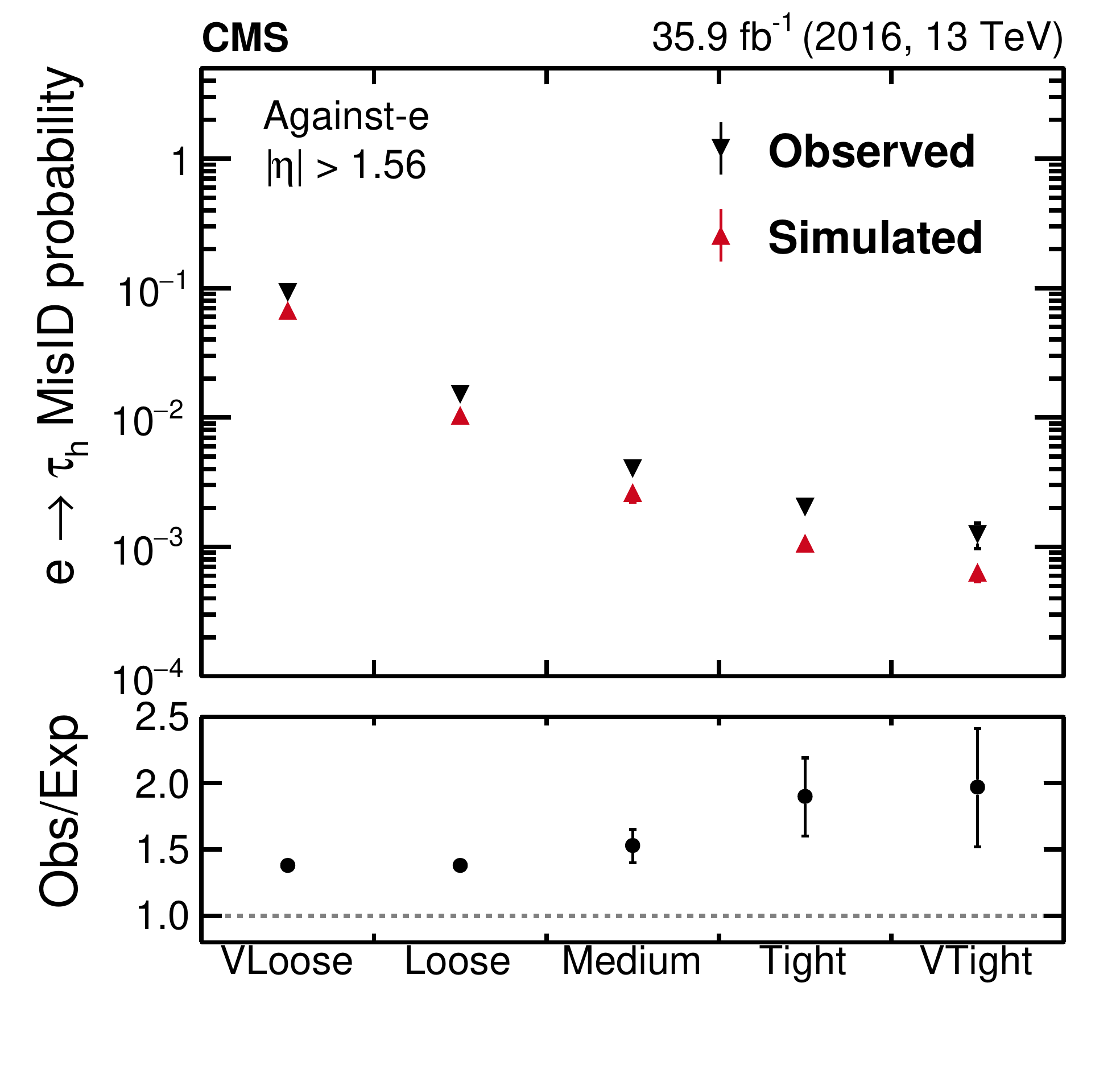}
\caption[dummy text]{Probability for electrons to pass different working points of the against-$\Pe$ discriminant, split into the barrel (left) and endcap (right) regions. For each working point, the $\Pe \mapsto \tauh$ misidentification probability is defined as the fraction of probes passing that working point relative to the total number of probes.
Vertical bars correspond to the statistical and the quadratic sum of the statistical and systematic uncertainties, respectively, for simulated and observed data.}
\label{fig:eTauFRResults}

\end{figure}

The fitted $\mvis$ distributions in the passing category are shown in Fig.~\ref{fig:eTauFR} for the medium and very tight WPs of the against-$\Pe$ discriminant in the barrel region of the ECAL, while the $\Pe \mapsto \tauh$ misidentification probabilities are displayed in Fig.~\ref{fig:eTauFRResults}.
In the barrel region, the measured misidentification probabilities in data exceed those in the simulations. The difference between data and simulation increases for the tight and very tight WPs of the discriminant, and a similar trend is observed for the probes in the endcap regions. The observed misidentification probabilities range from $\approx$5\% for the very loose WP to less than 0.1\% for the very tight WP in the barrel region, while in the endcap regions, the probabilities are larger, ranging between 0.1 and 10\%.

\subsection{Measurement of the \texorpdfstring{$\Pgm \mapsto \tauh$}{mu maps to tau[h]} probability}
\label{sec:muToTauFakeRate}

The $\Pgm \mapsto \tauh$ misidentification probability is also measured using a tag-and-probe method, following an approach similar to that used to measure the $\Pe\mapsto\tauh$ misidentification probability discussed in Section~\ref{sec:eToTauFakeRate}. For this, we select $\cPZ/\Pggx\to \Pgm\Pgm$ events, as described in Section~\ref{sec:validation_eventSelection_Zll}, and again divide these into two categories, depending on whether the probe passes or fails the specific working point of the against-$\Pgm$ discriminant.
The number of $\cPZ/\Pggx\to \Pgm\Pgm$ signal events in each category is then extracted from a simultaneous maximum likelihood fit to the mass of the tag-and-probe pair, in the range $70<\mvis<120\GeV$.
Separate fits are performed for probes in five $\abs{\eta}$ regions of $<$0.4, 0.4--0.8, 0.8--1.2, 1.2--1.7, and $>$1.7, corresponding to the geometry of the CMS muon spectrometer.

The normalization and distribution in $\mvis$ for signal and background processes are estimated as discussed in Section~\ref{sec:eToTauFakeRate}.
Systematic uncertainties are also similar, except that those related to electrons are replaced by those appropriate for the muons, such as the energy scale for the probe, which is changed by $\pm$1.5 and $\pm$3\% for the misidentified $\Pgm\mapsto\tauh$ and the genuine $\tauh$ candidates, respectively, with the resulting difference in the $\mvis$ template taken as an uncertainty in the differential distribution.
The uncertainty in the energy scale of the tag muon is negligible compared with the energy scale of the $\tauh$ candidates, and is therefore neglected.

\begin{figure}[!htbp]
\centering
\includegraphics[width=0.48\textwidth]{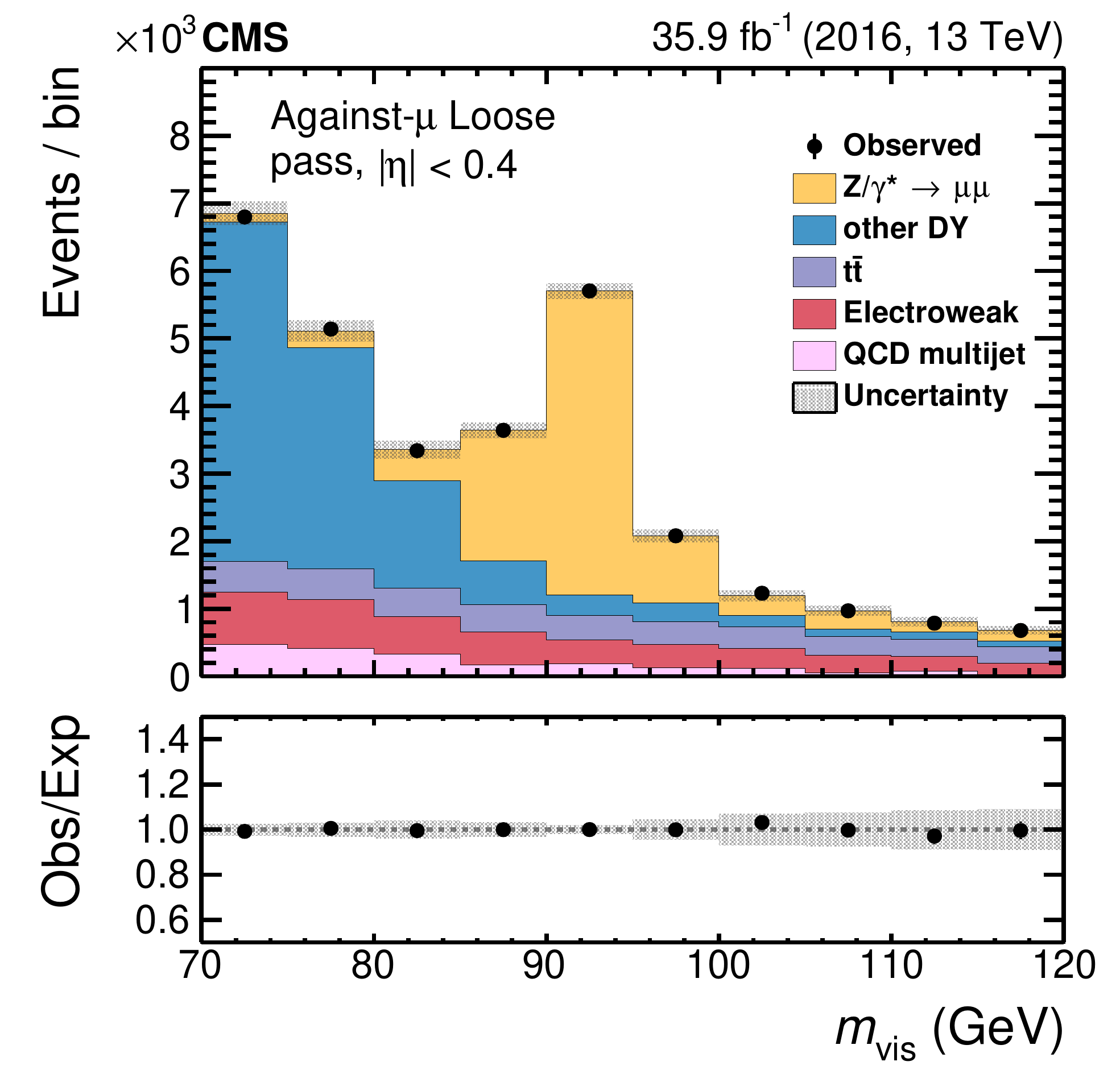}
\includegraphics[width=0.48\textwidth]{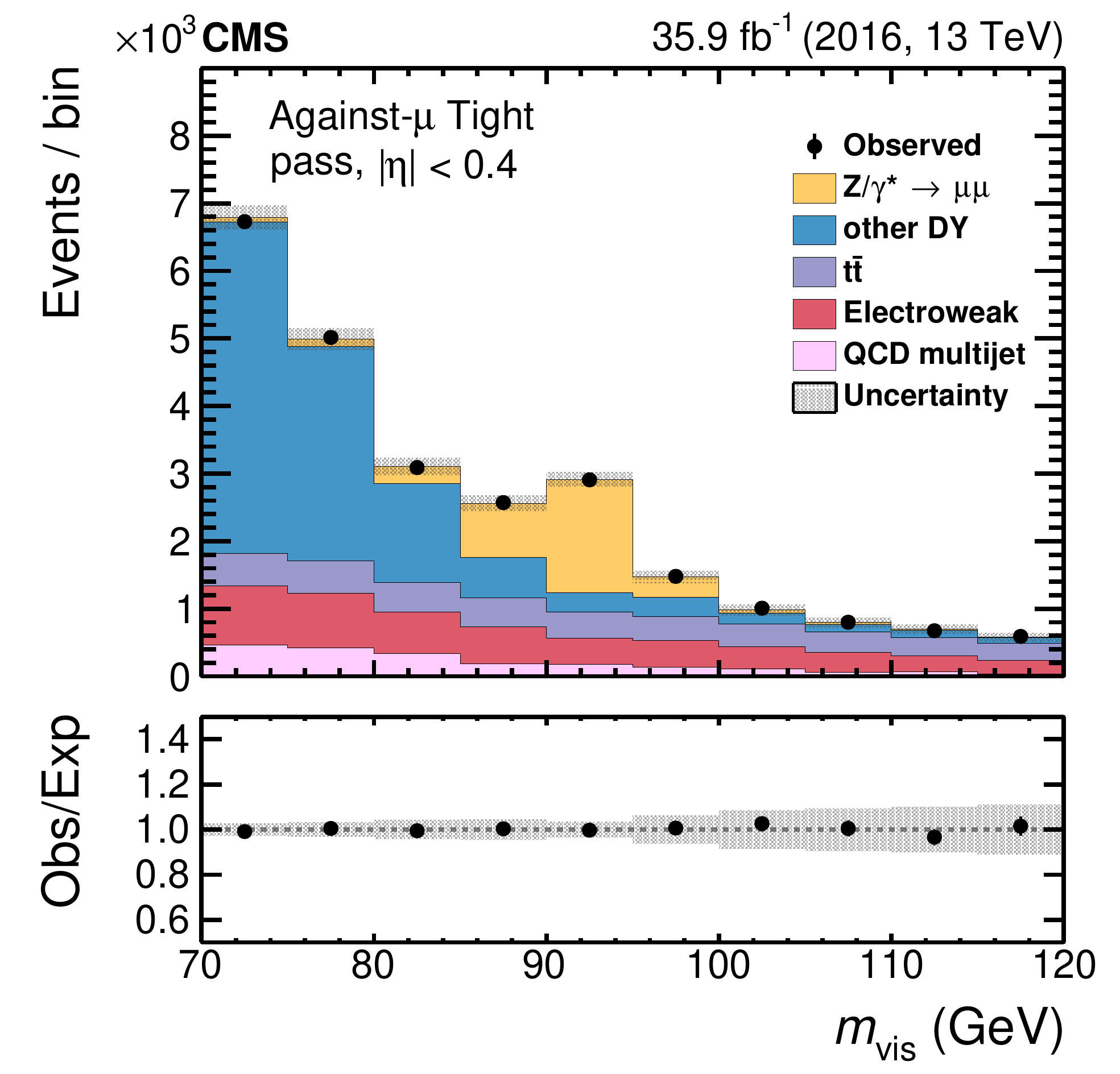}
\caption[dummy text]{Fitted distribution in $\mvis$ in the passing category for the loose (left) and tight (right) WPs of the against-$\Pgm$ discriminant in the region of $\abs{\eta}<0.4$. The electroweak background includes contributions from $\PW$+jets (dominating), diboson, and single top quark events.
Vertical bars correspond to the small (not visible) statistical uncertainties in the data points (68\% frequentist confidence intervals), while the shaded bands provide the quadratic sum of the statistical and systematic uncertainties after the fit.}
\label{fig:muTauFR_LooseLt0p4}

\end{figure}

Figure~\ref{fig:muTauFR_LooseLt0p4} shows the mass distribution in the $\Pgm\tauh$ pair after the maximum likelihood fit, for events where the probe muon is reconstructed as a $\tauh$ candidate, and passes the loose or tight WPs of the against-$\Pgm$ discriminant. The probes in these distributions lie within $\abs{\eta}<0.4$.

\begin{figure}[!htbp]
\centering
\includegraphics[width=0.48\textwidth]{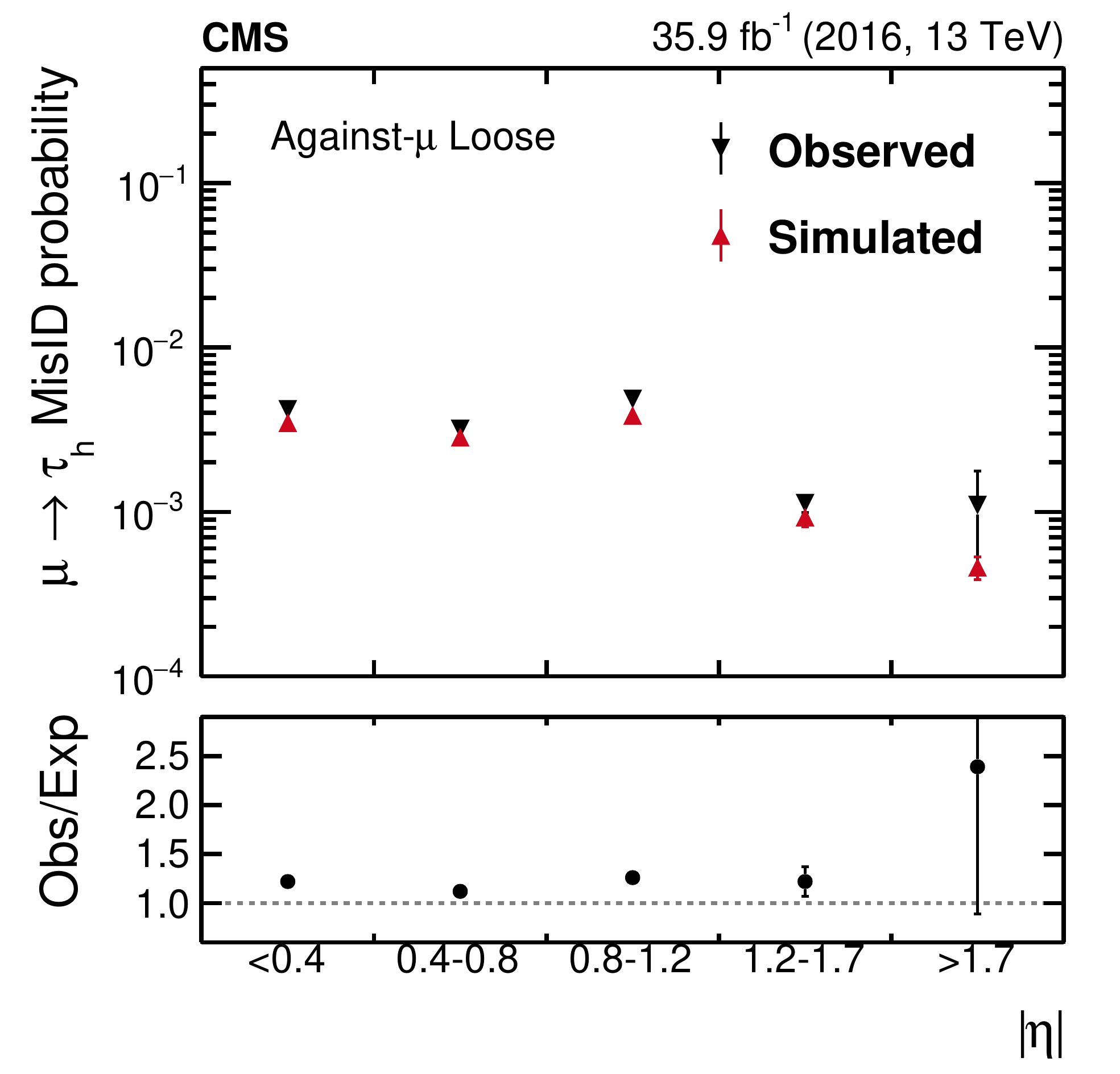}
\includegraphics[width=0.48\textwidth]{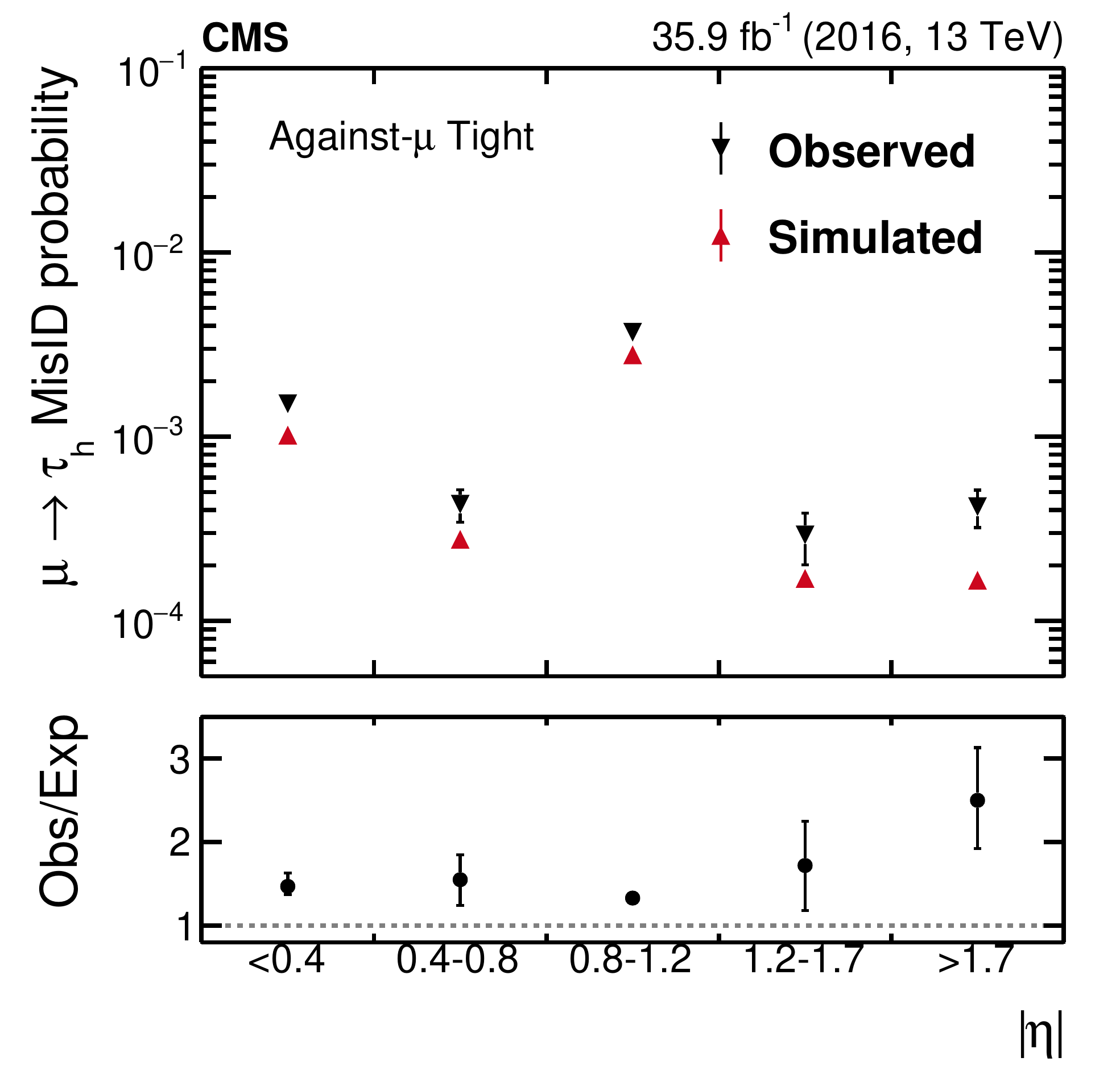}
\caption[dummy text]{Probability for muons to pass the loose (left) and tight (right) WPs of the against-$\Pgm$ discriminant, as a function of the $\abs{\eta}$ of the probe. For each working point, the $\Pgm \mapsto \tauh$ misidentification probability is defined as the fraction of probes passing that working point relative to the total number of probes.
Vertical bars correspond to the statistical and the quadratic sum of the statistical and systematic uncertainties, respectively, for simulated and observed data.}
\label{fig:muToTauFakeRateResults}

\end{figure}

The $\Pgm \mapsto \tauh$ misidentification rates are given for the loose and tight WPs of the against-$\Pgm$ discriminant in Fig.~\ref{fig:muToTauFakeRateResults}. For probes passing the WPs, the measured misidentification rates in data exceed the predictions, with the difference between data and simulation possibly increasing from small to large $\abs{\eta}$.
The observed trend is more significant for probes passing the tight WP. The observed misidentification probabilities for the loose WP are in the range of 0.1--0.5\%, with the highest probability lying in the $\abs{\eta}$ range between 0.8 and 1.2 which corresponds to transition between barrel and endcap regions of the muon spectrometer.
The probabilities for the tight WP range between 0.03 and 0.40\%, with the highest value again falling in the same $\abs{\eta}$ region.

\section{Measurement of the \texorpdfstring{$\tauh$}{tau[h]} energy scale}
\label{sec:TauEnergyScale}
The correction to the $\tauh$ energy scale is defined by the deviation of the average reconstructed $\tauh$ energy from the generator-level energy of the visible $\tauh$ decay products. The corresponding data-to-simulation correction is obtained from a fit of the distributions of observables sensitive to the energy scale, using samples of $\Pe\tauh$ and $\Pgm\tauh$ final states in $\cPZ/\Pggx$ events. The distributions sensitive to the energy scale are $m_{\tauh}$ and the mass of the $\ell\tauh$ system, $\mvis$. These are fitted, separately for the $\oneProngZeroPizero$,  $\oneProngOnePizero$, and $\threeProngZeroPizero$ decays to extract the correction factors between data and simulation.

The $\Pe\tauh$ and $\Pgm\tauh$ final states are selected as described in Section~\ref{sec:validation_eventSelection_ZTT}, except that the $\tauh$ candidates are required to pass the very tight WP of the MVA-based $\tauh$ isolation discriminant to further reduce backgrounds from jets misidentified as $\tauh$ candidates. Moreover, the requirement on $\mT$ is tightened to be less than 30\GeV, and the requirement on $D_\zeta$ is removed. Finally, the $\tauh$ candidate must satisfy the tight and loose, or very loose and tight WPs of the against-$\Pe$ and against-$\Pgm$ discriminants in the respective $\Pe\tauh$ or $\Pgm\tauh$ final states. Templates for events in which the reconstructed $\tauh$ is matched to some generated $\tauh$ are obtained by changing the reconstructed $\tauh$ energy between $-$6\% and $+$6\% in steps of 0.1\%, with the $\mvis$ and $m_{\tauh}$ recomputed at each step. The maximal energy shifts of $\pm$6\% are selected to be sufficiently away from the nominal value in the simulation such that the true value can be obtained between them.
While $m_{\tauh}$ displays higher sensitivity to the energy scale for the $\oneProngOnePizero$ and $\threeProngZeroPizero$ decay modes, it cannot be used in the $\oneProngZeroPizero$ decay mode, where only $\mvis$ is used. The backgrounds are modelled in the same way as described in Section~\ref{sec:tauIdEfficiency_TnP_Ztautau}, and the templates for processes in which there is no match between the reconstructed and generated $\tauh$ candidates are not changed as a function of the $\tauh$ energy scale.

\begin{figure}[!htbp]
\centering
\begin{tabular}{cc}
\multicolumn{2}{c}{\includegraphics[width=0.48\textwidth]{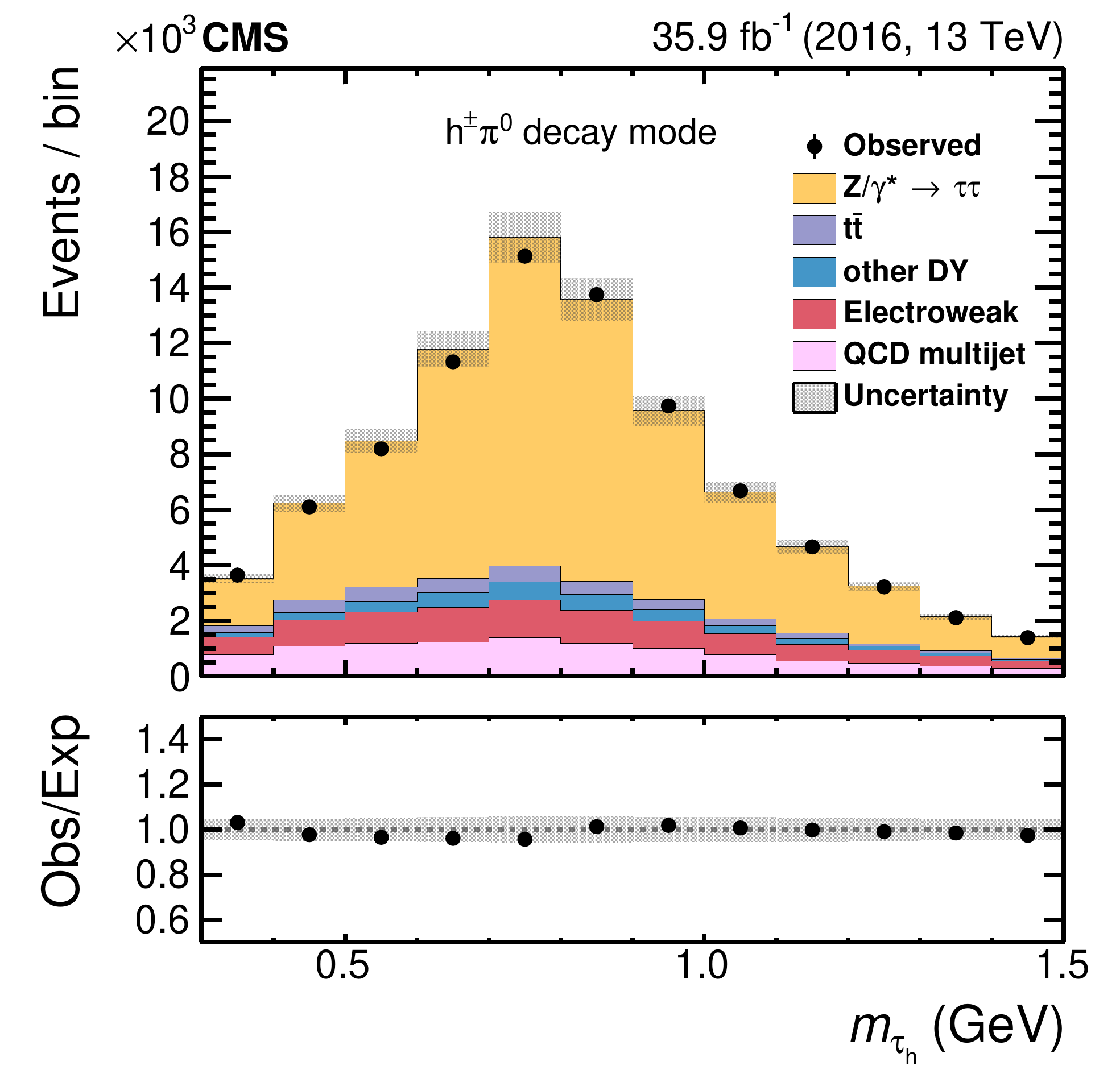}}\\
\includegraphics[width=0.48\textwidth]{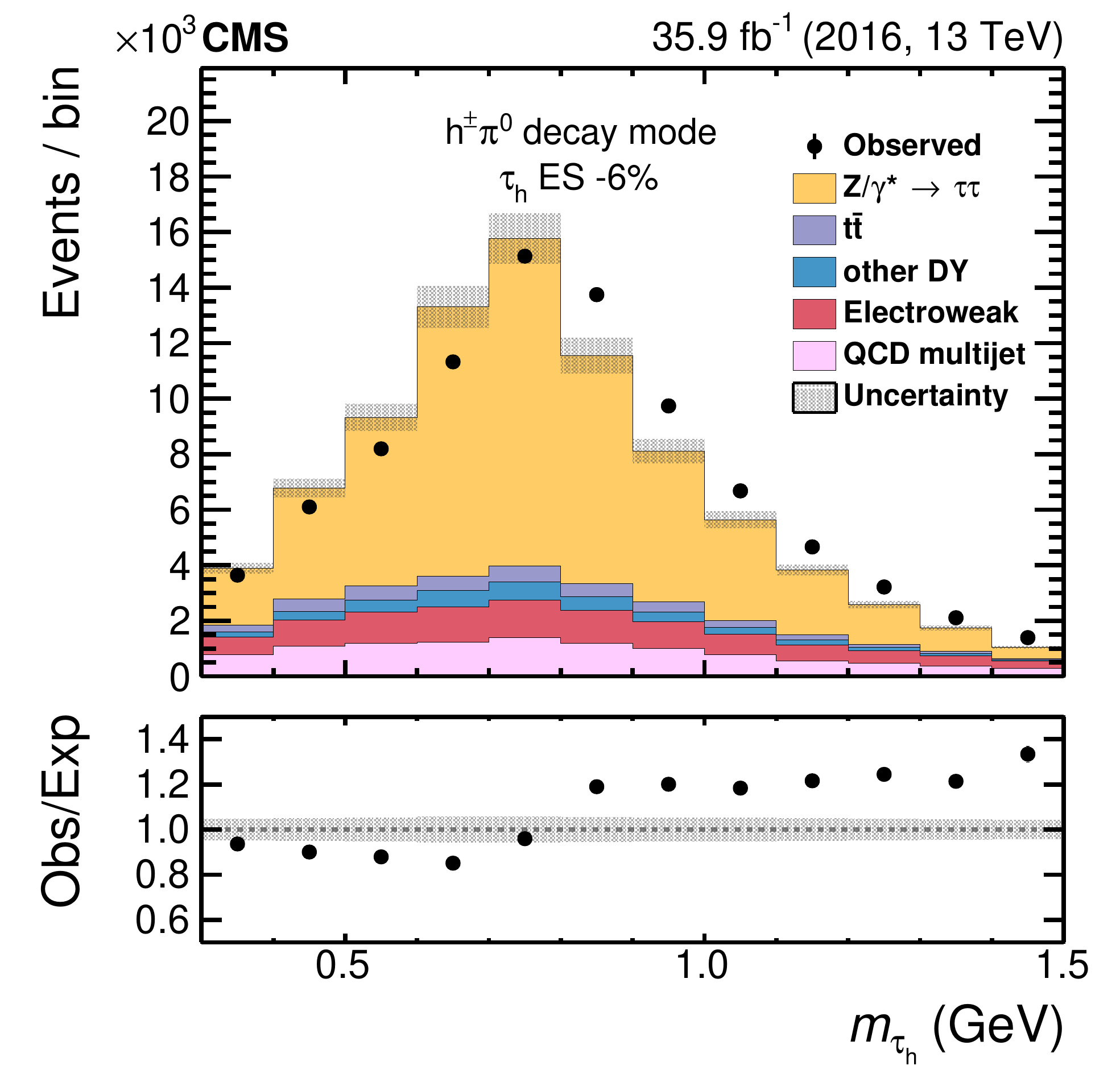} &
\includegraphics[width=0.48\textwidth]{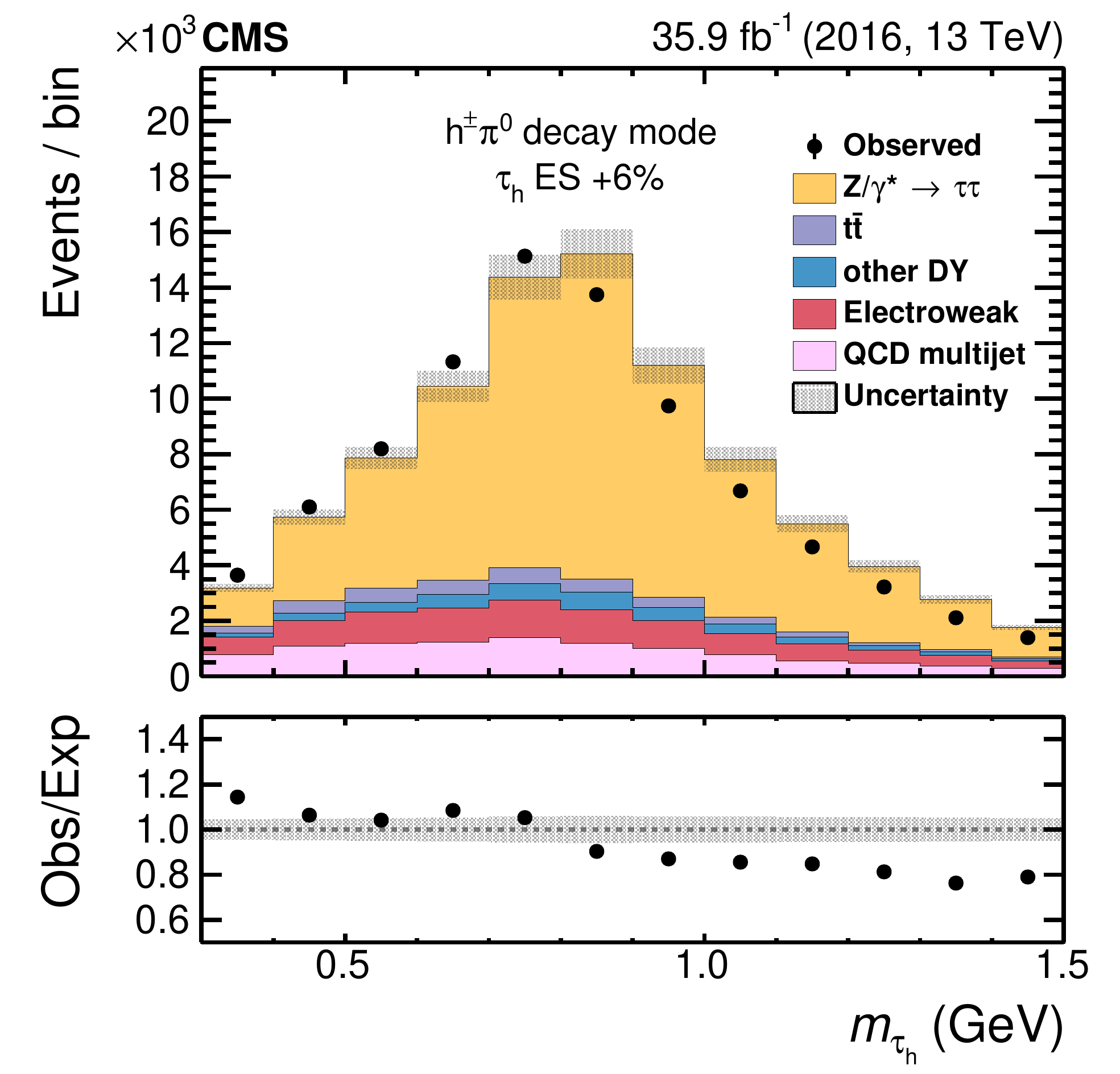} \\
\end{tabular}
\caption[dummy text]{The distributions in $m_{\tauh}$ for $\Pgm\tauh$ events in the $\oneProngOnePizero$ decay channel. The data are compared to predictions with different shifts applied to the $\tauh$ energy scale: 0\% (upper), $-$6\% (lower left), and $+$6\% (lower right). The electroweak background includes contributions from $\PW$+jets (dominating), diboson, and single top quark events. Vertical bars (smaller than the symbol size) correspond to the statistical uncertainty in the data points (68\% frequentist confidence intervals), while the shaded bands provide the expected systematic uncertainties.}
\label{fig:tauEnergyScale_visibleTauMass_ControlPlots}

\end{figure}

For illustration, the $m_{\tauh}$ templates corresponding to no shifts, and to shifts in $\tauh$ energy scale of $-$6 and $+$6\% are shown in Fig.~\ref{fig:tauEnergyScale_visibleTauMass_ControlPlots} for events selected in $\oneProngOnePizero$ decay mode.  The data are compared to predictions for these three energy scales.

A likelihood ratio method is used to extract the $\tauh$ energy scale for each decay mode.
In addition to those listed in Section~\ref{sec:systematics}, the following sources of systematic uncertainties are considered: uncertainties in the identification of $\tauh$ candidates, determined in Section~\ref{sec:Tau_identification_efficiency}, are split into those that are uncorrelated ($\approx$2\%) and correlated ($\approx$4.5\%) between the $\Pe\tauh$ and $\Pgm\tauh$ final states. The rates for electrons, muons, and jets misidentified as $\tauh$ candidates have uncertainties of 12, 25, and 20\%, respectively. Moreover, uncertainties in the energy scale of electrons (1\% in the barrel and 2.5\% in the endcaps) and muons (5\%) identified as $\tauh$ are taken into account in their differential distributions.
The results obtained from fits to $\mvis$ and $m_{\tauh}$ distributions for each decay mode in the $\Pe\tauh$ and $\Pgm\tauh$ final states are found to be compatible with each other, and their combination is given in Table~\ref{tab:tauEnergyScale_values}. The measurement is limited by systematic rather than statistical uncertainties.

\begin{table}[!hbtp]
        \centering
                \topcaption
                [Numerical results of the tau energy scale measurements.]
                {The data-to-simulation correction for the $\tauh$ energy scale from the combination of measurements performed in the $\Pe\tauh$ and $\Pgm\tauh$ final states separately using $m_{\tauh}$ and $\mvis$ distributions. The correction is relative to the reconstructed energy from simulation, expressed in \%.}
                \begin{tabular}{l D{,}{}{-1} D{,}{}{-1}}
                               Decay mode & \multicolumn{1}{c}{$m_{\tauh}$} & \multicolumn{1}{c}{$\mvis$} \\
                        \hline
                        $\oneProngZeroPizero$ & , - & -0.5, \pm 0.5 \\
                        $\oneProngOnePizero$ & 0.9, \pm 0.3 & 1.1, \pm 0.3 \\
                        $\threeProngZeroPizero$ & 0.6, \pm 0.2 & 0.6, \pm 0.3 \\
                        \hline
                \end{tabular}
        \label{tab:tauEnergyScale_values}
\end{table}

Additional studies performed using the $\Pgm\tauh$ final state are carried out to assess the stability of the measurement.
To gauge the impact of fluctuations caused by the limited number of MC events relative to the data, the simulated events used to model $\cPZ/\Pggx$ decays are split into four samples of equal size, and the measurement is performed using each of these four subsamples. The resulting fluctuations in the measured $\tauh$ energy scale are up to 1\%. Similarly, the effect of the contamination from backgrounds that arise from misidentification of the $\tauh$ is checked by changing the selection criteria, and found to be 0.5\%. The choice of the binning is investigated by changing the number of bins up and down by a factor 2. The results are compatible to within 1\%. Finally, the effect of the range of the fit is evaluated for the $\mvis$ template by increasing it by 10\GeV in either direction, resulting in changes compatible within 0.5\% of the original measurement. Although these checks do not guarantee that similar levels of fluctuation exist in the original measurement (especially, the assessment of the limited number of MC events),
an additional uncertainty of 1.0\% is added in quadrature to the uncertainty detailed in Table~\ref{tab:tauEnergyScale_values},
to reflect our limited knowledge of the true fluctuations. This results in a total uncertainty of $<$1.2\%.

\section{Performance of \texorpdfstring{$\tauh$}{tau[h]} identification in the high-level trigger}
\label{sec:Tau_Trigger_Performance}

The $\tauh$ reconstruction and identification algorithm described in Section~\ref{sec:Tau_Trigger} for the HLT was used to define a set of triggers for 2016 data taking. These triggers cover all final states of interest, namely, $\Pgt$ lepton pair production in $\Pgt_{\Pe}\tauh$, $\Pgt_{\Pgm}\tauh$, and $\tauh\tauh$ decays, $\tauh$ associated with \ptmiss ($\tauh\ptmiss$), and single $\tauh$ with large~\pt.

There are two types of HLT decision trees that use $\tauh$ candidates and which are aimed at
two different classes of final states, those that comprise other than $\tauh$ candidates in the event,
\eg, $\Pe\tauh$, $\Pgm\tauh$, $\tauh\ptmiss$, and those that include only $\tauh$ candidates,
\eg, $\tauh\tauh$.
The first type of trigger is based on L1 seeds that require the presence of an electron,
 a muon, or large $\ptmiss$, possibly together with a $\tauh$ candidate.
These triggers also have their corresponding selections
in $\Pe$, $\Pgm$, or $\ptmiss$ in the HLT, thereby greatly reducing the event rates processed
at later stages.
This allows reconstruction of $\tauh$ candidates directly through the resource-intensive L3 step,
wherein the PF sequence underpinning $\tauh$ reconstruction
is run using the full-detector acceptance.
In the second type of trigger, only $\tauh$ candidates are required to be identified at L1, without
additional lepton or $\ptmiss$ selections.
At HLT, since the L3 step would be too time consuming to run at the L1 output rates, the
L2 and L2.5 filtering steps are executed first. The efficiency of the L2 and L2.5 filter is
$>$95\% per $\tauh$ candidate.
In addition, this class of triggers has HLT $\tauh$ reconstruction used only in regions
of the detector centered on the direction of the L1 $\tauh$ candidates, further reducing thereby the processing time.

The triggers for $\Pgt$ pair production are aimed mainly to select efficiently the SM $\PH\to\Pgt\Pgt$ decays that require respective \pt thresholds of 20--25 and 30--35\GeV for $\Pgt_{\Pe}$ or $\Pgt_{\Pgm}$ and $\tauh$ final states.
In addition, trigger rates at an instantaneous luminosity of $\lumi=1.4\times 10^{34}\percms$ and PU close to 40 interactions per bunch crossing, typical for pp collisions in late 2016, were required not to exceed rates of about 10--15 and 50--65\unit{Hz} for the $\Pe\tauh$ or $\Pgm\tauh$ and $\tauh\tauh$ triggers, respectively.

The $\Pgm\tauh$ trigger is constructed as follows. First, we require the presence of a muon candidate with $\pt>18\GeV$ at L1. Then, an isolated muon, seeded by the L1 candidate, with  $\pt>19\GeV$ is selected at the HLT. Subsequently, an unseeded L3 $\tauh$ candidate is selected with $\pt>20\GeV$ that passes the loose charged-particle isolation WP. The isolation is relaxed linearly by 10\%/\GeVns for $\pt^{\tauh}>50\GeV$. Finally, the L3 $\tauh$ candidate must be separated from the muon by $\Delta R>0.3$.
At $\lumi=1.4\times 10^{34}\percms$, the rate for the $\Pgm\tauh$ trigger corresponds to $\approx$20\unit{Hz}.

To adapt to different conditions in instantaneous luminosity delivered by the LHC in 2016, ranging from ${\approx}3\times 10^{33}\percms$ to $1.4\times 10^{34}\percms$, and to provide highest efficiency possible within the limited rate budget, two variants of $\Pe\tauh$ triggers were developed.
The first one is similar to the $\Pgm\tauh$ trigger in that an isolated electromagnetic ($\Pe$ or $\Pgg$) object with $\pt>22\GeV$ is required at L1, and is used to initiate the reconstruction of an isolated electron at the HLT that is required to have $\pt>24\GeV$. A seedless L3 $\tauh$ candidate, not overlapping with the electron, is required to have $\pt>20\GeV$, and to pass the loose charged-particle isolation WP (linearly relaxed by 10\%/\GeVns for $\pt^{\tauh}>50\GeV$).
This trigger covered instantaneous luminosities of up to $9\times 10^{33}\percms$.

The second, a more stringent version of the $\Pe\tauh$ trigger, adds the requirement of an L1 $\tauh$ candidate to accompany the L1 electromagnetic object. First, the \pt threshold on the L1 $\tauh$ was set to 20\GeV, and, as the instantaneous luminosity increased, was raised to 26\GeV, and eventually the L1 isolation condition was also applied.
In the latter configuration at the HLT, the \pt threshold for the L3 $\tauh$candidate was adjusted to 30\GeV.
In the utilized ranges of instantaneous luminosity for which the $\Pe\tauh$ triggers were designed, the trigger rates remained below 15\unit{Hz}.

The $\tauh\tauh$ triggers require a pair of isolated L1 $\tauh$ candidates, with \pt above a threshold in the range of 28--36\GeV. The threshold is dynamically adjusted to maintain a constant rate of events passing L1,
independent of the instantaneous luminosity.
Even after satisfying the L1 requirements, the event rate is still too high to run the L3 $\tauh$ reconstruction.
The L3 reconstruction is therefore used only if at least two $\tauh$ candidates pass the L2 and L2.5 stages, as discussed in Section~\ref{sec:Tau_Trigger}.
At L3, the candidates have to have $\pt>35\GeV$, and pass the medium WP of the charged isolation (the charged isolation was replaced by the combined isolation
at $\lumi>1.3\times 10^{34}\percms$). The isolation is linearly relaxed by 6\%/\GeVns for $\pt^{\tauh}>73\GeV$.
Two such candidates must be present in the event, and must be separated by $\Delta R>0.5$.
At $\lumi=1.4\times 10^{34}\percms$, the rate of $\tauh\tauh$ triggers was below 60\unit{Hz}.

The benchmark process that guided the design of the $\tauh\ptmiss$ trigger is the decay of a charged resonance  $\mathrm{X}^\pm\to\Pgt\Pgn$, \eg, for $\mathrm{X}^\pm=\PH^\pm$ or $\PW'$, with a mass $m_\mathrm{X}>200\GeV$.
At L1, this trigger requires $\ptmiss$ in excess of 80--100\GeV, again with the threshold dynamically adjusted as a function of instantaneous luminosity to keep the rate of events passing L1 constant.
At the HLT, the selected events must further satisfy the condition of $\ptmiss>90\GeV$. After this, the L3 $\tauh$ reconstruction step is executed, and events are finally saved when an L3 $\tauh$ candidate with $\pt>50\GeV$, passing the loose WP of charged isolation (relaxed by 6\%/\GeVns for $\pt^{\tauh}>100\GeV$) is found, with its leading charged hadron satisfying $\pt>30\GeV$.
At $\lumi=1.4\times 10^{34}\percms$, the rate for the $\tauh\ptmiss$ trigger is about 20\unit{Hz}.

Finally, a high-\pt single-\tauh trigger was developed for searches for high-mass resonances decaying into at least one $\Pgt$ lepton, for example $\PW'$, $\PH^{\pm}$, or the heavy $\mathrm{A}$ or $\PH$ boson in MSSM.
This trigger was designed to cover portions of the phase space not covered by the more usual cross-triggers ($\tauh\tauh$, $\tauh\ptmiss$, $\Pe\tauh$, and $\Pgm\tauh$), \eg,~$\PH^{\pm}$ events with an energetic $\tauh$ but small $\ptmiss$.
The trigger that fulfilled those conditions required an isolated L1 $\tauh$ candidate with $\pt>120\GeV$. The $\tauh$ reconstruction at the HLT consists of steps taken in L2, L2.5, and L3.
The L3 requires one $\tauh$ candidate with $\pt>140\GeV$, and with a leading charged hadron with $\pt>50\GeV$. The L3 $\tauh$ candidate has to pass also the tight WP of charged isolation, which is linearly relaxed by 2\%/$\GeVns$ for $\pt^{\tauh}>275\GeV$, and is discarded for $\pt^{\tauh}>500\GeV$.
Rates of about 30\unit{Hz} were allocated to this trigger.

The basic features of triggers with $\tauh$ candidates used to record $\Pp\Pp$ collisions in 2016 are summarized in Table~\ref{tab:triggerpaths}.
The efficiencies of the $\tauh$ part of the triggers listed in Table~\ref{tab:triggerpaths} are measured via the tag-and-probe technique as a function of the offline-reconstructed $\pt^{\tauh}$, using data enriched in $\tauh$~leptons from $\cPZ/\Pggx\to\Pgt\Pgt\to\Pgm\tauh$ decays.
To single out this sample, the selections for the $\Pgm\tauh$ final state described in Section~\ref{sec:validation_eventSelection_ZTT}, together with the requirement of $\mT<30\GeV$ and the additional condition of $40<\mvis<80\GeV$, are applied to data previously collected through single-muon triggers.
Furthermore, to provide an efficiency measurement that is specific to the selections used in $\PH\to\Pgt\Pgt$ analyses, the $\tauh$ candidates must pass the tight WP of the MVA-based isolation discriminant.
The residual contamination from other objects misidentified as $\tauh$ is subtracted statistically using SS events passing the same selections. The purity of the final sample exceeds 95\%.

\begin{table}[!hbtp]
  \centering
  \topcaption{Triggers with $\tauh$ candidates used to record pp collisions in 2016: the final state (Channel), HLT \pt thresholds and $\tauh$ isolation working point, L1 \pt thresholds, peak instantaneous luminosity ($\lumi_\text{peak}$) in the period of operation as main trigger, and integrated luminosity ($\int\!\!\lumi$) collected with the trigger. The $\tauh\tauh$ and $\tauh\ptmiss$ triggers are seeded by sets of L1 triggers with thresholds dynamically adjusted as a function of the instantaneous luminosity to maintain a constant L1 rate, given by the ranges in \pt. The trigger \pt thresholds and isolation criteria were successively tightened over the data-taking period to keep the rate of events passing HLT approximately constant with increasing instantaneous luminosity.}
  \resizebox{\textwidth}{!}{
    \begin{tabular}{lllcc}
      \rule{0pt}{2.5ex}
      \rule{0pt}{2.5ex}
      Channel & HLT object and WP & L1 object & \multicolumn{1}{l}{$\lumi_\text{peak}$ ($\!\percms$)} & \multicolumn{1}{l}{$\int\!\!\lumi$ ($\!\fbinv$)} \\
      \hline
      \rule{0pt}{2.5ex}
      \multirow{1}{*}{$\Pgm\tauh$}
      & $\pt^\Pgm>19\GeV$, $\pt^{\tauh}>20\GeV$, loose~iso & $\pt^\Pgm>18\GeV$ & $1.5\times 10^{34}$ & 35.9 \\
      \hline
      \rule{0pt}{2.5ex}
      \multirow{3}{*}{$\Pe\tauh$}
      & $\pt^\Pe>24\GeV$, $\pt^{\tauh}>20\GeV$, loose~iso & $\pt^{\Pe/\Pgg}>22\GeV$                             & $0.9\times 10^{34}$ &  7.5 \\
      & $\pt^\Pe>24\GeV$, $\pt^{\tauh}>20\GeV$, loose~iso & $\pt^{\Pe/\Pgg}>22\GeV$, $\pt^{\tauh}>20\GeV$      & $1.3\times 10^{34}$ & 10.2 \\
      & $\pt^\Pe>24\GeV$, $\pt^{\tauh}>30\GeV$, loose~iso & $\pt^{\Pe/\Pgg}>22\GeV$, $\text{iso}\,\pt^{\tauh}>26\GeV$ & $1.5\times 10^{34}$ & 18.2 \\
      \hline
      \rule{0pt}{2.5ex}
      \multirow{2}{*}{$\tauh\tauh$}
      & $2\times\pt^{\tauh}>35\GeV$, medium~iso           & $2\times\text{iso}\,\pt^{\tauh}>28\text{--}36\GeV$ & $1.3\times 10^{34}$ & 27.3 \\
      & $2\times\pt^{\tauh}>35\GeV$, medium~comb.~iso & $2\times\text{iso}\,\pt^{\tauh}>28\text{--}36\GeV$ & $1.5\times 10^{34}$ &  8.6 \\
      \hline
      \rule{0pt}{2.5ex}
      \multirow{2}{*}{$\tauh\ptmiss$}
      & $\ptmiss>90\GeV$, & \multirow{2}{*}{$\ptmiss>80\text{--}100\GeV$} & \multirow{2}{*}{$1.5\times 10^{34}$} & \multirow{2}{*}{35.9} \\
      & $\pt^{\tauh}>50\GeV$, $\pt^{\mathrm{h^{\pm}}}>30\GeV$, loose~iso & & & \\
      \hline
      \rule{0pt}{2.5ex}
      \multirow{1}{*}{$\tauh$}
      & $\pt^{\tauh}>140\GeV$, $\pt^{\mathrm{h^{\pm}}}>50\GeV$, tight~iso & $\pt^{\tauh}>120\GeV$ & $1.5\times 10^{34}$ & 33.1 \\
      \hline
    \end{tabular}
  }
  \label{tab:triggerpaths}
\end{table}

To provide an unbiased measurement of the efficiency of the single-$\tauh$ part of the $\tauh\tauh$
and $\tauh\ptmiss$ triggers, special $\Pgm\tauh$ triggers were put in
place. The special triggers have one part that is required to match the
nominal single-muon trigger used to select events; the other part is required to pass
the $\tauh$ trigger identification for the trigger of interest.

\begin{figure}[!htbp]
\centering
\subfigure{\includegraphics[width=0.48\textwidth]{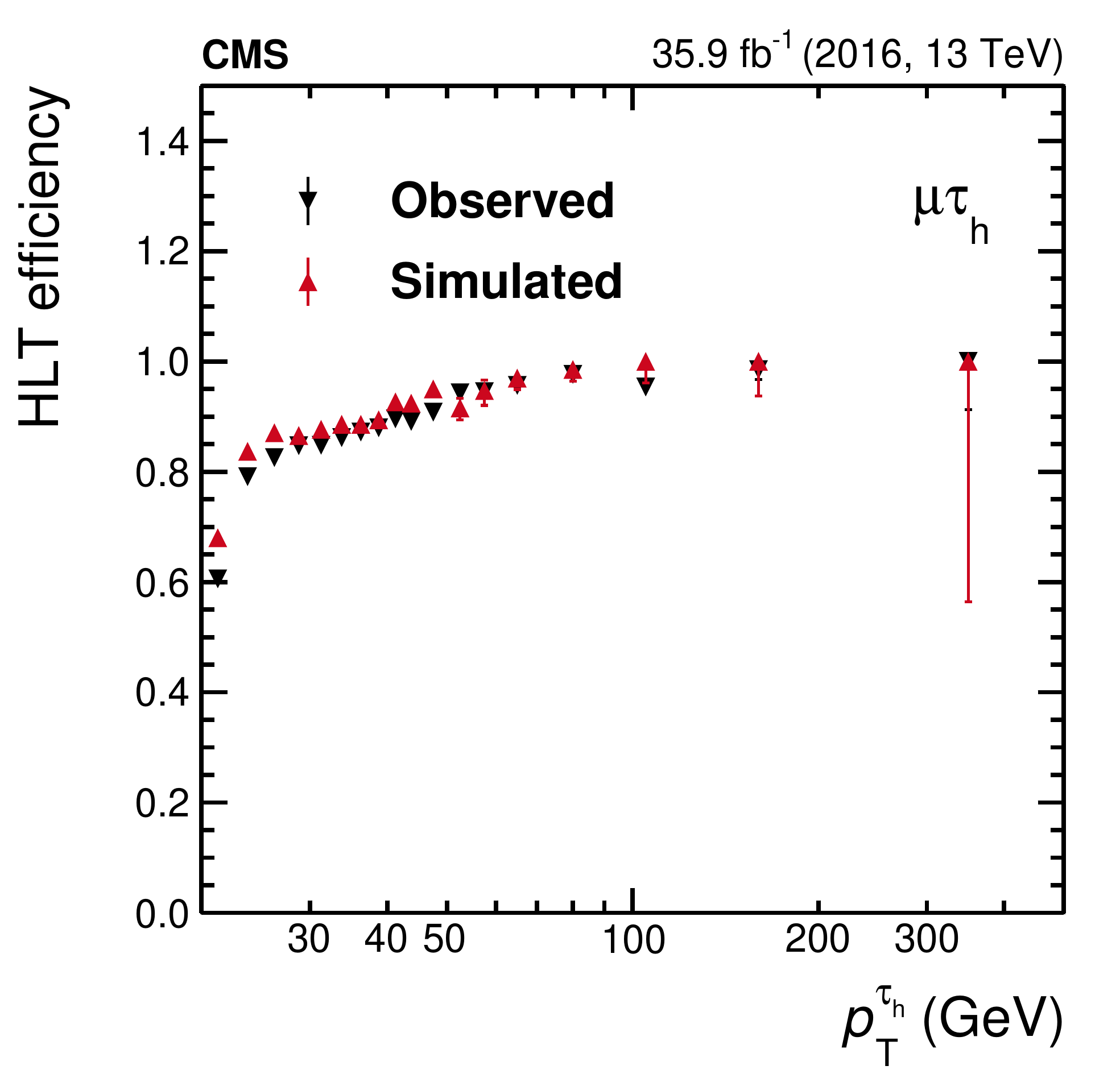}}
\subfigure{\includegraphics[width=0.48\textwidth]{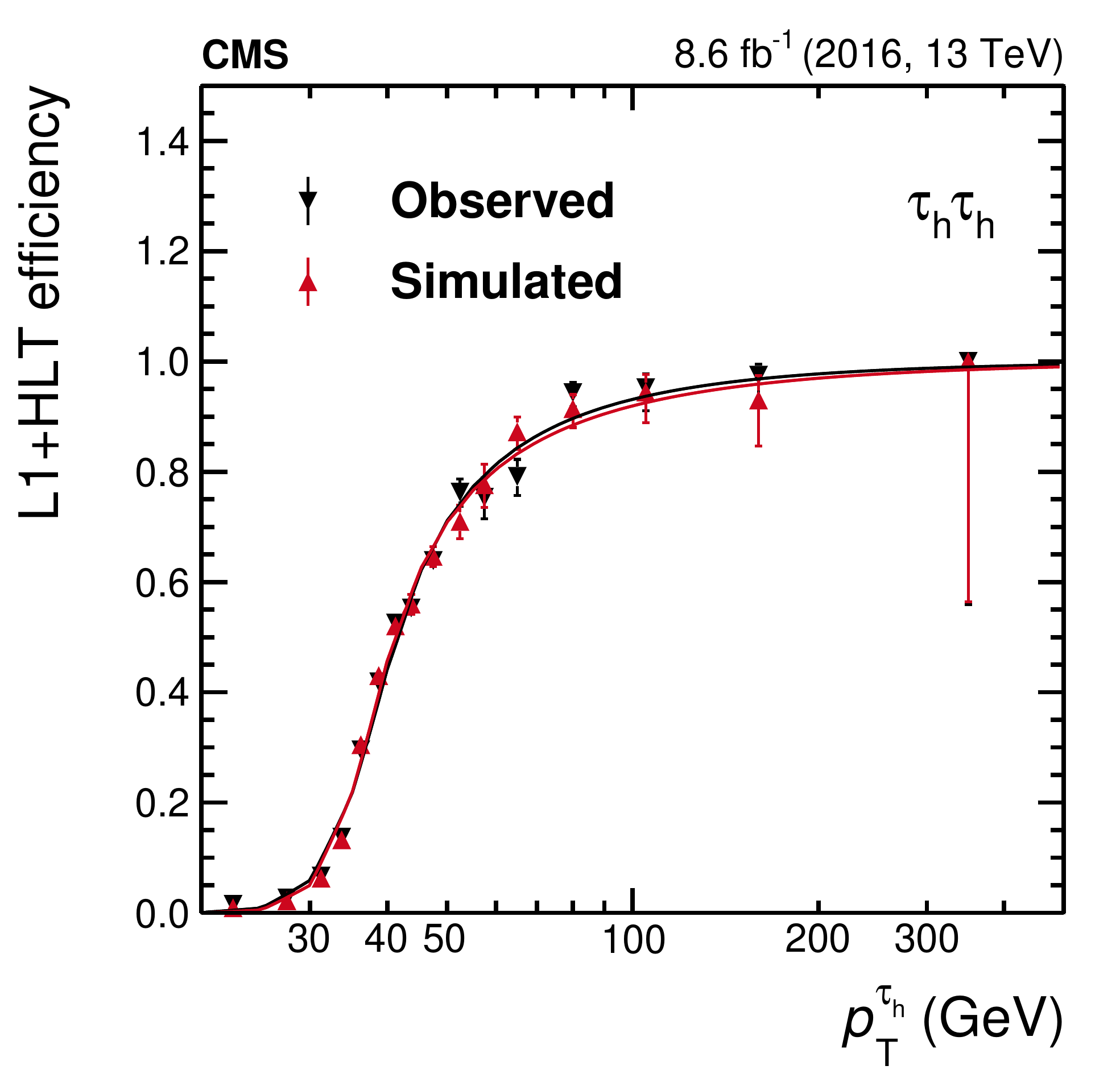}}
\caption{Single-$\tauh$ efficiency of the $\Pgm\tauh$ (left) and $\tauh\tauh$ (right) triggers.
The efficiency is computed per single $\tauh$, using the tag-and-probe method as a function of the offline-reconstructed $\pt^{\tauh}$. Observed data are compared to simulated $\cPZ/\Pggx\to\Pgt\Pgt$ events selected through the same procedure. Vertical bars correspond to the statistical uncertainties. The plot on the right has data points fitted using a cumulative (integral) distribution of the Crystal Ball function~\cite{SLAC-R-236}.}
\label{fig:efficiency_tautrg}

\end{figure}

The passing $\tauh$ probes correspond to those that pass the $\tauh$ part of the special trigger, \ie, the trigger is satisfied and its $\tauh$ part geometrically matches ($\Delta R<0.5$) the selected offline $\tauh$.
The efficiency of the $\tauh$ part of the $\Pgm\tauh$ and $\tauh\tauh$ triggers, measured using collision data relative to the DY simulation, is shown in Fig.~\ref{fig:efficiency_tautrg}. For the $\tauh\tauh$ trigger,
we use only the portion of the 2016 data that contains the trigger employing the combined isolation. In both cases, simulation agrees well with the data.
Data-to-simulation agreement is similar for the other triggers discussed in this section.

Figure~\ref{fig:efficiency_tautrg} shows that the nominal \pt threshold of the $\tauh$ triggers corresponds to an efficiency of 50\%, as expected for trigger and offline objects with the same energy scale.
The slow turn-on originates from two effects: in the \pt range above about twice the trigger threshold, it is caused by the relaxed isolation selection applied at HLT, but not in the offline selection;
in the range just above the trigger threshold, it is caused by an asymmetric energy response of the HLT $\tauh$ candidate relative to its offline counterpart. The asymmetry is due to a more inclusive selection of constituents of the $\tauh$ candidate at the HLT than offline. The second effect is clearly visible in the $\Pgm\tauh$ trigger with unseeded L3 $\tauh$ reconstruction, while for the $\tauh\tauh$ trigger it is smeared out by the resolution of the L1- and L2-$\tauh$ candidates (relative to offline), which is much worse than the resolution of the L3 candidates.

\section{Summary}
\label{sec:summary}

The ``hadron-plus-strips'' algorithm developed at the CMS experiment to reconstruct and identify $\Pgt\to\text{hadrons}+\Pnut$ decays in proton-proton collisions at $\sqrt{s}=7$ and 8\TeV, as presented in Ref.~\cite{TAU-14-001}, has been improved. The changes include a dynamic strip reconstruction, the reconstruction of highly boosted $\Pgt$ lepton pairs, and the introduction of additional variables in the multivariate-analysis discriminants used to reject jets and electrons. The isolation discriminants have also been optimized to cope with the large pileup of events in $\sqrt{s}=13\TeV$ proton-proton runs.

The performance of the improved algorithm has been measured using 35.9\fbinv of data recorded during 2016 at $\sqrt{s}=13\TeV$.
The $\tauh$ identification efficiency in data at low, intermediate, and high transverse momenta, as well as for highly Lorentz-boosted $\Pgt$ lepton pairs, is similar to that expected from Monte Carlo simulation, while differences of 10--20\% are found between data and simulation for the $\text{jet} \mapsto \tauh$ misidentification probability.
 The $\Pe \mapsto \tauh$ and $\Pgm \mapsto \tauh$ misidentification probabilities are smaller than those of the previous version of the algorithm under the same running conditions, while maintaining a high efficiency for the selection of genuine $\tauh$ candidates. The corresponding data-to-simulation scale factors have also been determined.
The energy scale of $\tauh$ candidates is measured, and its response relative to Monte Carlo simulation is found to be close to unity.
Finally, a specialized $\tauh$ reconstruction and identification algorithm has been used in the high-level trigger, and its performance has been presented.

\clearpage
\begin{acknowledgments}
\hyphenation{Bundes-ministerium Forschungs-gemeinschaft Forschungs-zentren Rachada-pisek} We congratulate our colleagues in the CERN accelerator departments for the excellent performance of the LHC and thank the technical and administrative staffs at CERN and at other CMS institutes for their contributions to the success of the CMS effort. In addition, we gratefully acknowledge the computing centres and personnel of the Worldwide LHC Computing Grid for delivering so effectively the computing infrastructure essential to our analyses. Finally, we acknowledge the enduring support for the construction and operation of the LHC and the CMS detector provided by the following funding agencies: the Austrian Federal Ministry of Education, Science and Research and the Austrian Science Fund; the Belgian Fonds de la Recherche Scientifique, and Fonds voor Wetenschappelijk Onderzoek; the Brazilian Funding Agencies (CNPq, CAPES, FAPERJ, FAPERGS, and FAPESP); the Bulgarian Ministry of Education and Science; CERN; the Chinese Academy of Sciences, Ministry of Science and Technology, and National Natural Science Foundation of China; the Colombian Funding Agency (COLCIENCIAS); the Croatian Ministry of Science, Education and Sport, and the Croatian Science Foundation; the Research Promotion Foundation, Cyprus; the Secretariat for Higher Education, Science, Technology and Innovation, Ecuador; the Ministry of Education and Research, Estonian Research Council via IUT23-4 and IUT23-6 and European Regional Development Fund, Estonia; the Academy of Finland, Finnish Ministry of Education and Culture, and Helsinki Institute of Physics; the Institut National de Physique Nucl\'eaire et de Physique des Particules~/~CNRS, and Commissariat \`a l'\'Energie Atomique et aux \'Energies Alternatives~/~CEA, France; the Bundesministerium f\"ur Bildung und Forschung, Deutsche Forschungsgemeinschaft, and Helmholtz-Gemeinschaft Deutscher Forschungszentren, Germany; the General Secretariat for Research and Technology, Greece; the National Research, Development and Innovation Fund, Hungary; the Department of Atomic Energy and the Department of Science and Technology, India; the Institute for Studies in Theoretical Physics and Mathematics, Iran; the Science Foundation, Ireland; the Istituto Nazionale di Fisica Nucleare, Italy; the Ministry of Science, ICT and Future Planning, and National Research Foundation (NRF), Republic of Korea; the Ministry of Education and Science of the Republic of Latvia; the Lithuanian Academy of Sciences; the Ministry of Education, and University of Malaya (Malaysia); the Ministry of Science of Montenegro; the Mexican Funding Agencies (BUAP, CINVESTAV, CONACYT, LNS, SEP, and UASLP-FAI); the Ministry of Business, Innovation and Employment, New Zealand; the Pakistan Atomic Energy Commission; the Ministry of Science and Higher Education and the National Science Centre, Poland; the Funda\c{c}\~ao para a Ci\^encia e a Tecnologia, Portugal; JINR, Dubna; the Ministry of Education and Science of the Russian Federation, the Federal Agency of Atomic Energy of the Russian Federation, Russian Academy of Sciences, the Russian Foundation for Basic Research, and the National Research Center ``Kurchatov Institute''; the Ministry of Education, Science and Technological Development of Serbia; the Secretar\'{\i}a de Estado de Investigaci\'on, Desarrollo e Innovaci\'on, Programa Consolider-Ingenio 2010, Plan Estatal de Investigaci\'on Cient\'{\i}fica y T\'ecnica y de Innovaci\'on 2013-2016, Plan de Ciencia, Tecnolog\'{i}a e Innovaci\'on 2013-2017 del Principado de Asturias, and Fondo Europeo de Desarrollo Regional, Spain; the Ministry of Science, Technology and Research, Sri Lanka; the Swiss Funding Agencies (ETH Board, ETH Zurich, PSI, SNF, UniZH, Canton Zurich, and SER); the Ministry of Science and Technology, Taipei; the Thailand Center of Excellence in Physics, the Institute for the Promotion of Teaching Science and Technology of Thailand, Special Task Force for Activating Research and the National Science and Technology Development Agency of Thailand; the Scientific and Technical Research Council of Turkey, and Turkish Atomic Energy Authority; the National Academy of Sciences of Ukraine, and State Fund for Fundamental Researches, Ukraine; the Science and Technology Facilities Council, UK; the US Department of Energy, and the US National Science Foundation.

Individuals have received support from the Marie-Curie programme and the European Research Council and Horizon 2020 Grant, contract No. 675440 (European Union); the Leventis Foundation; the A. P. Sloan Foundation; the Alexander von Humboldt Foundation; the Belgian Federal Science Policy Office; the Fonds pour la Formation \`a la Recherche dans l'Industrie et dans l'Agriculture (FRIA-Belgium); the Agentschap voor Innovatie door Wetenschap en Technologie (IWT-Belgium); the F.R.S.-FNRS and FWO (Belgium) under the ``Excellence of Science - EOS'' - be.h project n. 30820817; the Ministry of Education, Youth and Sports (MEYS) of the Czech Republic; the Lend\"ulet (``Momentum'') Programme and the J\'anos Bolyai Research Scholarship of the Hungarian Academy of Sciences, the New National Excellence Program \'UNKP, the NKFIA research grants 123842, 123959, 124845, 124850 and 125105 (Hungary); the Council of Scientific and Industrial Research, India; the HOMING PLUS programme of the Foundation for Polish Science, cofinanced from European Union, Regional Development Fund, the Mobility Plus programme of the Ministry of Science and Higher Education, the National Science Center (Poland), contracts Harmonia 2014/14/M/ST2/00428, Opus 2014/13/B/ST2/02543, 2014/15/B/ST2/03998, and 2015/19/B/ST2/02861, Sonata-bis 2012/07/E/ST2/01406; the National Priorities Research Program by Qatar National Research Fund; the Programa de Excelencia Mar\'{i}a de Maeztu, and the Programa Severo Ochoa del Principado de Asturias; the Thalis and Aristeia programmes cofinanced by EU-ESF, and the Greek NSRF; the Rachadapisek Sompot Fund for Postdoctoral Fellowship, Chulalongkorn University, and the Chulalongkorn Academic into Its 2nd Century Project Advancement Project (Thailand); the Welch Foundation, contract C-1845; and the Weston Havens Foundation (USA).

\end{acknowledgments}

\clearpage

\bibliography{auto_generated}

\cleardoublepage \appendix\section{The CMS Collaboration \label{app:collab}}\begin{sloppypar}\hyphenpenalty=5000\widowpenalty=500\clubpenalty=5000\vskip\cmsinstskip
\textbf{Yerevan Physics Institute, Yerevan, Armenia}\\*[0pt]
A.M.~Sirunyan, A.~Tumasyan
\vskip\cmsinstskip
\textbf{Institut f\"{u}r Hochenergiephysik, Wien, Austria}\\*[0pt]
W.~Adam, F.~Ambrogi, E.~Asilar, T.~Bergauer, J.~Brandstetter, M.~Dragicevic, J.~Er\"{o}, A.~Escalante~Del~Valle, M.~Flechl, R.~Fr\"{u}hwirth\cmsAuthorMark{1}, V.M.~Ghete, J.~Hrubec, M.~Jeitler\cmsAuthorMark{1}, N.~Krammer, I.~Kr\"{a}tschmer, D.~Liko, T.~Madlener, I.~Mikulec, N.~Rad, H.~Rohringer, J.~Schieck\cmsAuthorMark{1}, R.~Sch\"{o}fbeck, M.~Spanring, D.~Spitzbart, A.~Taurok, W.~Waltenberger, J.~Wittmann, C.-E.~Wulz\cmsAuthorMark{1}, M.~Zarucki
% ...
\vskip\cmsinstskip
% ...
1:  Also at Vienna University of Technology, Vienna, Austria\\
% ...
\end{sloppypar}
\end{document}